\documentclass[fleqn,usenatbib]{mnras}

\usepackage{newtxtext,newtxmath}
\usepackage[T1]{fontenc}

\DeclareRobustCommand{\VAN}[3]{#2}
\let\VANthebibliography\thebibliography
\def\thebibliography{\DeclareRobustCommand{\VAN}[3]{##3}\VANthebibliography}

\usepackage{orcidlink}
\usepackage{graphicx}   
\usepackage{amsmath}    
\usepackage{xcolor}
\usepackage[normalem]{ulem}

\newcommand{\Lp}{L_{\mathrm p}}
\newcommand{\Lth}{L_{30}}
\newcommand{\mysim}{\mathord{\sim}}
\newcommand{\mylesssim}{\mathord{\lesssim}}
\newcommand{\mygtrsim}{\mathord{\gtrsim}}
\newcommand{\myapprox}{\mathord{\approx}}

\title[Viewing-angle diffusion models]{An analytic diffusion model for viewing-angle effects in explosive transients}

\author[van der Boom \& Kushnir]{
Daniella van der Boom$^{\orcidlink{0009-0008-4872-3116} 1}$\thanks{E-mail: daniella.van-der-boom@weizmann.ac.il}
and Doron Kushnir$^{1}$
\\
$^{1}$Department of Particle Physics and Astrophysics, Weizmann Institute of Science, Rehovot 76100, Israel
}

\date{Accepted XXX. Received YYY; in original form ZZZ}
\pubyear{\the\year{}}

\begin{document}
\label{firstpage}
\pagerange{\pageref{firstpage}-\pageref{lastpage}}
\maketitle

\begin{abstract}
The tight observed relation between the peak bolometric luminosity, $\Lp$, and the width ratio $\Lth/\Lp$ of Type Ia supernovae, with $\Lth$ the luminosity 30\,d after explosion, makes viewing-angle effects a useful test of explosion models. We develop an analytic diffusion model for homologously expanding asymmetric ejecta with uniform density and grey opacity, based on a Green's-function solution and an Eddington surface-emission prescription, and obtain angle-dependent light curves for a displaced spherical source and for oblate ejecta. We validate them against Monte Carlo (MC) radiative transfer and derive geometric approximations to the peak-luminosity anisotropy. For a displaced spherical source, the quasi-steady approximation, equivalent to stationary diffusion in frozen ejecta, gives $\Lp(\theta)/\Lp^{\rm avg}\simeq1+(17/8)\,z_{\rm off}\cos\theta$, where $z_{\rm off}$ is the displacement in units of the ejecta radius, $\theta$ the viewing angle from the displacement axis, and $\Lp^{\rm avg}$ the angle-averaged peak. This dipole is $\myapprox10$ per cent below the finite-time value in the tested configurations. For oblate ejecta, the leading term is quadrupolar and is set by a single shape parameter. Effective models of 19 off-centre sub-Chandrasekhar explosions recover the MC peak times, peak luminosities, and width ratios to median errors of 3, 4, and 6 per cent, and the extent in $\Lp$ of the MC tracks, but underestimate the width-ratio variation with viewing angle by a factor of 2--3. An oblate construction captures the scale of peak luminosity variations in six equal-mass white-dwarf collisions. The model tracks remain inconsistent with the observed distribution, and for asymmetric models the transverse viewing-angle variation adds to the tension.
\end{abstract}



\section{Introduction}
\label{sec:introduction}

\begin{figure*}
    \centering
    \includegraphics[width=\linewidth]{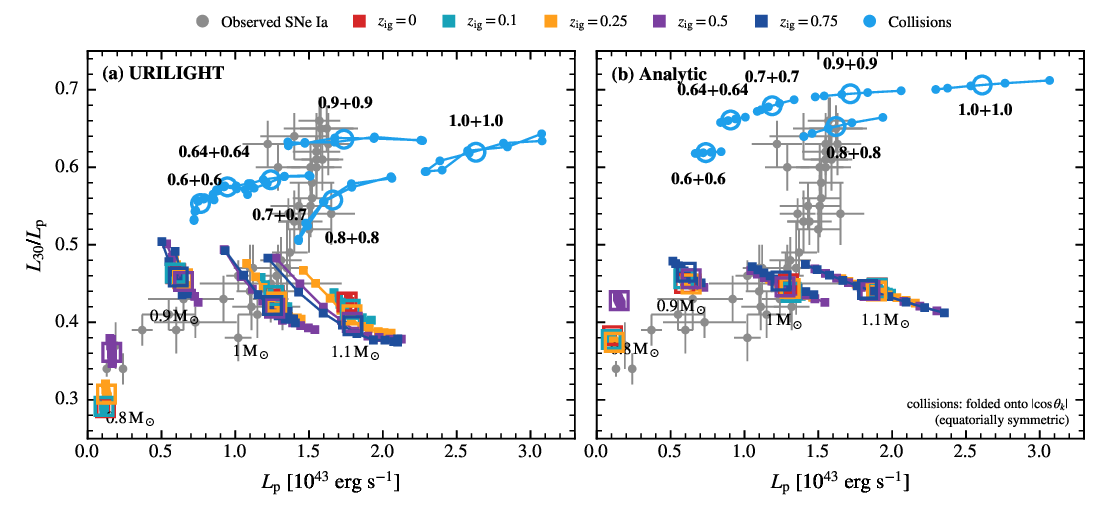}
    \caption{Viewing-angle tracks in the luminosity--width plane from (a) \textsc{URILIGHT} Monte Carlo RT \citep{Wygoda_2019a} and (b) the analytic toy models of this work, over the observed SNe~Ia of \citet{sharon_2025} (grey, $1\sigma$ error bars). Squares: the $19$ off-centre ignition models (Table~\ref{tab:example_table}), coloured by the ignition offset $z_{\rm ig}$ (in WD radii) and labelled by ejecta mass $M$; circles: the six equal-mass WD collisions (Table~\ref{tab:collision_table}), labelled by the WD masses in $M_\odot$. Small symbols are the ten bins in $\cos\theta_k$, where $\theta_k$ is the angle between the line of sight and the symmetry axis ($\cos\theta_k=+1$ faces the displaced ${}^{56}$Ni; the collision axis for the collisions), joined into a track; the large open symbol is the angle-averaged light curve. The \textsc{URILIGHT} collision tracks nearly retrace themselves (approximate equatorial symmetry), and the exactly symmetric analytic ones are folded onto $|\cos\theta_k|$. Toward the ${}^{56}$Ni-facing pole the off-centre tracks become brighter and narrower, moving across the observed sequence. The same tracks are shown in magnitude space in Fig.~\ref{fig:Lp_L30Lp_obs_mag}.}
    \label{fig:Lp_L30Lp_obs}
\end{figure*}

The peak bolometric luminosity $\Lp$ of Type~Ia supernovae (SNe~Ia) correlates tightly with the width ratio $\Lth/\Lp$, where $\Lth$ is the bolometric luminosity $30$ days after explosion \citep{sharon_2025}. This early-time relation complements the relation between the synthesized $^{56}$Ni mass, $M_{{}^{56}\rm Ni}$, and the $\gamma$-ray escape time, $t_0$ \citep{Wygoda_2019a,Sharon_2020a}, which requires late-time observations ($t\mysim100$~d) and, for multidimensional models, radiative-transfer (RT) calculations whose uncertainties grow as the ejecta become optically thin (different RT codes applied to identical ejecta diverge increasingly with time, even in the bolometric light curve; \citealt{Blondin_2022}). Because $\Lp$ and $\Lth/\Lp$ correlate closely with $M_{{}^{56}\rm Ni}$ and $t_0$, the early-time relation carries similar information about the explosion while using only the first $30$~d, when the ejecta are optically thick, and the RT codes agree much better. An asymmetric explosion occupies a track in the luminosity--width plane as the viewing angle changes. The off-centre ignition models in Fig.~\ref{fig:Lp_L30Lp_obs} move across the observed sequence: viewing the nickel-rich side makes them brighter and reduces $\Lth/\Lp$. We seek an analytic relation between this variation and the source displacement or the ejecta shape.

Fig.~\ref{fig:Lp_L30Lp_obs} reproduces the observed relation together with the viewing-angle tracks of the models studied here. The observed sample separates into two regimes: faster-declining events with $\Lth/\Lp<0.5$, which form a rising sequence spanning $\Lp\myapprox0.1$--$1.4\times10^{43}\,\mathrm{erg\,s^{-1}}$, and slower, brighter events with $\Lth/\Lp>0.5$, which cluster at $\Lp\myapprox1.2$--$1.65\times10^{43}\,\mathrm{erg\,s^{-1}}$; the sub-Chandrasekhar models remain at $\Lth/\Lp\mylesssim0.5$ at every viewing angle, whereas most of the collision models occupy $\Lth/\Lp\mygtrsim0.5$. The angle-averaged properties of these models are in tension with the observations \citep{sharon_2025}, and their viewing-angle tracks are long: $\Lp$ changes by tens of per cent between the extreme viewing angles of the asymmetric models of both families (Sections~\ref{sec:toy_model}--\ref{sec:collisions}). Whether such transverse variations are consistent with the observed distribution is a question of the residuals about the relation (Section~\ref{sec:discussion}); answering it requires both reliable RT calculations of the spread and an understanding of its physical origin, which properties of the ejecta control it and how strongly. This is the main motivation for the present work.

We focus on the two families of multidimensional models analysed by \citet{sharon_2025}. The first is the off-centre ignition of sub-Chandrasekhar white dwarfs (WDs), motivated by the double-detonation scenario, in which a helium-shell detonation drives a converging shock that ignites the carbon--oxygen (CO) core off centre \citep{Nomoto_1982a,Nomoto_1982b,Livne_1990,Woosley_Weaver_1994,Fink_2010,Shen_2014}, possibly for thin helium shells \citep{Bildsten_2007,Fink_2007,Moore_2013,Shen_Moore_2014,Townsley_2019,Boos_2021}. The models used by \citet{sharon_2025}, and here, are those of \citet{Schinasi_2024}, who isolate the effect of the off-centre ignition by igniting bare CO WDs at prescribed points along the symmetry axis; the ejecta are approximately spherical, while the ${}^{56}$Ni distribution is displaced along the axis. The second family is the head-on collision of two WDs \citep{Rosswog_2009,Raskin_2010,Kushnir_2013}, for which \citet{sharon_2025} used the two-dimensional models of \citet{Kushnir_2013}; here the ejecta themselves are flattened along the collision axis, together with the ${}^{56}$Ni-rich region. We represent the two cases by a displaced spherical source inside a spherical boundary and by concentric oblate source and boundary spheroids, respectively.

Numerical RT studies have established substantial viewing-angle effects in asymmetric supernova ejecta, for ellipsoidal ejecta \citep{Hoflich_1991,Sim_2007}, two-dimensional detonating failed-deflagration models \citep{Kasen_2007} and head-on WD collisions \citep{sharon_2025}, as well as for aspherical kilonova ejecta \citep{Darbha_2020}. In particular, \citet{Sim_2007b} found nearly linear angular variations of the peak magnitude and rise time for an offset ${}^{56}$Ni region in uniform-density ejecta. Analytic radioactive light-curve models describe the competition between heating, diffusion and expansion \citep{Colgate_McKee_1969,Arnett_1982,Pinto_Eastman_2000a,Pinto_Eastman_2000b}: Arnett's solution retains the fundamental diffusion mode, with the corollary that the luminosity at maximum equals the instantaneous deposition rate, and \citet{Pinto_Eastman_2000a} showed with the full set of eigenmodes how the location of the ${}^{56}$Ni shapes the rise. The time-weighted energy constraint of \citet{Katz_2013} is independent of geometry and of the diffusion approximation, and was used by \citet{Khatami_Kasen_2019} to relate the peak time and the peak luminosity, while \citet{Kushnir_2019} mapped diffusion in homologous expansion onto a static problem in general geometry and obtained explicit spherical solutions by convolving the eigenmode impulse response with the deposition history. \citet{Darbha_2020} also give a simple analytic estimate of directional light curves from the projected photospheric area. Here we derive explicit eigenfunction diffusion solutions for the two asymmetric geometries.

We solve diffusion in homologously expanding uniform-density ejecta for a displaced spherical source and for concentric oblate source and boundary geometries, and an Eddington surface-emission law projects the emergent flux toward an observer. The leading quasi-steady signal is dipolar for a displaced sphere and quadrupolar for oblate ejecta; its accuracy at peak is tested against the time-dependent solutions. We caution that the displacement $z_{\rm off}$ is an effective parameter of a uniform-density toy model rather than a measure of the ${}^{56}$Ni centroid of a stratified ejecta (Section~\ref{sec:toy_construction}). Sections~\ref{sec:diffusion_framework}--\ref{sec:limb_darkening} develop and validate the framework. Sections~\ref{sec:toy_model}--\ref{sec:collisions} test effective representations of numerical explosions, separating agreement in peak luminosity from the larger errors in width-ratio variation: as Fig.~\ref{fig:Lp_L30Lp_obs} shows, the analytic tracks reproduce the location, direction, and extent in $\Lp$ of the \textsc{URILIGHT} tracks, and the dominant systematic mismatch is in the width-ratio variation. Section~\ref{sec:discussion} discusses the implications and limits; the derivations and complete model results are given in the appendices.

\section{The Analytic Diffusion Framework}
\label{sec:diffusion_framework}
Following \citet{Kushnir_2019}, we describe the diffusion of radiation energy within homologously expanding ejecta using velocity coordinates $\mathbf{v}=\mathbf{x}/t$. By introducing scaled variables
\begin{equation}
\bar{e} = t^4 e, \qquad
\bar{\rho}_v = t^3 \rho, \qquad
\bar{\epsilon} = t^3 \epsilon, \qquad
\bar{t} = \tfrac{1}{2} t^2,
\label{eq:scaled_vars}
\end{equation}
where $e$ is the radiation energy density per unit volume, $\rho$ the mass density, $\epsilon$ the local energy generation rate per unit volume, and $\kappa$ the effective opacity; the explicit time dependence from adiabatic expansion is absorbed. The problem transforms into a static diffusion equation. For the uniform velocity-space density and constant grey opacity adopted in the analytic solutions below, the diffusion coefficient $D\equiv c/(3\kappa\bar\rho_v)$ is constant:
\begin{equation}
\frac{\partial \bar{e}}{\partial \bar{t}}
-D\,\nabla_{\mathbf{v}}^2 \bar{e}
= \bar{\epsilon}, \qquad D\equiv\frac{c}{3\kappa\bar\rho_v}.
\end{equation}
The scaled global energy generation rate $\bar{Q}$, luminosity $\bar{L}$, and total radiation energy $\bar{E}$ are related to their physical counterparts by $\bar{Q}=Q$, $\bar{L}=L$, and $\bar{E}=Et$.

To solve for an arbitrary heating source, we define the impulse response. When energy deposition occurs over a negligibly short interval at $\bar{t}_\delta$, the source term is a delta function,
\begin{equation}
\bar{\epsilon}(\mathbf{v}, \bar{t}) =
\bar{\varepsilon}_\delta(\mathbf{v})\,\delta(\bar{t} - \bar{t}_\delta).
\end{equation}
Here $\bar{\varepsilon}_\delta(\mathbf{v})$ is the spatial distribution of injected energy in velocity space, normalized to unit scaled energy, $\int\bar{\varepsilon}_\delta\,d^3v=1$, so that the impulse response below is the response to unit deposited energy (in Appendix~\ref{app:analytical_derivation} the impulse carries an arbitrary scaled energy $\bar E_\delta$, which divides out of the normalized response). For $\bar{t}>\bar{t}_\delta$, the system evolves according to the homogeneous diffusion equation with initial condition $\bar{e}(\mathbf{v},\bar{t}_\delta)=\bar{\varepsilon}_\delta(\mathbf{v})$. This impulse solution serves as the Green's function of the system. For a continuous, time-dependent heating rate $Q(t)$ with a fixed spatial distribution, the total luminosity is constructed via the convolution
\begin{equation}
L(\bar{t})=\int_0^{\bar{t}}d\bar{t}'\, Q(\bar{t}')\,L_\delta(\bar{t}-\bar{t}'),
\label{eq:convolution}
\end{equation}
where $L_\delta(\bar{t})$ is the impulse-response luminosity.

The analytic solutions below assume a uniform density in velocity space, a constant grey opacity, and a sharp ejecta edge, on which we impose the absorbing (Dirichlet) condition $\bar e=0$, the standard optically thick approximation for a free surface. The diffusion operator on this bounded domain has a discrete set of eigenfunctions, so the solution is an eigenfunction expansion in which each mode decays exponentially in $\bar t$ at a rate set by its eigenvalue; the general solution and the two geometries are derived in Appendix~\ref{app:analytical_derivation}, and the quasi-steady angular pattern reached under continuous heating in Appendix~\ref{app:ld_multipoles}. Both the diffusion equation and the Eddington closure introduced below describe optically thick transport. For uniform-density ejecta the centre-to-surface optical depth is $\tau(t)=3\kappa M/(4\pi v_o^2t^2)$, which for the fiducial parameters of Section~\ref{sec:mc_validation} ($M\myapprox1\,M_\odot$, $\kappa=0.2\,\mathrm{cm^2\,g^{-1}}$, $v_o\myapprox1.29\times10^{4}\,\mathrm{km\,s^{-1}}$) is $\myapprox34$ at $15$~d, $\myapprox8$ at $30$~d and unity at $\myapprox90$~d. The description is therefore accurate around maximum light and degrades gradually as the ejecta approach transparency; what fails first is the angular structure, since the RT viewing-angle contrast must decay as photons free-stream, whereas the diffusion solution holds it near its quasi-steady pattern, while the angle-averaged luminosity is expected to remain more accurate because both descriptions approach $L\to Q$ asymptotically (an asymptotic statement, not a guarantee at every epoch; Appendix~\ref{app:validation_details} and Section~\ref{sec:toy_comparison} quantify the residuals).

\subsection{Radioactive Heating}
\label{sec:heating}

The total energy deposition rate is $Q(t)=Q_\gamma(t)f_{\rm dep}(t)+Q_{\rm pos}(t)$, where $Q_\gamma(t)$ is the direct $\gamma$-ray component, $Q_{\rm pos}(t)$ accounts for positron kinetic energy, and $f_{\rm dep}(t)$ is the $\gamma$-ray deposition fraction. The positron kinetic energy is assumed to be deposited locally and in full. The radioactive heating rates are \citep[following][]{Wygoda_2019a}, with lifetimes from \citet{Junde_1999} and mean decay energies from \citet{Ambwani_1988},
\begin{equation}
Q_\gamma(t)=\frac{M_{{}^{56}\rm Ni}}{M_\odot}
\left[6.45\,e^{-t/8.76\,{\rm d}} + 1.38\,e^{-t/111.4\,{\rm d}}\right]
\times10^{43}\,\text{erg\,s}^{-1},
\end{equation}
\begin{equation}
Q_{\rm pos}(t)=4.64\,\frac{M_{{}^{56}\rm Ni}}{M_\odot}
\left[e^{-t/111.4\,{\rm d}}-e^{-t/8.76\,{\rm d}}\right]
\times10^{41}\,\text{erg\,s}^{-1},
\end{equation}
where $M_{{}^{56}\rm Ni}$ is the initial ${}^{56}$Ni mass. We adopt the
$\gamma$-ray deposition fraction from \citet{Jeffery_1999},
\begin{equation}
f_{\rm dep}(t)=1-e^{-t_0^2/t^2},
\label{eq:f_dep}
\end{equation}
where $t_0$ is the characteristic $\gamma$-ray escape time. The limit $t_0\rightarrow\infty$ corresponds to complete deposition, $f_{\rm dep}=1$.

\subsection{Viewing-Angle Luminosity: Eddington Grey-Atmosphere Closure}
\label{sec:viewing_angle}

The diffusion solution provides the impulse surface flux $F_\delta(\mathbf{r}_s, t)$ emerging from each point $\mathbf{r}_s$ on the sharp ejecta boundary $\partial \mathcal{S}$. To determine the luminosity seen by a distant observer at a specific orientation, we relate the local flux to the emergent specific intensity using the Eddington grey-atmosphere closure. Assuming a quasi-static, semi-infinite grey atmosphere, the specific intensity at surface point $\mathbf{r}_s$ is
\begin{equation}
I_\delta(\mathbf{r}_s, \mu, t) =
\frac{3}{4\pi} F_\delta(\mathbf{r}_s, t)
\left( \mu + \frac{2}{3} \right),
\end{equation}
where $\mu = \hat{\mathbf{n}} \cdot \hat{\mathbf{k}}$ is the direction cosine between the outward surface normal $\hat{\mathbf{n}}$ and the unit vector toward the observer $\hat{\mathbf{k}}$. Relativistic Doppler shifts and aberration are neglected in accordance with the non-relativistic diffusion treatment ($\mathbf{v}\ll c$).

We consider a distant observer in the direction
\begin{equation}
\hat{\mathbf{k}} = \sin\theta_k\,\hat{\mathbf{x}} + \cos\theta_k\,\hat{\mathbf{z}},
\end{equation}
where $\theta_k$ is the viewing angle relative to the symmetry axis. Azimuthal symmetry allows us to set $\phi_{k}=0$ without loss of generality.

The isotropic equivalent luminosity inferred by this observer is obtained by integrating the emergent intensity over the visible ejecta surface:
\begin{equation}
L_{\delta,\text{iso}}(t, \theta_k) =
4\pi \int_{\partial \mathcal{S}_{\text{vis}}}
I_\delta(\mathbf{r}_s, \mu, t)\,(\hat{\mathbf{n}} \cdot \hat{\mathbf{k}})\,dA,
\end{equation}
where $\partial \mathcal{S}_{\text{vis}}$ is the visible hemisphere defined by $\mu>0$. Here $L_{\delta,\text{iso}}$ is the luminosity a distant observer would infer under the assumption of isotropic emission; for aspherical ejecta or an off-centre source, it is not the true bolometric luminosity, and only its average over all viewing directions recovers the latter (Section~\ref{sec:consistency}).

\label{sec:spherical_geom}

For spherical ejecta, in spherical coordinates $(r,\theta,\phi)$, the outward unit normal at the ejecta boundary $r=R(t)=v_o t$ is simply $\hat{\mathbf{n}}=\hat{\mathbf{e}}_r$. The direction cosine toward an observer at $\hat{\mathbf{k}}=(\sin\theta_k,0,\cos\theta_k)^T$ is
\begin{equation}
\mu(\theta, \phi;\, \theta_k) =
\sin\theta\cos\phi\sin\theta_k + \cos\theta\cos\theta_k,
\end{equation}
and the surface element is $dA = R^2(t)\sin\theta\,d\theta\,d\phi$. The visibility condition $\mu>0$ defines a perfect geometric hemisphere centred on the observer's line of sight.

\label{sec:consistency}

The standard, angle-averaged impulse bolometric luminosity is recovered by averaging over all observer orientations:
\begin{equation}
L_\delta^{\rm avg}(t) = \frac{1}{4\pi}
\int_{4\pi} L_{\delta,\text{iso}}(t, \theta_k)\,d\Omega_k.
\end{equation}
Substituting the definition of $L_{\delta, \rm iso}$ and the relation
$F_\delta=\int(\hat{\mathbf{n}}\cdot\hat{\mathbf{k}})I_\delta\,d\Omega_k$
recovers the standard total impulse luminosity:
\begin{equation}
L_\delta^{\rm avg}(t) = \oint_{\partial \mathcal{S}} F_\delta(\mathbf{r}_s, t)\,dA
= L_\delta(t).
\end{equation}
This confirms that the directional integration is globally consistent with diffusion theory: $L_{\rm iso}$ accounts for local surface-flux variations and the geometric projection of aspherical ejecta, while its orientation average remains constrained by the total energy emerging from $\partial\mathcal{S}$.

\subsection{Parameters of the analytic solution}
\label{sec:analytic_params}

Before applying the framework to numerical ejecta, we state the parameters that completely specify the solution for the two geometries used in this work; we refer to them as the toy-model parameters throughout. The ${}^{56}$Ni region is confined to a sharply bounded sub-volume within which the initial energy-deposition density is uniform.

For the off-centre spherical case, consider spherical ejecta of outer velocity $v_o$ containing a spherical ${}^{56}$Ni source of radius $x_sv_o$ displaced along the symmetry axis by $z_{\rm off}v_o$. The solution is fully specified by five parameters: the fractional source radius $x_s\in(0,1]$; the fractional axial displacement $z_{\rm off}$; the diffusion time-scale $\tilde{t}$, defined through $\tilde{t}^{\,2}\equiv v_o^2/D$; the ${}^{56}$Ni mass $M_{{}^{56}\rm Ni}$; and the $\gamma$-ray escape time $t_0$. The source is contained within the ejecta only for $x_s+|z_{\rm off}|\le1$. Negative $z_{\rm off}$ is handled by reflection: the solution for $-z_{\rm off}$ follows from that for $+|z_{\rm off}|$ by $\theta\to\pi-\theta$, so that the odd multipoles change sign, and the radial bounds of Appendix~\ref{app:offcentre_diffusion} are written for $z_{\rm off}\ge0$. The impulse response depends on the geometry through $(x_s,z_{\rm off})$ and on time only through the ratio $t/\tilde{t}$, so $\tilde{t}$ is the single time-scale of the homogeneous diffusion problem. The remaining two parameters enter exclusively through the convolution with the heating rate (Equation~\ref{eq:convolution}): $M_{{}^{56}\rm Ni}$ as the overall normalization of $Q(t)$, and $t_0$ through the deposition fraction $f_{\rm dep}(t)$ (Equation~\ref{eq:f_dep}). 

For oblate spheroidal ejecta the two spherical shape parameters are replaced by three: the ejecta boundary coordinate $\xi_s$, equivalently the axis ratio $v_{\rm pol}/v_{\rm eq}=\tanh\xi_s$; the source boundary coordinate $\xi_1$; and the ratio $f$ between the focal length of the source spheroid and the focal length $a$ of the ejecta spheroid. The diffusion time-scale is then defined by $\tilde{t}^{\,2}\equiv a^2/D$. The source is concentric with the ejecta and centred on the equatorial plane, so that $z_{\rm off}=0$ and the viewing-angle dependence arises entirely from the aspherical boundary and heating source. 

\section{Validation against Monte Carlo Radiative Transfer}
\label{sec:mc_validation}

We first test the transport treatment on idealized ejecta with the same density, opacity, source geometry, and deposition in both calculations, so that any difference measures the accuracy of the diffusion solution and the grey-atmosphere closure alone.

The reference calculations are performed with the \textsc{URILIGHT} code \citep{Wygoda_2019a, Wygoda_2019b}, a multidimensional MC RT code based on the approximations of the \textsc{SEDONA} code \citep{Kasen_2006}. \textsc{URILIGHT} assumes homologous expansion, local thermodynamic equilibrium (LTE) ionization and excitation, and Sobolev expansion opacities, with a thermalization parameter of $\epsilon_{\rm th}=0.8$. Because it neglects non-thermal ionization, it is expected to provide reliable bolometric light curves primarily during the first $t\le30$ days after explosion \citep{sharon_2025}. This limitation applies only to the explosion models of Sections~\ref{sec:models}-\ref{sec:collisions}, where the opacity is derived from the LTE ionization and excitation state, and not to the idealized calculations presented here, which adopt a constant grey opacity.

Both comparisons below are performed with a complete $\gamma$-ray deposition, $f_{\rm dep}=1$, imposed consistently in the analytic calculation and in \textsc{URILIGHT}. In addition, we verified that applying the Green's-function solution to a spatially resolved source reproduces the light curves of the above-mentioned \textsc{URILIGHT} calculation, confirming that the impulse-response formulation can be used to construct the solution for arbitrary source distributions.

%
\subsection{Off-centre spherical source}
\label{sec:mc_validation_sphere}
\begin{figure}
    \centering
    \includegraphics[width=0.9\linewidth]{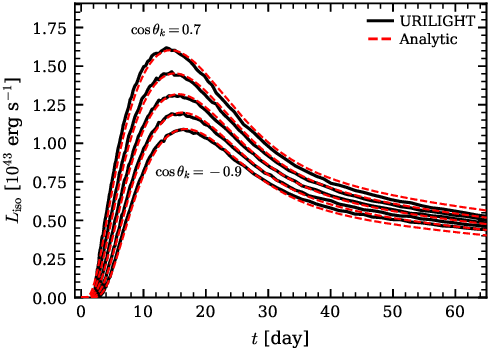}
    \caption{Validation of the analytic off-centre diffusion solution (red dashed) against \textsc{URILIGHT} (solid black) for idealized uniform-density spherical ejecta with the parameters of Section~\ref{sec:mc_validation_sphere} and complete $\gamma$-ray deposition. The isotropic-equivalent luminosity $L_{\rm iso}$ is shown for five of the ten viewing-angle bins ($\cos\theta_k=-0.9,-0.5,-0.1,+0.3,+0.7$), with $\cos\theta_k=+1$ facing the displaced ${}^{56}$Ni source; the curves brighten monotonically from $-0.9$ (labelled below its peak) to $+0.7$ (labelled above it).}

    \label{fig:diff_sol_zoff_01_fdep1}
\end{figure}

Fig.~\ref{fig:diff_sol_zoff_01_fdep1} compares the analytic light curves with
\textsc{URILIGHT} for an off-centre spherical ${}^{56}$Ni source with $x_s=0.8435$, $z_{\rm off}=0.0999$, $\tilde{t}=31.503\,\mathrm{d}$ and $M_{{}^{56}\rm Ni}=0.604\,M_\odot$, with $f_{\rm dep}=1$; the diffusion time-scale follows from $M=1.006\,M_\odot$, $v_o=1.291\times10^{4}\,\mathrm{km\,s^{-1}}$ (corresponding to $\bar{\rho}_v = 0.112\,M_\odot\,(10^4\,\mathrm{km\,s^{-1}})^{-3}$ and $\kappa=0.2\,\mathrm{cm^2\,g^{-1}}$). These are the exact parameters of the run; its solver configuration file is distributed with the data. The \textsc{URILIGHT} luminosities are collected in ten viewing-angle bins, uniform in $\cos\theta_k$ over $[-1,1]$ (bin centres $\cos\theta_k=\pm0.1,\pm0.3,\pm0.5,\pm0.7,\pm0.9$), and the analytic solution is evaluated at the bin centres; this binning is used throughout the paper.

The observables are measured separately in each viewing-angle bin, $\mu_k=\cos\theta_k$. The peak luminosity $\Lp(\mu_k)=L_{\rm iso}(t_{\rm p}(\mu_k),\mu_k)$ is evaluated at the peak time of that bin, $t_{\rm p}(\mu_k)$, which depends on the viewing angle (for the calculation of Fig.~\ref{fig:diff_sol_zoff_01_fdep1}, $t_{\rm p}$ decreases from $16.4$~d at $\cos\theta_k=-0.9$ to $13.2$~d at $+0.9$ in \textsc{URILIGHT}, and from $16.8$ to $14.0$~d in the analytic solution), whereas $\Lth(\mu_k)=L_{\rm iso}(30\,{\rm d},\mu_k)$ is evaluated at the same time after explosion for all bins. For each calculation we also form the angle-averaged light curve, $\langle L\rangle(t)=\tfrac12\int_{-1}^{1}L_{\rm iso}(t,\mu_k)\,d\mu_k$. For the analytic solution, this is the angle-integrated luminosity $L(t)$ of Section~\ref{sec:consistency}; for \textsc{URILIGHT} it is the luminosity of all escaping packets, which equals the mean of the ten bin luminosities because the bins subtend equal solid angles. The angle-averaged peak luminosity and width ratio, $\Lp^{\rm avg}=\max_t\langle L\rangle$ and $(\Lth/\Lp)^{\rm avg}=\langle L\rangle(30\,{\rm d})/\Lp^{\rm avg}$, are measured from $\langle L\rangle(t)$. Because the bins peak at different times, the mean of the ten per-bin peak luminosities exceeds $\Lp^{\rm avg}$, by $0.9$ per cent for the \textsc{URILIGHT} calculation of Fig.~\ref{fig:diff_sol_zoff_01_fdep1} and by less than $0.7$ per cent for every ejecta model of Sections~\ref{sec:toy_model} and~\ref{sec:collisions}. For the analytic solution, a second, smaller difference arises: $L_{\rm iso}$ is evaluated at the bin centres rather than averaged over each bin, so the equal-weight mean of the ten bins differs from the exact angle integral. For $\Lp(\mu)/\Lp^{\rm avg}=1+\sum_lb_lP_l(\mu)$ (notation of Fig.~\ref{fig:limb_darkening}) the dipole term cancels and, to leading order, the mean is lower by $0.005\,b_2$. This is at most $0.14$ per cent for the ejecta models ($b_2\le0.28$) and about $0.25$ per cent for the oblate test of Fig.~\ref{fig:diff_sol_oblate_fdep1}. The viewing-angle variation of an observable $Y$ is measured by the outer-bin extent, $\Delta Y=|Y(0.9)-Y(-0.9)|$ for a displaced source and $|Y(0.9)-Y(0.1)|$ for ejecta symmetric about the equatorial plane, normalized by $Y^{\rm avg}$ where stated. For $Y=\Lp$ the two bins are therefore compared at their own peak times, $\Delta\Lp=|L_{\rm iso}(t_{\rm p}(0.9),0.9)-L_{\rm iso}(t_{\rm p}(-0.9),-0.9)|$, and not at a common time. The analytic light curves reproduce the MC results over the full range of viewing angles. Across the ten bins the largest $\Lp$ exceeds the smallest by $56$ per cent, and $\Delta\Lp/\Lp^{\rm avg}=45$ per cent; the peak luminosity is matched to better than $1.2$ per cent (mean signed error $-0.1$ per cent) and the width ratio $\Lth/\Lp$ to better than $4.2$ per cent (mean signed error $+1.1$ per cent) at every inclination. The largest early discrepancies arise from the MC initialization, its wavelength truncation, and the neglected light-travel time; dedicated tests that reduce the early residual to a few per cent are given in Appendix~\ref{app:validation_details}.

\subsection{Oblate spheroidal ejecta}
\label{sec:mc_validation_oblate}
Fig.~\ref{fig:diff_sol_oblate_fdep1} is the corresponding test of the oblate solution of Appendix~\ref{app:oblate_diffusion}. The idealized ejecta have uniform density and opacity, total mass $M=1.401\,M_\odot$, and an axis ratio $\tanh\xi_s\approx0.73$ ($\xi_s=0.9245$, with $v_{\rm eq}=1.24\times10^{4}\,\mathrm{km\,s^{-1}}$ and $v_{\rm pol}=9.05\times10^{3}\,\mathrm{km\,s^{-1}}$). The ${}^{56}$Ni source is moderately non-confocal ($\xi_1=0.7049$, $f=0.6966$) and carries $M_{{}^{56}\rm Ni}=0.2540\,M_\odot$, and the diffusion time-scale is $\tilde{t}=30.426\,\mathrm{d}$ (for $\kappa=0.2\,\mathrm{cm^2\,g^{-1}}$).  As above, $f_{\rm dep}=1$ is imposed in both calculations. Here, the source is centred on the equatorial plane and the configuration is symmetric about it, so the viewing-angle dependence is produced entirely by the aspherical boundaries, whereas in Fig.~\ref{fig:diff_sol_zoff_01_fdep1} it is produced entirely by the displacement of the source within a spherical boundary. The two tests are
therefore complementary. 

The analytic solution reproduces the MC light curves to better than $3.8$ per cent in $\Lp$ (mean signed error $+2.0$ per cent) and better than $1.5$ per cent in $\Lth/\Lp$ (mean signed error $+0.4$ per cent) across the ten viewing angles, over which $\Lp$ varies by $74$ per cent.

\begin{figure}
    \centering
    \includegraphics[width=0.9\linewidth]{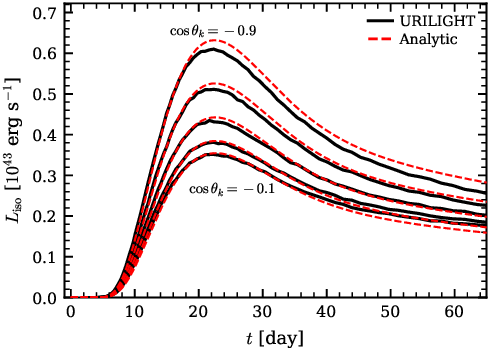}
    \caption{Validation of the analytic oblate diffusion solution (red dashed) against \textsc{URILIGHT} (solid black) for idealized uniform-density oblate ejecta with the parameters of Section~\ref{sec:mc_validation_oblate} and complete $\gamma$-ray deposition. $L_{\rm iso}$ is shown for the same five bins as in Fig.~\ref{fig:diff_sol_zoff_01_fdep1}; the configuration is symmetric about the equatorial plane, so the curves brighten with $|\cos\theta_k|$, from edge-on ($\cos\theta_k=-0.1$, labelled below its peak) to pole-on ($-0.9$, labelled above it).}
    \label{fig:diff_sol_oblate_fdep1}
\end{figure}

Over the optically thick epochs, $t\myapprox10$--$40$~d, the analytic solution reproduces the MC light curves to better than $5$ (off-centre) and $12$ (oblate) per cent at every viewing angle, over viewing-angle spreads in $\Lp$ of $56$ and $74$ per cent; at later times the residual is antisymmetric in viewing angle, because the MC contrast decays as the ejecta become optically thin while the analytic contrast approaches its quasi-steady pattern (Appendix~\ref{app:validation_details}).

\section{Viewing-Angle Dependence of the Peak Luminosity}
\label{sec:limb_darkening}

The viewing-angle dependence can be separated into two steps: the angular structure of the diffusive surface flux, and the projection of that flux toward an observer through the Eddington surface-emission prescription of Section~\ref{sec:viewing_angle}.  Appendix~\ref{app:limb_darkening} derives both steps and the quasi-steady limit used below; here we collect the resulting closed forms. Equivalently, the quasi-steady angular pattern can be obtained from a stationary diffusion problem in physical coordinates, with the ejecta frozen at a given epoch (Appendix~\ref{app:ld_multipoles}).

For spherical ejecta, let $F(\theta,t)$ denote the full surface flux after convolution with the heating history. Expanding $F(\theta,t)=\sum_lF_l(t)P_l(\cos\theta)$ and defining $a_l\equiv F_l/F_0$, the observer integral gives, exactly at every instant,
\begin{equation}
L_{\rm iso}(t,\theta_k)=L(t)\left[1+\sum_{l\ge1}w_l\,a_l(t)\,
P_l(\cos\theta_k)\right],
\label{eq:ld_series}
\end{equation}
where
\begin{equation}
w_l=\frac{3}{2}\int_0^1P_l(\mu)\,\mu\left(\mu+\frac{2}{3}\right)d\mu,
\label{eq:ld_weights}
\end{equation}
so that $w_l=\{1,17/24,13/40,1/16,-1/48,\dots\}$.  These numbers contain only the observer-projection prescription; all information about the source and ejecta geometry is contained in the multipoles $a_l$. These projection identities are exact within the surface-emission prescription; the quasi-steady approximation enters below, when the finite-time multipoles $a_l(t)$ are replaced by their zero-frequency values (Appendix~\ref{app:ld_multipoles}).

For a uniform spherical source displaced by $z_{\rm off}v_o$ inside a spherical
outer boundary, Appendix~\ref{app:ld_multipoles} shows that the quasi-steady
flux multipoles are
\begin{equation}
a_l^{\rm qs}=(2l+1)z_{\rm off}^{\,l}.
\end{equation}
The leading anisotropy is therefore the dipole,
$a_1^{\rm qs}=3z_{\rm off}$; the quadrupole is
$a_2^{\rm qs}=5z_{\rm off}^2$ and is second order in the displacement.
Approximating the finite-time peak multipoles by their
quasi-steady values, $a_l(t_{\rm peak})\simeq a_l^{\rm qs}$, as discussed in
Appendix~\ref{app:ld_multipoles}, gives
\begin{equation}
\frac{\Lp(\theta_k)}{\Lp^{\rm avg}}\myapprox
1+\sum_{l\ge1}w_l(2l+1)z_{\rm off}^{\,l}\,
P_l(\cos\theta_k),
\label{eq:ld_sphere}
\end{equation}
or, retaining the first two terms,
\begin{equation}
\frac{\Lp(\theta_k)}{\Lp^{\rm avg}}\myapprox
1+\frac{17}{8}\,z_{\rm off}\cos\theta_k
+\frac{13}{8}\,z_{\rm off}^2P_2(\cos\theta_k).
\label{eq:ld_sphere_num}
\end{equation}
To this order the viewing-angle dependence is controlled by $z_{\rm off}$ alone; in particular, the source radius $x_s$ does not enter the quasi-steady multipoles. The quasi-steady replacement is the main approximation: measured against the finite-time dipole of the same analytic solutions, the coefficient $17/8$ is low by $7$--$15$ per cent (median $11$ per cent; Appendix~\ref{app:ld_peak}, where the deficit is defined as $1-b_1^{\rm closed}/b_1$, Equations~\ref{eq:ld_bl_def} and~\ref{eq:ld_bl_closed}), nearly independently of $z_{\rm off}$ and of the remaining parameters, so that the normalized anisotropy depends primarily on the displacement over the tested range rather than being strictly independent of the other parameters.

Equation~\eqref{eq:ld_sphere_num} is the main result of this paper, and its most important feature is the steepness of the leading term: at the outermost bin centres, $\cos\theta_k=\pm0.9$, the dipole term alone gives $\Delta\Lp/\Lp^{\rm avg}=2\times0.9\times\tfrac{17}{8}|z_{\rm off}|\simeq3.83\,|z_{\rm off}|$, so that a displacement of only $z_{\rm off}=0.1$ gives an outer-bin spread of $38$ per cent in the peak luminosity ($42.5$ per cent, $(17/4)\,z_{\rm off}$, between  $\cos\theta_k=\pm1$). This produces appreciable viewing-angle variation even for modest source offsets; assessing its consistency with the observed relation requires the residual comparison discussed in Section~\ref{sec:discussion}. How this effective displacement relates to the ${}^{56}$Ni distribution of actual ejecta, and to nebular line shifts, is discussed in Sections~\ref{sec:toy_construction} and~\ref{sec:discussion}.

\begin{figure*}
    \centering
    \includegraphics[width=0.9\linewidth]{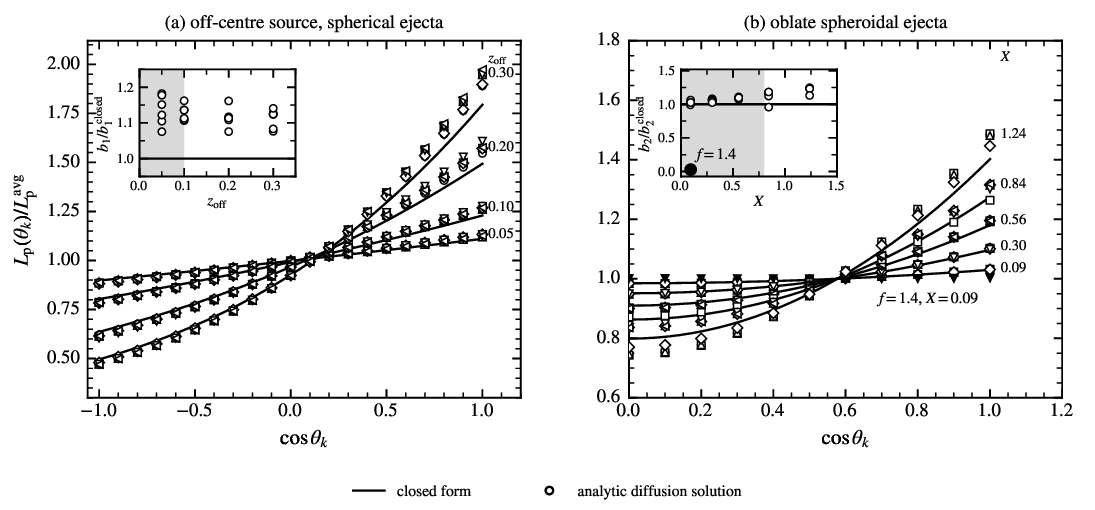}
    \caption{Normalized peak luminosity $\Lp(\theta_k)/\Lp^{\rm avg}$ against viewing angle in the analytic diffusion models (no \textsc{URILIGHT} calculations enter). Open markers are full finite-time analytic solutions; curves are the quasi-steady closed forms. Left: off-centre spherical sources at several $z_{\rm off}$ (Equation~\ref{eq:ld_sphere}); at each $z_{\rm off}$ the markers vary the remaining parameters over $0.15\le x_s\le0.90$, $16.4\le\tilde t\le36.6$~d, $0.15\le M_{{}^{56}\rm Ni}/M_\odot\le1.20$ and $20\le t_0\le45$~d (and $t_0=\infty$). Right: oblate spheroidal ejecta at several $X$ (Equation~\ref{eq:ld_oblate}), with $0.77\le\xi_s\le2.35$, $0.20\le\xi_1\le2.10$, $0.30\le f\le1.40$, $11.6\le\tilde t\le40.1$~d, $0.20\le M_{{}^{56}\rm Ni}/M_\odot\le1.50$ and $20\le t_0\le60$~d (and $t_0=\infty$); only $\cos\theta_k\ge0$ is shown. Insets show the ratio of the measured Legendre coefficient $b_l$ of $\Lp(\theta_k)/\Lp^{\rm avg}=\sum_lb_lP_l(\cos\theta_k)$ (Equation~\ref{eq:ld_bl_def}) to its closed form, $b_1/(3w_1z_{\rm off})$ against $z_{\rm off}$ and $b_2/(w_2X)$ against $X$. The labelled $f=1.4$ run at $X=0.09$ (filled symbol, at a ratio of $0.03$) is the exception discussed in the text. Models sharing $z_{\rm off}$ (or $X$) collapse onto nearly the same curve; the systematic offsets are the finite-time and shape corrections of Appendix~\ref{app:ld_peak}. Shaded regions mark $z_{\rm off}\le0.1$ and $X\le0.80$.} 
    \label{fig:limb_darkening}
\end{figure*}

We now turn to oblate ejecta, treating both the boundary and the source as weakly deformed spheres; the exact two-spheroid solution of Appendix~\ref{app:oblate_diffusion} is valid at any flattening, and Fig.~\ref{fig:limb_darkening} quantifies the range over which the two agree. For a smooth convex boundary, the observer integral weights the surface flux by the area associated with each outward-normal direction, $dA=d\Omega_{\hat{\mathbf n}}/K$, where $K$ is the Gaussian curvature of the boundary in velocity space (the physical curvature is $K/t^2$; the factor is common to all multipoles and cancels in their normalized ratios). Defining $\Phi\equiv F/K$ as a function of the polar angle $\Theta$ of the outward normal, Equation~\eqref{eq:ld_series} carries over with $a_l$ replaced by $\Phi_l/\Phi_0$ (Appendix~\ref{app:ld_projection}). Equatorial symmetry removes the odd multipoles, so the leading anisotropy is the quadrupole. Writing the boundary and the source as $v_{\rm b}(\theta)=v_o[1+\varepsilon P_2(\cos\theta)]$ and $v_{\rm src}(\theta)=x_sv_o[1+\varepsilon_sP_2(\cos\theta)]$ with $|\varepsilon|,|\varepsilon_s|\ll1$, Appendix~\ref{app:ld_multipoles} gives, to first order, three quadrupole contributions: the radioactive source, $3\varepsilon_sx_s^2$; the deformed diffusion boundary, $-4\varepsilon$; and the surface area per unit solid angle (the curvature), which contributes a second term equal to the boundary one. With $\varepsilon=-2\delta/3+O(\delta^2)$ and $\delta\equiv1-v_{\rm pol}/v_{\rm eq}$,
\begin{equation}
\begin{split}
\frac{\Phi_2}{\Phi_0}
&=3\varepsilon_sx_s^2-4\varepsilon-4\varepsilon\\
&\simeq\frac{16}{3}\delta-f^2\left(1-q^2\right)
\equiv X,
\qquad q\equiv\frac{v_{\rm pol}}{v_{\rm eq}},
\end{split}
\label{eq:ld_Phi2_summary}
\end{equation}
where the source term has been written in terms of the oblate-geometry parameters, $3\varepsilon_sx_s^2\simeq-f^2(1-q^2)$. With $w_2=13/40$, and replacing the finite-time peak quadrupole by its quasi-steady value as above,
\begin{equation}
\begin{split}
\frac{\Lp(\theta_k)}{\Lp^{\rm avg}}
&\myapprox1+\frac{13}{40}\,X\,P_2(\cos\theta_k)\\
&\myapprox1+\left[\frac{26}{15}\,\delta
-\frac{13}{40}\,f^2\left(1-q^2\right)\right]P_2(\cos\theta_k).
\end{split}
\label{eq:ld_oblate}
\end{equation}
For oblate ejecta ($\varepsilon<0$) the boundary and curvature terms brighten the poles, whereas an oblate radioactive source ($\varepsilon_s<0$) darkens them. The source enters only through $f$; its size and its boundary coordinate $\xi_1$ do not appear, and for a confocal source ($f=1$), $X\simeq10\delta/3$. For the outer bins of this equatorially symmetric geometry, Equation~\eqref{eq:ld_oblate} gives $\Delta\Lp/\Lp^{\rm avg}=w_2|X|[P_2(0.9)-P_2(0.1)]=(39/100)\,|X|$, and $(39/80)\,|X|$ between pole and equator. Equation~\eqref{eq:ld_oblate} combines the first-order shape expansion with the quasi-steady replacement. For the six collision toy models tested here ($X\le0.80$), the closed form underestimates the peak spread of the full spheroidal solution by $3$--$11$ per cent (Section~\ref{sec:collisions}), the combined finite-time and shape error of those configurations; it is not a uniform accuracy guarantee for every small-$X$ source, as the example below shows.

Fig.~\ref{fig:limb_darkening} compares the closed forms with the full analytic diffusion solutions: in the spherical case the two differ only through the quasi-steady replacement, whereas in the oblate case Equation~\eqref{eq:ld_oblate} is additionally first order in the flattening, so we restrict the comparison to $X<1.5$, which comfortably contains the six collision-model fits of Appendix~\ref{app:collision} (all have $X\le0.80$). This restriction is not by itself a sufficient accuracy criterion for arbitrary $f$: for one strongly non-confocal run ($f=1.4$, $\xi_s=1.73$, $\xi_1=1.20$) at small $X=0.09$, the source and boundary terms of Equation~\eqref{eq:ld_Phi2_summary} nearly cancel, and the measured quadrupole is only a few per cent of $w_2X$ although the normalized luminosities differ from the closed curve by less than $3$ per cent; a small luminosity residual can mask a large fractional anisotropy error (this run is marked in the inset of Fig.~\ref{fig:limb_darkening}). In both panels, models sharing the same $z_{\rm off}$ (or the same $X$) collapse onto nearly the same viewing-angle curve despite variations in the remaining toy-model parameters: the off-centre problem is predominantly dipolar and controlled by $z_{\rm off}$, while the oblate problem is quadrupolar and controlled by $X$. 

Two further finite-time approximations are measured in Appendix~\ref{app:ld_peak}: the quasi-steady quadrupole of a slightly oblate boundary with $f\le1$ is within $4.5$ per cent of the finite-time one (median $1$ per cent), and evaluating all viewing angles at the angle-averaged peak time introduces an error that is second order in the anisotropy but reaches $18$ per cent for $z_{\rm off}=0.4$ (Table~\ref{tab:ld_frozen}). We correct for none of these, so the closed forms contain no calibrated number.


%
\section{The Off-Centre Ignition Toy Model}
\label{sec:toy_model}

We now test whether effective uniform-density ejecta can reproduce the angle-dependent light curves of stratified explosion models. For each off-centre ignition model (Section~\ref{sec:models}) we construct an effective toy model whose parameters are extracted from the ejecta (Section~\ref{sec:toy_construction}), compare the analytic light curves with \textsc{URILIGHT} across all viewing angles (Section~\ref{sec:toy_comparison}), and characterize the systematic mismatch in the width ratio (Section~\ref{sec:width_mismatch}).

\subsection{The off-centre ignition models}
\label{sec:models}
We use the $19$ bare-CO, off-centre ignition models of \citet{Schinasi_2024}, introduced in Section~\ref{sec:introduction}. The progenitors are solar-metallicity CO WDs (uniform composition $X({}^{12}\mathrm{C})=X({}^{16}\mathrm{O})=0.4925$, $X({}^{22}\mathrm{Ne})=0.015$), spanning the mass range $0.8-1.1\,M_\odot$, and the explosions are computed with the \textsc{FLASH} code in two-dimensional axisymmetric geometry using a $38$-isotope reaction network. These two-dimensional models are axisymmetric, so that observable properties depend only on the inclination angle $\theta_k$ between the observer's line of sight and the symmetry axis of the ejecta ($\theta_k=0^\circ$ and $180^\circ$ correspond to viewing directions along the axis). The ignition location is parametrized by the dimensionless quantity $z_{\mathrm{ig}}$, defined as the displacement of the ignition point along the symmetry axis normalized by the WD radius. Across the model grid, we consider $z_{\mathrm{ig}} = 0,\,0.1,\,0.25,\,0.5$, and $0.75$, although not every WD mass is available at all five offsets; the final sample contains 19 ejecta models (Table~\ref{tab:example_table}). The case $z_{\mathrm{ig}}=0$ corresponds to central ignition. Off-centre ignition produces an asymmetric ${}^{56}$Ni distribution, and the synthesized ${}^{56}$Ni mass ranges from $M_{{}^{56}\rm Ni}\approx0.04\,M_\odot$ for the $0.8\,M_\odot$ models to $\myapprox0.8\,M_\odot$ for the $1.1\,M_\odot$ models (Table~\ref{tab:example_table}).

\subsection{Toy-model construction}
\label{sec:toy_construction}

The five toy-model parameters of Section~\ref{sec:analytic_params} are extracted from each ejecta as follows, starting with the ejecta mass and characteristic velocity.

\begin{figure} 
\centering 
\includegraphics[width=0.9\linewidth]{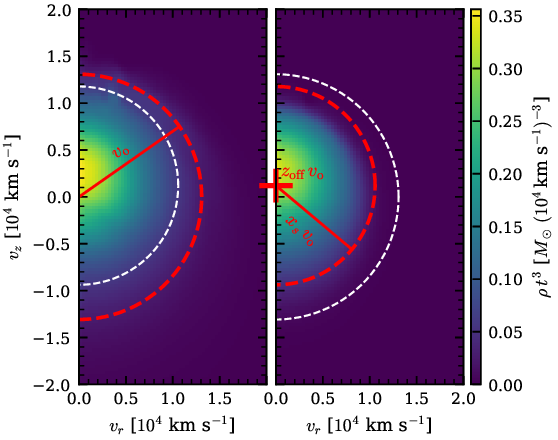} \caption{Effective sphere geometry of the toy-model construction (Section~\ref{sec:toy_construction}) for the $1\,M_\odot$, $z_{\rm ig}=0.5$ off-centre ignition model ($\# 13$, Table~\ref{tab:example_table}), shown in velocity space, in the $(v_r,v_z)$ plane. Both panels show the velocity-space density $\rho t^3$ in units of $M_\odot\,(10^4\,\mathrm{km\,s^{-1}})^{-3}$, with velocities in $10^4\,\mathrm{km\,s^{-1}}$, on a common colour scale, common axis ranges and common ticks, cropped at $2.0\times10^4\,\mathrm{km\,s^{-1}}$. Left: total mass density in velocity space $\rho t^3$, with the effective outer-velocity sphere of radius $v_o$ (red dashed). Right: the ${}^{56}$Ni mass density $X_{{}^{56}\rm Ni}\,\rho t^3$, where $X_{{}^{56}\rm Ni}$ is the ${}^{56}$Ni mass fraction, with the effective ${}^{56}$Ni sphere of radius $x_s v_o$ centred at an axial offset $z_{\rm off}v_o$ (red dashed). The thin white dashed lines show the other panel's sphere, for comparison: the outer-velocity sphere ($v_o$) in the right panel, the ${}^{56}$Ni sphere ($x_s v_o$, $z_{\rm off}v_o$) in the left panel. For this model, $v_o=1.31\times10^{4}\,\text{km s}^{-1}$, $x_s=0.81$, and $z_{\rm off}=0.09$ (${}^{56}$Ni centroid $z_{{\rm com},{}^{56}\rm Ni}=0.025$; Table~\ref{tab:example_table}).}
\label{fig:subchandra_effective_geometry} 
\end{figure}

The toy model is constructed to have the same total ejecta mass, $M$, and total ${}^{56}$Ni mass, $M_{{}^{56}\rm Ni}$, as the original ejecta. We first define an effective outer ejecta radius through the characteristic velocity
\begin{equation}
v_o \equiv \left(\frac{3M}{4\pi \bar{\rho}_v}\right)^{1/3},
\label{eq:vo}
\end{equation}
where
\begin{equation}
\bar{\rho}_v =
\frac{\sum_i \rho_{v,i} m_i}{\sum_i m_i}
\end{equation}
is the mass-weighted mean density in velocity space, each $\rho_{v,i}$ being the velocity-space density $t^3\rho_i^{\rm phys}$ of cell $i$ of mass $m_i$. The corresponding uniform-density ejecta sphere is illustrated in the left panel of Fig.~\ref{fig:subchandra_effective_geometry}. Because $v_o$ is set by the mean velocity-space density, the construction matches the ejecta mass but not its kinetic energy: the uniform sphere carries $E_k=\tfrac{3}{10}Mv_o^2$, which for the $19$ models of Table~\ref{tab:example_table} falls short of the actual ejecta kinetic energy by $7-16$ per cent (median $11$ per cent).

We characterize the effective geometry of the radioactive material by fitting the ${}^{56}$Ni density distribution in velocity space with a displaced hard-sphere model. The fit has three free parameters: the source displacement $z_{\rm off}$, the dimensionless source radius $x_s$, and the density normalization of the hard sphere $A$. The fit is performed on the two-dimensional axisymmetric ejecta that enter the \textsc{URILIGHT} calculations, on their uniform $(v_r,v_z)$ mesh. We minimize the unweighted sum of squared residuals $\mathcal{S}$
\begin{equation}
\mathcal{S}=\sum_{i}\left[A\,H_i(x_s,z_{\rm off})-\frac{\rho_{v,{}^{56}{\rm Ni},i}}{\max\rho_{v,{}^{56}\rm Ni}}\right]^2,
\label{eq:hard_sphere_fit}
\end{equation}
where the sum runs over all cells $i$, $\rho_{v,{}^{56}{\rm Ni},i}$ is the ${}^{56}$Ni density in velocity space, $H_i=1$ if the cell centre lies within the displaced sphere and $0$ otherwise, and $A$ is the normalization. Every cell carries the same weight, so the residuals are weighted by area in the $(v_r,v_z)$ plane, not by volume or mass. The minimum is found by differential evolution. The displacement $z_{\rm off}$ is defined as the offset of the equivalent ${}^{56}$Ni sphere's centre along the symmetry axis, normalized by the characteristic velocity $v_o$. This construction is illustrated in the right panel of Fig.~\ref{fig:subchandra_effective_geometry}.

The fitted values of $x_s$ and the density normalization do not, in general, reproduce the correct ${}^{56}$Ni mass, because the fit is based solely on the spatial distribution of the ${}^{56}$Ni density and does not enforce mass conservation. We therefore retain only the fitted value of $z_{\rm off}$, taken from the three-parameter fit without refitting after $x_s$ is fixed. The source radius is instead fixed by requiring the toy model to contain the correct ${}^{56}$Ni mass,
\begin{equation}
x_s=\left(\frac{M_{{}^{56}\rm Ni}}{M}\right)^{1/3},
\end{equation}
which follows from mass conservation if the source is a pure ${}^{56}$Ni region at the uniform ejecta density (unit ${}^{56}$Ni mass fraction inside the source). The fitted displacements are small, $z_{\rm off}\le0.12$, across the entire grid (Table~\ref{tab:example_table}, which also lists the corresponding velocities). We note that the displaced source is fully contained within the analytic ejecta only when $x_s+|z_{\rm off}|\le1$.

We compare the fitted displacement with the ${}^{56}$Ni centre of mass computed directly from each ejecta model, $z_{{\rm com},{}^{56}\rm Ni}=\langle v_z\rangle_{\rm Ni}/v_o$, where $\langle v_z\rangle_{\rm Ni}$ is the ${}^{56}$Ni-mass-weighted mean axial velocity (Table~\ref{tab:example_table}). Both are close to zero for the four centrally ignited models. For the $14$ models with $z_{\rm off}>0.01$, the fitted displacement exceeds the centroid in all but one case, with a median ratio $z_{\rm off}/z_{{\rm com},{}^{56}\rm Ni}=2.8$ (range $0.85$--$8.2$), and the two scale differently with the ignition offset: $z_{{\rm com},{}^{56}\rm Ni}$ stays within $0.013$--$0.025$ for all models with $z_{\rm ig}\ge0.25$, whereas $z_{\rm off}$ increases to $0.12$. The difference arises because the ${}^{56}$Ni density is not uniform within the ${}^{56}$Ni-rich region: its high-density part is displaced toward $+z$, while lower-density ${}^{56}$Ni extends to higher velocities at $z<0$, and the hard-sphere fit follows the high-density part whereas the centroid weights all of the mass (Appendix~\ref{app:extended_results} quantifies this). Thus, $z_{\rm off}$ is an effective parameter that partially accounts for the density stratification of the ${}^{56}$Ni distribution, which the uniform-density toy model does not include; there is no universal conversion between the two.

Nebular line shifts trace the emissivity-weighted velocities of the emitting ions, which for some lines are the stable iron-group material rather than the ${}^{56}$Ni \citep[e.g.][]{Maeda_2010b}. They therefore test the ejecta models rather than the fitted $z_{\rm off}$ \citep[e.g.][]{Motohara_2006, Maeda_2010, Silverman_2013, Dong_2015, Black_2016, Graham_2017, Diamond_2018, Maguire_2018, Vallely_2020, Kumar_2023, Kwok_2023, DerKacy_2023}.

The diffusion time-scale, $\tilde{t}^2\equiv v_o^2/D$, is computed using the diffusion coefficient $D=c/(3\kappa\bar{\rho}_v)$, where $\kappa$ is the mass-weighted Rosseland mean opacity of the ejecta at $\tau>2$, evaluated at the peak of the angle-averaged \textsc{URILIGHT} light curve, following \citet{sharon_2025}:
\begin{equation}
\kappa=\frac{\sum_{\tau_i>2}m_i\,\kappa_{{\rm R},i}(t_{\rm p}^{\rm avg})}{\sum_{\tau_i>2}m_i},
\label{eq:kappa_eff}
\end{equation}
where $m_i$ and $\kappa_{{\rm R},i}$ are the mass and Rosseland mean opacity of cell $i$ and $\tau_i$ its optical depth from the outer edge. The optical depth is computed from a spherically averaged density and opacity profile by integrating inward from the outer boundary.

The characteristic $\gamma$-ray escape time $t_0$, as defined by Equation~\eqref{eq:f_dep}, is computed following \citet{Wygoda_2019a}, using the average column density of the ejecta as seen by the ${}^{56}$Ni elements and an effective $\gamma$-ray opacity of $\kappa_{\rm eff}=0.025\,\mathrm{cm}^2\,\mathrm{g}^{-1}$.

We emphasize that none of these parameters is calibrated to reproduce the \textsc{URILIGHT} light curves: four of the five are determined directly from the initial ejecta model, while $\tilde t$ requires the RT calculation. This is because the effective opacity $\kappa$ (and hence $D$) is evaluated from the ejecta state at peak luminosity and therefore requires the \textsc{URILIGHT} calculation to have been completed. It is thus measured a posteriori from the simulation rather than fitted to the light curves. In practice, however, the measured values span the narrow range $\kappa\myapprox0.10$--$0.18\,\mathrm{cm^2\,g^{-1}}$ across the grid (Table~\ref{tab:example_table}). Section~\ref{sec:geometry} shows explicitly (by varying $\tilde t$, which is equivalent to varying $\kappa$) that the normalized viewing-angle variability, the main quantity of interest here, is insensitive to this choice, although the absolute $\Lp$ and $\Lth/\Lp$ are not. The full parameter table for all models is given in Table~\ref{tab:example_table}.

\subsection{Comparison with \textsc{URILIGHT} on the numerical ejecta}
\label{sec:toy_comparison}
\begin{figure}
    \centering
    \includegraphics[width=0.9\linewidth]{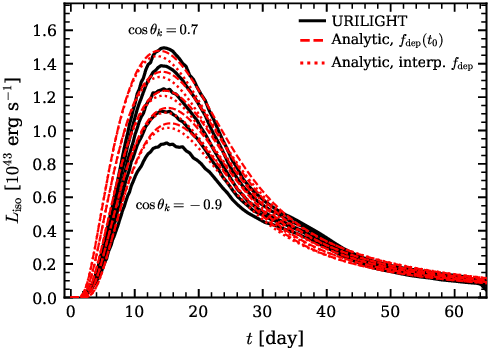}
    \caption{Bolometric light curves of the effective spherical off-centre toy model (red) compared with \textsc{URILIGHT} (solid black) for the $1\,M_\odot$, $z_{\rm ig}=0.5$ off-centre ignition model (\#13; parameters in Table~\ref{tab:example_table}). $L_{\rm iso}$ is shown for the same five bins as in Fig.~\ref{fig:diff_sol_zoff_01_fdep1}, with $\cos\theta_k=+1$ facing the ${}^{56}$Ni offset; the curves brighten monotonically from $-0.9$ (labelled below its peak) to $+0.7$ (labelled above it). The two analytic curves differ only in the $\gamma$-ray deposition: the dashed curve uses Equation~\eqref{eq:f_dep} with the tabulated $t_0$, while the dotted curve uses the $f_{\rm dep}(t)$ from the \textsc{URILIGHT} $\gamma$-ray transport of the same ejecta, so their difference isolates the error of Equation~\eqref{eq:f_dep}. The discrepant pair at the bottom is the bin facing away from the offset ($\cos\theta_k=-0.9$).}
    \label{fig:1Msun_ejecta109_sol}
\end{figure}

Whereas Section~\ref{sec:mc_validation} tests the diffusion solution in isolation, the comparison here tests the full toy-model construction of Section~\ref{sec:toy_construction}. We extract the five effective toy-model parameters from an actual sub-Chandrasekhar ejecta, the $\#13$ $1\,M_\odot$, $z_{\rm ig}=0.5$ model of Table~\ref{tab:example_table}, and compare the resulting analytic light curves directly with \textsc{URILIGHT} at each viewing angle. Fig.~\ref{fig:1Msun_ejecta109_sol} shows this comparison for five of the ten bins, with two analytic variants that differ only in the $\gamma$-ray deposition, Equation~\eqref{eq:f_dep} with the single escape time $t_0$ or the $f_{\rm dep}(t)$ tabulated by the \textsc{URILIGHT} $\gamma$-ray transport of the same ejecta, which separates the error of the deposition prescription from that of the diffusion model; unless stated otherwise, the comparisons below refer to the former.

For this model $\Lp$ is reproduced to better than $\myapprox5$ per cent at nine of the ten inclinations ($13$ per cent at $\cos\theta_k=-0.9$, facing away from the offset), whereas $\Lth/\Lp$ is overestimated at nine of the ten inclinations, by up to $10$ per cent toward the offset; this overestimate is not specific to this model (Section~\ref{sec:width_mismatch}).

At early times ($t\lesssim7$~d) the analytic luminosity exceeds the MC one by tens of per cent, mainly because the uniform sphere misrepresents the outer ejecta, and at late times ($t\ge30$~d) the angle-averaged luminosity is reproduced to within $11$ per cent while the analytic viewing-angle contrast stays near its quasi-steady value as the MC contrast decays; both residuals are quantified in Appendix~\ref{app:extended_results}.

Fig.~\ref{fig:zoff_L30Lp_Lp_M1sol} extends this comparison to the five $1\,M_\odot$ models, ordered by $z_{\rm off}$; across their $50$ model--angle combinations, the analytic-deposition toy model reproduces $\Lp$ with a median relative error of $\mysim3$ per cent and a maximum error of $\mysim13$ per cent, while the corresponding errors in $\Lth/\Lp$ have a median of $\mysim4$ per cent and a maximum of $\mysim12$ per cent. The latter error is predominantly systematic rather than scatter about the RT result: the toy model overpredicts $\Lth/\Lp$ in $43$ of the $50$ combinations. The bias is also strongly ordered with viewing angle, running from $\myapprox-5$ per cent at $\cos\theta_k=-0.9$ to $\myapprox+9$ per cent at $+0.9$ for the most displaced model, with a maximum of $\myapprox+10$ per cent at $\cos\theta_k\simeq0.1$--$0.5$.

The mean level and angular variation of $\Lth/\Lp$ are therefore reproduced with very different accuracy. The value obtained from the angle-averaged light curve is recovered to within $1.0-7.5$ per cent, whereas the outer-bin spread $\Delta(\Lth/\Lp)$ reaches only $27-43$ per cent of the \textsc{URILIGHT} value for the four models with $z_{\rm off}\ge0.02$. Replacing the analytic deposition prescription by the tabulated $f_{\rm dep}(t)$ generally improves the agreement, particularly in $\Lth/\Lp$, but leaves the viewing-angle dependence of both quantities essentially unchanged. The deposition prescription therefore contributes to the offset in the overall width ratio but does not explain the underprediction of the angular variation. We separate these two components of the discrepancy in Section~\ref{sec:width_mismatch}.

In the $\Lp$--$\Lth/\Lp$ plane (Fig.~\ref{fig:Lp_L30Lp_obs}a), the off-centre ignition models remain at $\Lth/\Lp\le0.50$ at every viewing angle, in the regime of the faster-declining observed events, and their tracks run across the observed sequence rather than along it, as found by \citet{sharon_2025}: toward the ${}^{56}$Ni-facing pole $\Lp$ increases while $\Lth/\Lp$ decreases. In \textsc{URILIGHT} the outer-bin extent of the tracks reaches $0.88\times10^{43}\,\mathrm{erg\,s^{-1}}$ in $\Lp$ and $0.10$ in $\Lth/\Lp$ (model \#19; $26$ per cent when normalized), and most model--angle combinations of the $1.1\,M_\odot$ models are brighter than the brightest observed event. The analytic toy model (Fig.~\ref{fig:Lp_L30Lp_obs}b) reproduces the direction of the tracks and their extent in $\Lp$: for the $0.9$--$1.1\,M_\odot$ models with $z_{\rm ig}>0$ (excluding model \#9, Section~\ref{sec:geometry}), the ratio of the analytic to the \textsc{URILIGHT} outer-bin extent in $\Lp$, $\Delta\Lp^{\rm ana}/\Delta\Lp^{\rm MC}$, which is the quantity visible in Fig.~\ref{fig:Lp_L30Lp_obs}, is $0.70$--$1.07$ for ten of these $11$ models and $0.53$ for model \#7, with a median of $0.79$ over the $11$ models. Its tracks are, however, flatter in $\Lth/\Lp$, by the factor $\mysim2$--$3$ discussed in Sections~\ref{sec:width_mismatch} and~\ref{sec:geometry}.

\begin{figure*}
    \centering
    \includegraphics[width=0.9\linewidth]{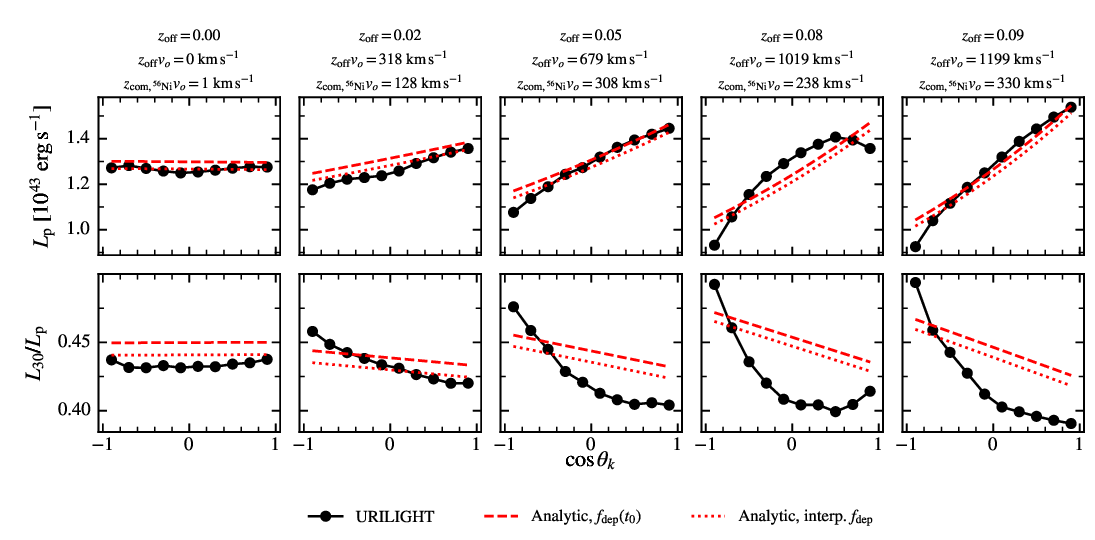}
    \caption{Peak luminosity (top) and width ratio (bottom) against viewing angle for the five $1\,M_\odot$ off-centre ignition models (\#10--\#14, Table~\ref{tab:example_table}), one column per model in order of increasing $z_{\rm off}$; the column titles give $z_{\rm off}$, the offset velocity $z_{\rm off}v_o$ and the ${}^{56}$Ni centroid velocity $z_{{\rm com},{}^{56}\rm Ni}v_o$ of the numerical ejecta (with a common $v_o=1.31\times10^{4}\,\mathrm{km\,s^{-1}}$). Solid black lines with circles: \textsc{URILIGHT}; red dashed and dotted: the analytic toy model with Equation~\eqref{eq:f_dep} and with the tabulated $f_{\rm dep}(t)$, respectively. All ten bins are shown, with $\cos\theta_k=+1$ facing the ${}^{56}$Ni offset; rows share their vertical axis.}
    \label{fig:zoff_L30Lp_Lp_M1sol}
\end{figure*}

\subsection{The width-ratio mismatch}
\label{sec:width_mismatch}
\begin{figure*}
    \centering
    \includegraphics[width=0.9\linewidth]{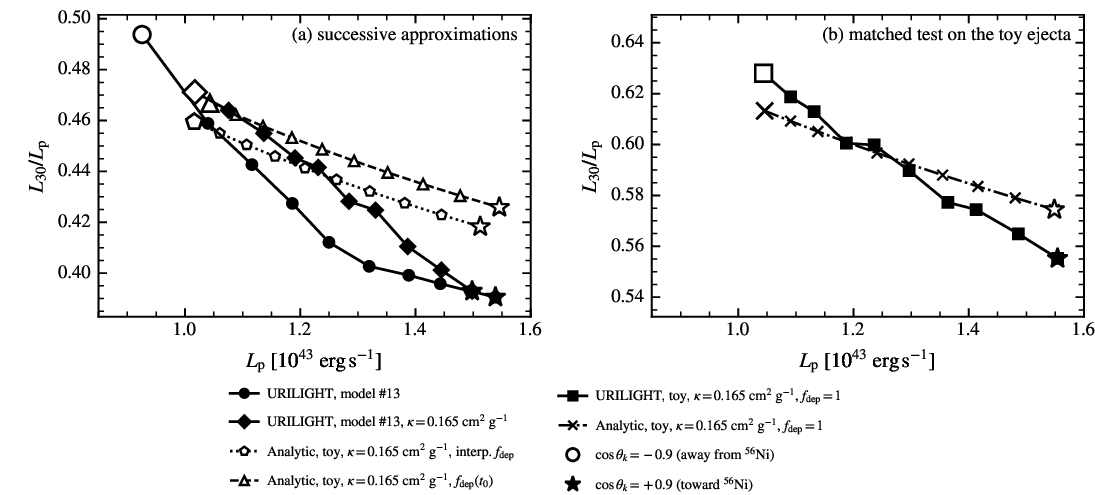}
    \caption{Viewing-angle sequences of model \#13 in the $\Lp$--$\Lth/\Lp$ plane for the ladder of calculations of Table~\ref{tab:ladder}, each running from $\cos\theta_k=-0.9$, facing away from the ${}^{56}$Ni offset (open marker), to $+0.9$, facing it (star). (a) Successive approximations, (0)--(i) on the original ejecta and (ii)--(iii) on the effective spherical ejecta: circles, \textsc{URILIGHT} with its full opacity and $\gamma$-ray transport; diamonds, \textsc{URILIGHT} with the constant $\kappa=0.165\,\mathrm{cm^2\,g^{-1}}$; pentagons (dotted), the analytic toy model with the tabulated $f_{\rm dep}$; triangles (dashed), the same with Equation~\eqref{eq:f_dep}. (b) The matched test (iv) on the effective toy ejecta with complete deposition: squares, \textsc{URILIGHT}; crosses (dot-dashed), the analytic solution. Solid lines with filled markers are \textsc{URILIGHT}, broken lines with open markers analytic; the panels share the $\Lp$ axis and cover equal ranges in $\Lth/\Lp$.}

    \label{fig:check_geom}
\end{figure*}

The width-ratio error has two components: a bias in its value and an underestimate of its viewing-angle variation. Fig.~\ref{fig:check_geom} and Table~\ref{tab:ladder} compare successive approximations for model \#13: (0) \textsc{URILIGHT} with its full opacity and $\gamma$-ray transport; (i) \textsc{URILIGHT} on the same two-dimensional ejecta with the toy model's grey opacity, $\kappa=0.165\,\mathrm{cm^2\,g^{-1}}$, held fixed in space and time; (ii) the analytic toy model of Section~\ref{sec:toy_construction} with the deposition fraction tabulated by the \textsc{URILIGHT} $\gamma$-ray transport; and (iii) the same with Equation~\eqref{eq:f_dep} and $t_0=31.1$~d. Steps (0)--(i) use the original ejecta and (ii)--(iii) the effective spherical ejecta, so step (ii) changes both the geometry and the transport treatment. A separate matched test (iv) runs \textsc{URILIGHT} on the effective toy ejecta itself, with the same grey opacity and complete $\gamma$-ray deposition, and compares it with the analytic solution for identical ejecta, as in Section~\ref{sec:mc_validation_sphere}.

\begin{table}
\centering
\caption{Outer-bin spreads (Section~\ref{sec:mc_validation_sphere}) of $\Lp$ and of $W\equiv\Lth/\Lp$ for model \#13 along the ladder of calculations of Section~\ref{sec:width_mismatch} (Fig.~\ref{fig:check_geom}). Successive approximations for model \#13: (0)--(i) use the original ejecta and (ii)--(iii) the effective spherical ejecta; the matched test (iv) compares \textsc{URILIGHT} (MC) and the analytic solution on the effective ejecta with complete deposition.}
\label{tab:ladder}
\small
\setlength{\tabcolsep}{4pt}
\begin{tabular}{lcc}
\hline
Calculation & $\Delta\Lp/\Lp^{\rm avg}$ & $\Delta W/W^{\rm avg}$ \\
\hline
(0) MC, full opacity, $\gamma$-ray transport & $48.4\%$ & $24.7\%$ \\
(i) MC, constant $\kappa$ & $38.2\%$ & $18.1\%$ \\
(ii) analytic toy, tabulated $f_{\rm dep}$ & $39.9\%$ & $9.4\%$ \\
(iii) analytic toy, $f_{\rm dep}(t_0)$ & $39.4\%$ & $9.2\%$ \\
\hline
(iv) MC, toy ejecta, $f_{\rm dep}=1$ & $40.1\%$ & $12.3\%$ \\
(iv) analytic, toy ejecta, $f_{\rm dep}=1$ & $39.4\%$ & $6.5\%$ \\
\hline
\end{tabular}
\end{table}

Relative to the full calculation (0), the constant opacity of step (i) shifts $\Lp$ by up to $10$ and $\Lth/\Lp$ by up to $7$ per cent (medians $-2$ and $+3$ per cent), with opposite signs at the two ends of the viewing-angle sequence, and reduces both spreads (Table~\ref{tab:ladder}). The contraction is expected: the effective opacity is measured at peak light, whereas by $t=30$~d the true effective opacity has dropped by a factor of $2$--$4$, and holding it fixed suppresses the angle-dependent timing of the transition to transparency. Step (ii) replaces this MC calculation by the analytic toy model with the same opacity and the \textsc{URILIGHT} deposition history, and therefore changes the geometry and the transport treatment together. Relative to (i), it reproduces $\Lp$ to within $3$ per cent and $\Lth/\Lp$ to within $6.5$ per cent at every viewing angle and recovers the spread in $\Lp$, but halves the spread in $\Lth/\Lp$. Step (iii) replaces the tabulated deposition by Equation~\eqref{eq:f_dep}; it raises both $\Lp$ and $\Lth/\Lp$ by $\myapprox2$ per cent at every viewing angle and leaves both spreads nearly unchanged. Since $t_0\myapprox30$--$36$~d for all $19$ models (Table~\ref{tab:example_table}), $\Lth$ is evaluated close to $t_0$, where Equation~\eqref{eq:f_dep} is least accurate. At the end of the sequence, the width ratio of (iii) exceeds that of (0) by up to $10.3$ per cent (median $9$ per cent over the ten bins). Because step (ii) changes geometry and transport at once, these comparisons do not isolate a separate geometry contribution. The matched toy-ejecta test (iv) shows that the diffusion approximation can account for much of the additional loss of width-ratio contrast in steps (ii) + (iii): for identical ejecta the analytic solution reproduces $\Lp$ to within $1$ per cent and $\Lth/\Lp$ to within $3.5$ per cent at every viewing angle and recovers the spread in $\Lp$, yet retains only about half of the \textsc{URILIGHT} spread in $\Lth/\Lp$ ($6.5$ against $12.3$ per cent). As the ejecta become transparent, the MC transport loses luminosity anisotropy at $30$~d more rapidly than the diffusive surface model, whose Dirichlet boundary and grey-atmosphere closure hold the contrast nearly fixed (Section~\ref{sec:mc_validation}). Because test (iv) uses complete deposition, and hence a different heating history from the left-panel sequence, it establishes the mechanism without uniquely separating the geometric reduction from the transport treatment in the original problem; a definitive partition would require the intermediate MC calculation on the effective geometry with matched heating, including its spatial distribution. The decomposition has been carried out for this model only, since each ladder requires dedicated \textsc{URILIGHT} runs; it is consistent with the factor $\mysim2$--$3$ by which the analytic spread in $\Lth/\Lp$ falls below the \textsc{URILIGHT} one across the grid (Fig.~\ref{fig:zoff_DL30Lp_DLp_Mi}).

\section{Isolating Geometric Effects}
\label{sec:geometry}

The question of this section is whether the remaining toy-model parameters change the normalized peak spread at fixed $z_{\rm off}$. Section~\ref{sec:limb_darkening} predicts that they do not to leading order: $x_s$, $\tilde t$, $t_0$ and $M_{{}^{56}\rm Ni}$ are absent from the quasi-steady closed form. In the numerical application, these parameters are uncertain to different degrees; most importantly, $\tilde t$ is not known a priori because it depends on the effective opacity. We therefore treat $x_s$, $\tilde t$ and $t_0$ as nuisance parameters and vary them independently over broad ranges about the values extracted from the $1\,M_\odot$ models, and measure the outer-bin spreads $\Delta\Lp/\Lp^{\rm avg}$ and $\Delta(\Lth/\Lp)/(\Lth/\Lp)^{\rm avg}$ defined in Section~\ref{sec:mc_validation_sphere}.

\begin{figure*}
    \centering
    \includegraphics[width=0.9\linewidth]{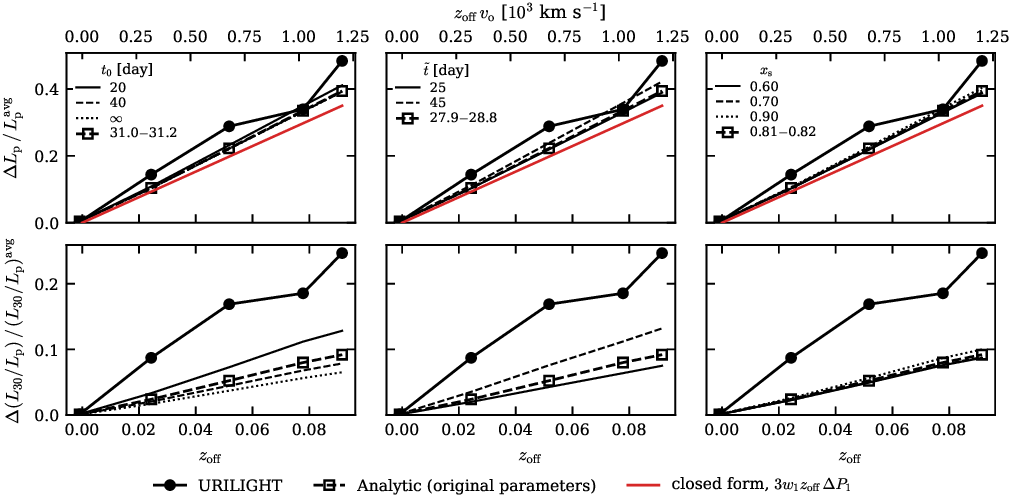}
    \caption{Outer-bin spreads $\Delta\Lp/\Lp^{\rm avg}$ (top) and $\Delta(\Lth/\Lp)/(\Lth/\Lp)^{\rm avg}$ (bottom; Section~\ref{sec:mc_validation_sphere}) against $z_{\rm off}$ for the $1\,M_\odot$ models. Each column varies one analytic parameter about the values of the models themselves: $t_0\in\{20,40,\infty\}$~d (left; $t_0\to\infty$ is complete deposition), $\tilde t\in\{25,45\}$~d (middle) and $x_s\in\{0.60,0.70,0.90\}$ (right). Thin black curves (line styles in the top-row legend) are the parameter variations, the curve with open squares is the analytic solution with the parameters of the models, and the curve with filled circles is the \textsc{URILIGHT} result. The red line in the top row is the closed form $\Delta\Lp/\Lp^{\rm avg}=3.83\,|z_{\rm off}|$ of Section~\ref{sec:limb_darkening}, not a fit; this lies $\myapprox11$ per cent below the analytic curves, the quasi-steady dipole deficit of Appendix~\ref{app:ld_peak}, and the \textsc{URILIGHT} points lie further above it (Section~\ref{sec:width_mismatch}). No closed form exists for $\Lth/\Lp$.}
\label{fig:zoff_DL30Lp_DLp_M1sol_exp}
\end{figure*}

Fig.~\ref{fig:zoff_DL30Lp_DLp_M1sol_exp} varies $t_0$, $\tilde t$ and $x_s$ at fixed $z_{\rm off}$. The analytic curves nearly collapse and remain approximately parallel to the leading prediction $\Delta\Lp/\Lp^{\rm avg}=3.83\,|z_{\rm off}|$, so substantial changes in the deposition time-scale, the diffusion time-scale or the source size produce only modest corrections to the normalized peak spread; within the analytic model the normalized width-ratio spread is likewise far more sensitive to $z_{\rm off}$ than to these parameters. The fifth parameter, $M_{{}^{56}\rm Ni}$, need not be varied: at fixed geometry and deposition history it multiplies the light curve and cancels from the normalized anisotropy, although a physical change in $M_{{}^{56}\rm Ni}$ can also modify $x_s$, $t_0$ and the effective opacity. These results confirm, in the full finite-time solution and against the RT calculations, that the normalized peak spread depends primarily on $z_{\rm off}$ over the tested ranges of the remaining parameters.

\begin{figure}
    \centering
    \includegraphics[width=0.9\linewidth]{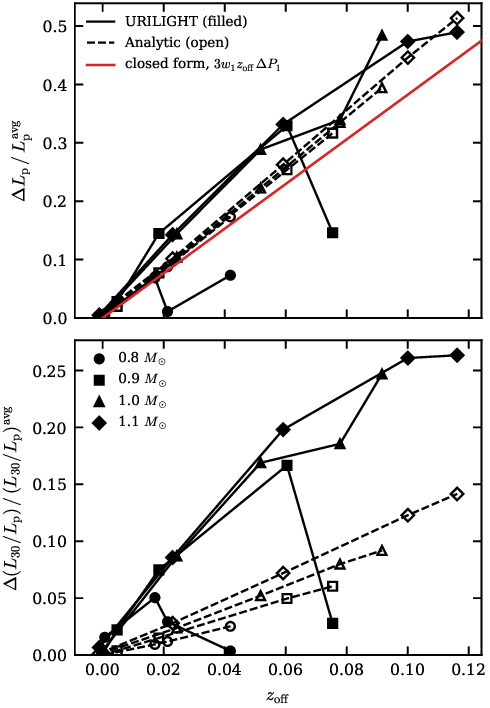}
    \caption{Normalized outer-bin spreads of the peak luminosity (top) and the width ratio (bottom) against the fitted source displacement for all $19$ off-centre ignition models. Marker shape identifies the ejecta mass (circles, squares, triangles and diamonds for $0.8$, $0.9$, $1.0$ and $1.1\,M_\odot$); filled markers and solid curves are \textsc{URILIGHT}, open markers and dashed curves the analytic models. The red line is the leading quasi-steady prediction $3.83\,|z_{\rm off}|$ (Section~\ref{sec:limb_darkening}). The analytic peak spread follows this relation with a finite-time correction, while the MC results include the non-monotonic cases discussed in Section~\ref{sec:geometry}; the analytic width-ratio spread is generally smaller than the MC value.}
    \label{fig:zoff_DL30Lp_DLp_Mi}
\end{figure}

Fig.~\ref{fig:zoff_DL30Lp_DLp_Mi} extends Fig.~\ref{fig:zoff_DL30Lp_DLp_M1sol_exp} across ejecta mass: instead of varying one analytic parameter at a time for the $1\,M_\odot$ models, it shows the same outer-bin spreads against $z_{\rm off}$ for all $19$ off-centre ignition models ($0.8$--$1.1\,M_\odot$), each with its own toy-model parameters (Table~\ref{tab:example_table}). The red line in the top panel is the same closed form, $\Delta\Lp/\Lp^{\rm avg}=3.83\,|z_{\rm off}|$, which has no free parameter and no dependence on ejecta mass. Three models are clear exceptions in Fig.~\ref{fig:zoff_DL30Lp_DLp_Mi}: models \#4, \#9 and \#14, for which the \textsc{URILIGHT} $\Lp$ is non-monotonic with viewing angle\footnote{Of the other $16$ models, nine (\#7, \#8, \#11--\#13 and \#16--\#19) are strictly monotonic. In the four central-ignition models (\#1, \#5, \#10 and \#15) $\Lp$ is lowest near the equator, with a maximum-to-minimum variation of $1.4$--$2.6$ per cent, and in models \#2, \#3 and \#6 it peaks at $\cos\theta_k=0.5$--$0.7$ rather than at the ${}^{56}$Ni-facing pole. In models \#2, \#3 and \#6, $\Lp$ departs from monotonicity by at most $1.3$ per cent, both between adjacent bins and from its maximum to the ${}^{56}$Ni-facing pole.}. In the $0.9\,M_\odot$, $z_{\rm ig}=0.75$ model (\#9), $\Lp$ peaks near $\cos\theta_k\simeq0.5$ and decreases toward the ${}^{56}$Ni-facing pole, so that its outer-bin spread ($14.6$ per cent) understates its full maximum-to-minimum variation ($28$ per cent) and falls well below the closed form; the decline from the maximum to the ${}^{56}$Ni-facing pole is $\myapprox12$ per cent in model \#9 and $\myapprox4$ per cent in models \#4 and \#14. Repeating the \textsc{URILIGHT} calculations for these models with the grey opacity of the corresponding toy model, held fixed in space and time, removes the non-monotonicity of models \#9 and \#14, reduces it to $1$ per cent in model \#4, and improves the agreement with the overall trend, but does not eliminate the discrepancy with the closed form. The $0.8\,M_\odot$ models also depart from the closed form; since $M_{{}^{56}\rm Ni}$ cancels from the normalized anisotropy, their low ${}^{56}$Ni mass is not by itself the reason, and we have not isolated it.

\section{Head-on White Dwarf Collisions}
\label{sec:collisions}

We now apply the oblate framework to the head-on WD collision models of \citet{Kushnir_2013}, whose ejecta are intrinsically oblate, in contrast to the approximately spherical ejecta of the off-centre ignition models. We consider six equal-mass collisions, from $0.6+0.6$ to $1.0+1.0\,M_\odot$, for which we compare angle-resolved \textsc{URILIGHT} light curves with the analytic solution. The two lighter equal-mass collisions, $0.5+0.5$ and $0.55+0.55\,M_\odot$, are excluded: both their total density and their ${}^{56}$Ni distribution are prolate, whereas the analytic solution of Appendix~\ref{app:oblate_diffusion} applies to oblate spheroids only (Appendix~\ref{app:collision_equal_mass}). For each of the six equal-mass collision models, we construct an effective oblate toy model in close analogy to Section~\ref{sec:toy_construction}; the construction, the resulting parameters, and the light-curve comparisons are given in Appendix~\ref{app:collision} (Table~\ref{tab:collision_table}). Across the ten viewing angles, the per-collision median absolute relative error in $\Lp$ is $0.8$--$5.5$ per cent. The effective oblate construction also captures the magnitude of the peak-luminosity spread, despite the departures of the actual ${}^{56}$Ni distribution from an oblate spheroid (Appendix~\ref{app:collision_models}). The width ratio has a systematic offset, but its variation with viewing angle is generally small, as discussed below.

\begin{figure}
    \centering
    \includegraphics[width=0.9\linewidth]{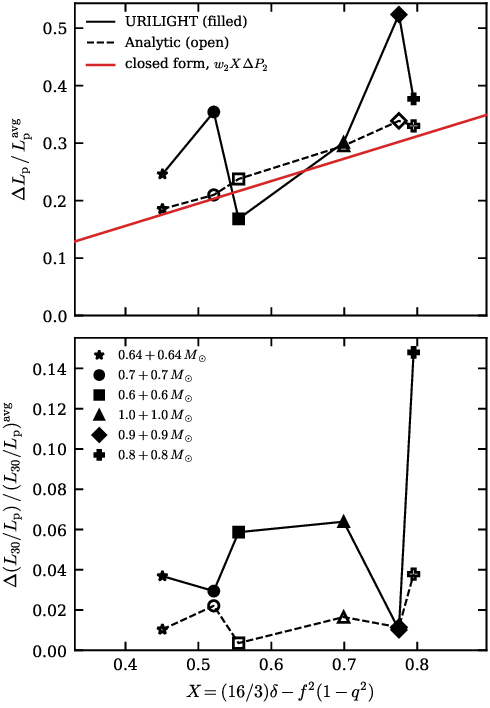}
    \caption{Normalized viewing-angle variability of the six equal-mass collisions (Table~\ref{tab:collision_table}) against the single oblate shape parameter $X=(16/3)\delta-f^2(1-q^2)$ of Equation~\eqref{eq:ld_Phi2_summary}. Top: the peak-luminosity spread $\Delta\Lp/\Lp^{\rm avg}$. Bottom: the width-ratio spread $\Delta(\Lth/\Lp)/(\Lth/\Lp)^{\rm avg}$. Filled markers joined by solid lines are \textsc{URILIGHT} MC calculations; the same markers left open and joined by dashed lines are the analytic toy model; and each marker shape is one collision. Each $\Delta$ is the outer-bin spread for equatorially symmetric ejecta, pole-on ($\cos\theta_k=0.9$) minus edge-on ($\cos\theta_k=0.1$) (Section~\ref{sec:mc_validation_sphere}). The red line in the top panel is the closed form of Section~\ref{sec:limb_darkening}, $w_2X\,\Delta P_2$ with $w_2=13/40$ and no calibrated factor (Appendix~\ref{app:ld_peak}), and $\Delta P_2=P_2(0.9)-P_2(0.1)=1.2$ the range of $P_2$ over the extreme folded bin centres, which is a prediction and not a fit. Marker shapes are assigned in order of $X$ rather than of WD mass. There is no corresponding closed-form expression for the width ratio, so the bottom panel shows no red line. The analytic solution generally predicts a smaller width-ratio spread, but the absolute variation is small, as illustrated by the collision tracks in Fig.~\ref{fig:Lp_L30Lp_obs} and quantified in the text.}
    \label{fig:collisions_delta_vs_X}
\end{figure}

Fig.~\ref{fig:collisions_delta_vs_X} shows the outer-bin spreads of all six collisions against the shape parameter $X$ of Section~\ref{sec:limb_darkening} (Table~\ref{tab:collision_table}), which plays the role of $z_{\rm off}$ for the off-centre models (Fig.~\ref{fig:zoff_DL30Lp_DLp_Mi}).

The oblate toy model captures the scale of the \textsc{URILIGHT} peak variations, but does not reproduce their ordering with $X$ (Table~\ref{tab:collision_spreads}): the analytic spread in $\Lp$ increases approximately with $X$, from $19$ to $34$ per cent, whereas the \textsc{URILIGHT} spreads range from $17$ to $52$ per cent with no such ordering. Fig.~\ref{fig:Lp_L30Lp_obs} shows the same: the collision tracks of panels (a) and (b) cover comparable ranges in $\Lp$. The tracks have only a small extent in $\Lth/\Lp$ in both calculations (Table~\ref{tab:collision_spreads}); the analytic solution generally predicts a smaller width-ratio spread, but the absolute differences are comparable to the median measurement uncertainty of $\Lth/\Lp$ in the observed sample, $0.03$ \citep{sharon_2025}. Fig.~\ref{fig:Lp_L30Lp_obs} also shows the systematic upward displacement of the analytic collision tracks in $\Lth/\Lp$, by $0.06$--$0.10$, which is distinct from their small variation with viewing angle.

\begin{table}
\centering
\caption{Outer-bin spreads of the six equal-mass collisions (pole-on, $\cos\theta_k=0.9$, minus edge-on, $\cos\theta_k=0.1$; Section~\ref{sec:mc_validation_sphere}), in order of increasing $X$: the normalized spread of $\Lp$ and the absolute difference in $\Lth/\Lp$, for \textsc{URILIGHT} (MC) and the analytic toy model. The last column is the offset of the analytic angle-averaged $\Lth/\Lp$ from the \textsc{URILIGHT} one.}
\label{tab:collision_spreads}
\setlength{\tabcolsep}{4pt}
\begin{tabular}{lccccccc}
\hline
 & & \multicolumn{2}{c}{$\Delta\Lp/\Lp^{\rm avg}$ [\%]} & \multicolumn{2}{c}{$\Delta(\Lth/\Lp)$} & \\
Collision & $X$ & MC & analytic & MC & analytic & bias \\
\hline
$0.64+0.64$ & 0.45 & 25 & 19 & 0.021 & 0.007 & $+0.09$ \\
$0.7+0.7$ & 0.52 & 35 & 21 & 0.017 & 0.015 & $+0.10$ \\
$0.6+0.6$ & 0.56 & 17 & 24 & 0.032 & 0.002 & $+0.07$ \\
$1.0+1.0$ & 0.70 & 30 & 30 & 0.040 & 0.012 & $+0.09$ \\
$0.9+0.9$ & 0.77 & 52 & 34 & 0.006 & 0.008 & $+0.06$ \\
$0.8+0.8$ & 0.79 & 38 & 33 & 0.083 & 0.025 & $+0.09$ \\
\hline
\end{tabular}
\end{table}

The closed form, Equation~\eqref{eq:ld_oblate} in Fig.~\ref{fig:collisions_delta_vs_X}, lies $3$--$11$ per cent below the analytic spread of these six configurations (median $7$ per cent; all have $X\le0.80$), the combined finite-time and shape error discussed in Appendix~\ref{app:ld_peak}. The remaining differences from \textsc{URILIGHT} primarily reflect the effective representation of the numerical ejecta by the toy model.

%
\section{Discussion and Conclusions}
\label{sec:discussion}

We derived viewing-angle-dependent diffusion light curves for homologously expanding, uniform-density ejecta with a displaced spherical source or concentric oblate source and boundary (Sections~\ref{sec:diffusion_framework} and~\ref{sec:limb_darkening}, Appendices~\ref{app:analytical_derivation} and~\ref{app:limb_darkening}). An Eddington surface-emission law converts the diffusive flux into the observer's luminosity. The leading quasi-steady peak anisotropy is dipolar for a displaced sphere, $\Lp(\theta)/\Lp^{\rm avg}\simeq1+(17/8)\,z_{\rm off}\cos\theta$, and quadrupolar for oblate ejecta, controlled by the single shape parameter $X$. The spherical coefficient is about $11$ per cent below the finite-time value in the tested configurations.

The idealized Monte Carlo tests support the treatment near maximum light (Section~\ref{sec:mc_validation}). Effective representations of the $19$ off-centre explosions of \citet{Schinasi_2024} recover peak time, peak luminosity and width ratio with median errors of $3$, $4$ and $6$ per cent, and their normalized peak spread follows the closed form in $z_{\rm off}$ (Sections~\ref{sec:toy_model} and~\ref{sec:geometry}). Six collision models also recover the scale of the peak-luminosity variation, although their MC spreads are not ordered by the effective shape parameter (Section~\ref{sec:collisions}). The main limitation is the angular variation of the width ratio, which the off-centre toy models typically underestimate by a factor of $2$--$3$; the model-\#13 comparisons identify contributions from the opacity and transport treatments without uniquely isolating the geometric reduction (Section~\ref{sec:width_mismatch}).

The effective displacement describes the radioactive-source geometry of the toy model; it is not a universal estimator of the ${}^{56}$Ni mass centroid, and nebular line shifts, which probe the emitting-ion distribution, must be compared with the ejecta models separately (Section~\ref{sec:toy_construction}). The model tracks remain inconsistent with the observed luminosity--width distribution of \citet{sharon_2025}, and the transverse viewing-angle variation adds to the tension for asymmetric models; a quantitative exclusion would require an orientation-weighted comparison in the residual coordinate of the observed relation, including measurement errors. The solutions provide an interpretation of the peak-luminosity anisotropy within the tested optically thick regime, and their solvers are publicly available (see Data availability).

\section*{Acknowledgements}
We thank Boaz Katz for valuable discussions. We also thank Avshalom Badash and Tal Wasserman for their helpful input. DK is supported by a research grant from The Abramson Family Center for Young Scientists, an ISF grant, the Minerva Stiftung, and the Pazi Foundation.

%
\section*{Data availability}
The analytic diffusion solvers developed and used in this work are publicly available at \url{https://github.com/danmvdb/viewing-angle-lc} (release v1.0, MIT licence, the version used for all analytic calculations of this paper). The light curves and light-curve observables used in the comparisons of this paper are provided as CSV files in the online supplementary material, which also contains typeset versions of the tables. For the $19$ off-centre ignition models (Table~\ref{tab:example_table}) and the six equal-mass collision models (Table~\ref{tab:collision_table}), they list, for each of the ten viewing-angle bins and for the angle average, the peak time $t_{\rm peak}$, the peak luminosity $\Lp$ and the ratio $\Lth/\Lp$ of \textsc{URILIGHT} and of the analytic calculations with both deposition prescriptions. They also contain the angle-resolved \textsc{URILIGHT} and analytic light curves $L_{\rm iso}(t)$ of Figs~\ref{fig:diff_sol_zoff_01_fdep1}, \ref{fig:diff_sol_oblate_fdep1}, \ref{fig:1Msun_ejecta109_sol} and~\ref{fig:M06_M06_sol}, together with their peak observables; the error-ladder calculations of Fig.~\ref{fig:check_geom}; the analytic parameter variations of Fig.~\ref{fig:zoff_DL30Lp_DLp_M1sol_exp}; the analytic calculations testing the closed-form peak anisotropy (Fig.~\ref{fig:limb_darkening}); and the weak-oblate relaxation calculations of Appendix~\ref{app:ld_peak}, together with the exact oblate solutions they are compared with. The solver configuration files of the two validation runs of Section~\ref{sec:mc_validation} are included with the CSV files.

\bibliographystyle{mnras}
\bibliography{example}

@article{Kushnir_2013,
doi = {10.1088/2041-8205/778/2/L37},
url = {https://doi.org/10.1088/2041-8205/778/2/L37},
year = {2013},
month = {nov},
publisher = {The American Astronomical Society},
volume = {778},
number = {2},
pages = {L37},
author = {Kushnir, Doron and Katz, Boaz and Dong, Subo and Livne, Eli and Fernández, Rodrigo},
title = {HEAD-ON COLLISIONS OF WHITE DWARFS IN TRIPLE SYSTEMS COULD EXPLAIN TYPE Ia SUPERNOVAE},
journal = {\apjl},
}

@article{Kushnir_2019,
doi = {10.3847/2515-5172/ab5064},
url = {https://doi.org/10.3847/2515-5172/ab5064},
year = {2019},
month = {oct},
publisher = {The American Astronomical Society},
volume = {3},
number = {10},
pages = {162},
author = {Kushnir, Doron and Katz, Boaz},
title = {Analytical Calculation of the Numerical Results of Khatami and Kasen for Transient Peak Time and Luminosity},
journal = {Research Notes of the AAS}
}

@article{Schinasi_2024,
    author = {Schinasi-Lemberg, Eden and Kushnir, Doron},
    title = {Off-centre ignition of sub-Chandrasekhar white dwarfs does not resolve the tension with the observed t0–MNi56 relation of type Ia supernovae},
    journal = {\mnras},
    volume = {536},
    number = {3},
    pages = {3041-3052},
    year = {2025},
    month = {12},
    issn = {0035-8711},
    doi = {10.1093/mnras/stae2770},
    url = {https://doi.org/10.1093/mnras/stae2770},
    eprint = {https://academic.oup.com/mnras/article-pdf/536/3/3041/61218698/stae2770.pdf},
}

@article{Kasen_2006,
doi = {10.1086/506190},
url = {https://doi.org/10.1086/506190},
year = {2006},
month = {nov},
publisher = {},
volume = {651},
number = {1},
pages = {366},
author = {Kasen, Daniel and Thomas, R. C. and Nugent, P.},
title = {Time-dependent Monte Carlo Radiative Transfer Calculations for Three-dimensional Supernova Spectra, Light Curves, and Polarization},
journal = {\apj},
}

@article{Wygoda_2019a,
    author = {Wygoda, Nahliel and Elbaz, Yonatan and Katz, Boaz},
    title = {Type Ia supernovae have two physical width–luminosity relations and they favour sub-Chandrasekhar and direct collision models – I. Bolometric},
    journal = {\mnras},
    volume = {484},
    number = {3},
    pages = {3941-3950},
    year = {2019},
    month = {01},
    issn = {0035-8711},
    doi = {10.1093/mnras/stz145},
    url = {https://doi.org/10.1093/mnras/stz145},
    eprint = {https://academic.oup.com/mnras/article-pdf/484/3/3941/27738414/stz145.pdf},
}

@article{Wygoda_2019b,
    author = {Wygoda, Nahliel and Elbaz, Yonatan and Katz, Boaz},
    title = {Type Ia supernovae have two physical width–luminosity relations and they favour sub-Chandrasekhar and direct collision models - II. Colour evolution},
    journal = {\mnras},
    volume = {484},
    number = {3},
    pages = {3951-3967},
    year = {2019},
    month = {01},
    issn = {0035-8711},
    doi = {10.1093/mnras/stz146},
    url = {https://doi.org/10.1093/mnras/stz146},
    eprint = {https://academic.oup.com/mnras/article-pdf/484/3/3951/27738382/stz146.pdf},
}

@ARTICLE{Pinto_Eastman_2000a,
       author = {{Pinto}, Philip A. and {Eastman}, Ronald G.},
        title = "{The Physics of Type IA Supernova Light Curves. I. Analytic Results and Time Dependence}",
      journal = {\apj},
         year = 2000,
        month = feb,
       volume = {530},
       number = {2},
        pages = {744-756},
          doi = {10.1086/308376},
       adsurl = {https://ui.adsabs.harvard.edu/abs/2000ApJ...530..744P}
}

@article{Pinto_Eastman_2000b,
doi = {10.1086/308380},
url = {https://doi.org/10.1086/308380},
year = {2000},
month = {feb},
publisher = {},
volume = {530},
number = {2},
pages = {757},
author = {Pinto, Philip A. and Eastman, Ronald G.},
title = {The Physics of Type Ia Supernova Light Curves. II.
Opacity and Diffusion},
journal = {\apj},
}

@misc{Jeffery_1999,
      title={Radioactive Decay Energy Deposition in Supernovae and the Exponential/Quasi-Exponential Behavior of Late-Time Supernova Light Curves}, 
      author={David J. Jeffery},
      year={1999},
      eprint={astro-ph/9907015},
      archivePrefix={arXiv},
      primaryClass={astro-ph},
}

@article{Junde_1999,
title = {Nuclear Data Sheets for A = 56},
journal = {Nuclear Data Sheets},
volume = {86},
number = {2},
pages = {315-424},
year = {1999},
issn = {0090-3752},
doi = {10.1006/ndsh.1999.0004},
url = {https://www.sciencedirect.com/science/article/pii/S0090375299900042},
author = {Huo Junde},
}

@article{sharon_2025,
    author = {Sharon, Amir and Kushnir, Doron and Wygoda, Nahliel},
    title = {All known Type Ia supernovae models fail to reproduce the observed bolometric luminosity–width correlation},
    journal = {\mnras},
    volume = {540},
    number = {4},
    pages = {3247-3262},
    year = {2025},
    month = {07},
    issn = {0035-8711},
    doi = {10.1093/mnras/staf808},
    url = {https://doi.org/10.1093/mnras/staf808},
    eprint = {https://academic.oup.com/mnras/article-pdf/540/4/3247/63509320/staf808.pdf},
}

@article{Sharon_2020a,
title = {All Known SN Ia Models Fail to Reproduce the Observed t$_{0}$--M$_{\rm Ni56}$ Correlation},
journal = {Research Notes of the AAS},
doi = {10.3847/2515-5172/abb9a3},
url = {https://doi.org/10.3847/2515-5172/abb9a3},
year = {2020},
month = {sep},
publisher = {The American Astronomical Society},
volume = {4},
number = {9},
pages = {158},
author = {Sharon, Amir and Kushnir, Doron},
}

@book{Flammer_1957,
  title={Spheroidal Wave Functions},
  author={Flammer, C.},
  isbn={9780598752079},
  lccn={57005621},
  year={1957},
  publisher={Stanford University Press}
}

@ARTICLE{Colgate_McKee_1969,
       author = {{Colgate}, Stirling A. and {McKee}, Chester},
        title = "{Early Supernova Luminosity}",
      journal = {\apj},
         year = 1969,
        month = aug,
       volume = {157},
        pages = {623},
          doi = {10.1086/150102},
       adsurl = {https://ui.adsabs.harvard.edu/abs/1969ApJ...157..623C}
}

@ARTICLE{Arnett_1982,
       author = {{Arnett}, W.~D.},
        title = "{Type I supernovae. I - Analytic solutions for the early part of the light curve}",
      journal = {\apj},
         year = 1982,
        month = feb,
       volume = {253},
        pages = {785-797},
          doi = {10.1086/159681},
       adsurl = {https://ui.adsabs.harvard.edu/abs/1982ApJ...253..785A}
}

@ARTICLE{Katz_2013,
       author = {{Katz}, Boaz and {Kushnir}, Doron and {Dong}, Subo},
        title = "{An exact integral relation between the Ni56 mass and the bolometric light curve of a type Ia supernova}",
      journal = {ArXiv e-prints},
         year = 2013,
        month = jan,
          doi = {10.48550/arXiv.1301.6766},
archivePrefix = {arXiv},
       eprint = {1301.6766},
 primaryClass = {astro-ph.HE}
}

@ARTICLE{Khatami_Kasen_2019,
       author = {{Khatami}, David K. and {Kasen}, Daniel N.},
        title = "{Physics of Luminous Transient Light Curves: A New Relation between Peak Time and Luminosity}",
      journal = {\apj},
         year = 2019,
        month = jun,
       volume = {878},
       number = {1},
          eid = {56},
        pages = {56},
          doi = {10.3847/1538-4357/ab1f09},
archivePrefix = {arXiv},
       eprint = {1812.06522},
 primaryClass = {astro-ph.HE},
       adsurl = {https://ui.adsabs.harvard.edu/abs/2019ApJ...878...56K}
}

@article{Darbha_2020,
doi = {10.3847/1538-4357/ab9a34},
url = {https://doi.org/10.3847/1538-4357/ab9a34},
year = {2020},
month = {jul},
publisher = {The American Astronomical Society},
volume = {897},
number = {2},
pages = {150},
author = {Darbha, Siva and Kasen, Daniel},
title = {Inclination Dependence of Kilonova Light Curves from Globally Aspherical Geometries},
journal = {\apj},
}

@article{Sim_2007,
    author = {Sim, S. A.},
    title = {Multidimensional simulations of radiative transfer in Type Ia supernovae},
    journal = {\mnras},
    volume = {375},
    number = {1},
    pages = {154-162},
    year = {2007},
    month = {02},
    issn = {0035-8711},
    doi = {10.1111/j.1365-2966.2006.11271.x},
    url = {https://doi.org/10.1111/j.1365-2966.2006.11271.x},
    eprint = {https://academic.oup.com/mnras/article-pdf/375/1/154/3092332/mnras0375-0154.pdf},
}

@article{Kasen_2007,
doi = {10.1086/516834},
url = {https://doi.org/10.1086/516834},
year = {2007},
month = {jun},
publisher = {},
volume = {662},
number = {1},
pages = {459},
author = {Kasen, Daniel and Plewa, Tomasz},
title = {Detonating Failed Deflagration Model of Thermonuclear Supernovae. II. Comparison to Observations},
journal = {\apj},
}

@article{Sim_2007b,
    author = {Sim, S. A. and Sauer, D. N. and Röpke, F. K. and Hillebrandt, W.},
    title = {Light curves for off-centre ignition models of Type Ia supernovae},
    journal = {\mnras},
    volume = {378},
    number = {1},
    pages = {2-12},
    year = {2007},
    month = {06},
    issn = {0035-8711},
    doi = {10.1111/j.1365-2966.2007.11795.x},
    url = {https://doi.org/10.1111/j.1365-2966.2007.11795.x},
    eprint = {https://academic.oup.com/mnras/article-pdf/378/1/2/3958841/mnras0378-0002.pdf},
}

@article{Townsley_2019,
doi = {10.3847/2041-8213/ab27cd},
url = {https://doi.org/10.3847/2041-8213/ab27cd},
year = {2019},
month = {jun},
publisher = {The American Astronomical Society},
volume = {878},
number = {2},
pages = {L38},
author = {Townsley, Dean M. and Miles, Broxton J. and Shen, Ken J. and Kasen, Daniel},
title = {Double Detonations with Thin, Modestly Enriched Helium Layers can Make Normal Type Ia Supernovae},
journal = {\apjl},
}

@article{Fink_2010,
    author  = {Fink, M. and R{\"o}pke, F. K. and Hillebrandt, W. and
               Seitenzahl, I. R. and Sim, S. A. and Kromer, M.},
    title   = {Double-detonation sub-Chandrasekhar supernovae:
               Can minimum helium shell masses detonate the core?},
    journal = {\aap},
    volume  = {514},
    pages   = {A53},
    year    = {2010},
    doi     = {10.1051/0004-6361/200913892}
}

@ARTICLE{Livne_1990,
       author = {{Livne}, Eli},
        title = "{Successive Detonations in Accreting White Dwarfs as an Alternative Mechanism for Type I Supernovae}",
      journal = {\apjl},
         year = 1990,
        month = may,
       volume = {354},
        pages = {L53},
          doi = {10.1086/185721},
       adsurl = {https://ui.adsabs.harvard.edu/abs/1990ApJ...354L..53L}
}

@article{Shen_2014,
doi = {10.1088/0004-637X/785/1/61},
url = {https://doi.org/10.1088/0004-637X/785/1/61},
year = {2014},
month = {mar},
publisher = {The American Astronomical Society},
volume = {785},
number = {1},
pages = {61},
author = {Shen, Ken J. and Bildsten, Lars},
title = {THE IGNITION OF CARBON DETONATIONS VIA CONVERGING SHOCK WAVES IN WHITE DWARFS},
journal = {\apj},
}

@article{Hoflich_1991,
       author = {{H{\"o}flich}, P.},
        title = "{Asphericity effects in scattering dominated photospheres}",
      journal = {\aap},
         year = 1991,
        month = feb,
       volume = {246},
       number = {2},
        pages = {481-489},
       adsurl = {https://ui.adsabs.harvard.edu/abs/1991A%26A...246..481H},
}

@article{Nomoto_1982a,
    author  = {Nomoto, Ken'ichi},
    title   = {Accreting White Dwarf Models for Type I Supernovae. I. Presupernova Evolution and Triggering Mechanisms},
    journal = {\apj},
    volume  = {253},
    pages   = {798--810},
    year    = {1982},
    doi     = {10.1086/159682}
}

@article{Nomoto_1982b,
    author  = {Nomoto, Ken'ichi},
    title   = {Accreting White Dwarf Models for Type I Supernovae. II. Off-Center Detonation Supernovae},
    journal = {\apj},
    volume  = {257},
    pages   = {780--792},
    year    = {1982},
    doi     = {10.1086/160031}
}

@article{Woosley_Weaver_1994,
    author  = {Woosley, S. E. and Weaver, T. A.},
    title   = {Sub-Chandrasekhar Mass Models for Type Ia Supernovae},
    journal = {\apj},
    volume  = {423},
    pages   = {371--379},
    year    = {1994},
    doi     = {10.1086/173813}
}

@article{Bildsten_2007,
    author  = {Bildsten, Lars and Shen, Ken J. and Weinberg, Nevin N. and Nelemans, Gijs},
    title   = {Faint Thermonuclear Supernovae from AM Canum Venaticorum Binaries},
    journal = {\apjl},
    volume  = {662},
    pages   = {L95--L98},
    year    = {2007},
    doi     = {10.1086/519489}
}

@article{Fink_2007,
    author  = {Fink, M. and Hillebrandt, W. and R{\"o}pke, F. K.},
    title   = {Double-detonation supernovae of sub-Chandrasekhar mass white dwarfs},
    journal = {\aap},
    volume  = {476},
    pages   = {1133--1143},
    year    = {2007},
    doi     = {10.1051/0004-6361:20078438}
}

@article{Moore_2013,
    author  = {Moore, Kevin and Townsley, Dean M. and Bildsten, Lars},
    title   = {The Effects of Curvature and Expansion on Helium Detonations on White Dwarf Surfaces},
    journal = {\apj},
    volume  = {776},
    pages   = {97},
    year    = {2013},
    doi     = {10.1088/0004-637X/776/2/97}
}

@article{Shen_Moore_2014,
    author  = {Shen, Ken J. and Moore, Kevin},
    title   = {The Initiation and Propagation of Helium Detonations in White Dwarf Envelopes},
    journal = {\apj},
    volume  = {797},
    pages   = {46},
    year    = {2014},
    doi     = {10.1088/0004-637X/797/1/46}
}

@article{Boos_2021,
    author  = {Boos, S. J. and Townsley, D. M. and Shen, K. J. and Caldwell, S. and Miles, B. J.},
    title   = {Multidimensional Parameter Study of Double Detonation Type Ia Supernovae Originating from Thin Helium Shell White Dwarfs},
    journal = {\apj},
    volume  = {919},
    pages   = {126},
    year    = {2021},
    doi     = {10.3847/1538-4357/ac07a2}
}

@article{Rosswog_2009,
    author  = {Rosswog, S. and Kasen, D. and Guillochon, J. and Ramirez-Ruiz, E.},
    title   = {Collisions of White Dwarfs as a New Progenitor Channel for Type Ia Supernovae},
    journal = {\apjl},
    volume  = {705},
    pages   = {L128--L132},
    year    = {2009},
    doi     = {10.1088/0004-637X/705/2/L128}
}

@article{Raskin_2010,
    author  = {Raskin, C. and Scannapieco, E. and Rockefeller, G. and Fryer, C. and Diehl, S. and Timmes, F. X.},
    title   = {56Ni Production in Double-degenerate White Dwarf Collisions},
    journal = {\apj},
    volume  = {724},
    pages   = {111--125},
    year    = {2010},
    doi     = {10.1088/0004-637X/724/1/111}
}

@ARTICLE{Ambwani_1988,
       author = {{Ambwani}, Kapil and {Sutherland}, Peter},
        title = "{Gamma-ray spectra and energy deposition for type IA supernovae}",
      journal = {\apj},
         year = 1988,
        month = feb,
       volume = {325},
        pages = {820-827},
          doi = {10.1086/166052},
       adsurl = {https://ui.adsabs.harvard.edu/abs/1988ApJ...325..820A},
}

@ARTICLE{Maeda_2010,
       author = {{Maeda}, K. and {Benetti}, S. and {Stritzinger}, M. and {R{\"o}pke}, F.~K. and {Folatelli}, G. and {Sollerman}, J. and {Taubenberger}, S. and {Nomoto}, K. and {Leloudas}, G. and {Hamuy}, M. and {Tanaka}, M. and {Mazzali}, P.~A. and {Elias-Rosa}, N.},
        title = "{An asymmetric explosion as the origin of spectral evolution diversity in type Ia supernovae}",
      journal = {\nat},
         year = 2010,
        month = jul,
       volume = {466},
       number = {7302},
        pages = {82-85},
          doi = {10.1038/nature09122},
       adsurl = {https://ui.adsabs.harvard.edu/abs/2010Natur.466...82M},
}

@ARTICLE{Maguire_2018,
       author = {{Maguire}, K. and {Sim}, S.~A. and {Shingles}, L. and {Spyromilio}, J. and {Jerkstrand}, A. and {Sullivan}, M. and {Chen}, T. -W. and {Cartier}, R. and {Dimitriadis}, G. and {Frohmaier}, C. and {Galbany}, L. and {Gutierrez}, C.~P. and {Hosseinzadeh}, G. and {Howell}, D.~A. and {Inserra}, C. and {Rudy}, R. and {Sollerman}, J.},
        title = "{Using late-time optical and near-infrared spectra to constrain Type Ia supernova explosion properties}",
      journal = {\mnras},
         year = 2018,
        month = jul,
       volume = {477},
       number = {3},
        pages = {3567-3582},
          doi = {10.1093/mnras/sty820},
       adsurl = {https://ui.adsabs.harvard.edu/abs/2018MNRAS.477.3567M},
}

@ARTICLE{Dong_2015,
       author = {{Dong}, Subo and {Katz}, Boaz and {Kushnir}, Doron and {Prieto}, Jose L.},
        title = "{Type Ia supernovae with bimodal explosions are common - possible smoking gun for direct collisions of white dwarfs}",
      journal = {\mnras},
         year = 2015,
        month = nov,
       volume = {454},
       number = {1},
        pages = {L61-L65},
          doi = {10.1093/mnrasl/slv129},
       adsurl = {https://ui.adsabs.harvard.edu/abs/2015MNRAS.454L..61D},
}

@ARTICLE{Kwok_2023,
       author = {{Kwok}, L.~A. and {Jha}, S.~W. and {Temim}, T. and {Fox}, O.~D. and {Larison}, C. and {Camacho-Neves}, Y. and {Brenner Newman}, M.~J. and {Pierel}, J.~D.~R. and {Foley}, R.~J. and {Andrews}, J.~E. and {Badenes}, C. and {Barna}, B. and {Bostroem}, K.~A. and {Deckers}, M. and {Fl{\"o}rs}, A. and {Garnavich}, P. and {Graham}, M.~L. and {Graur}, O. and {Hosseinzadeh}, G. and {Howell}, D.~A. and {Hughes}, J.~P. and {Johansson}, J. and {Kendrew}, S. and {Kerzendorf}, W.~E. and {Maeda}, K. and {Maguire}, K. and {McCully}, C. and {O'Brien}, J.~T. and {Rest}, A. and {Sand}, D.~J. and {Shahbandeh}, M. and {Strolger}, L.-G. and {Szalai}, T. and {Ashall}, C. and {Baron}, E. and {Burns}, C.~R. and {DerKacy}, J.~M. and {Evans}, T.~M. and {Fisher}, A. and {Galbany}, L. and {Hoeflich}, P. and {Hsiao}, E. and {de Jaeger}, T. and {Karamehmetoglu}, E. and {Krisciunas}, K. and {Kumar}, S. and {Lu}, J. and {Maund}, J. and {Mazzali}, P.~A. and {Medler}, K. and {Morrell}, N. and {Phillips}, M.~M. and {Shappee}, B.~J. and {Stritzinger}, M. and {Suntzeff}, N. and {Telesco}, C. and {Tucker}, M. and {Wang}, L.},
        title = "{A JWST Near- and Mid-infrared Nebular Spectrum of the Type Ia Supernova 2021aefx}",
      journal = {\apjl},
         year = 2023,
        month = feb,
       volume = {944},
       number = {1},
        pages = {L3},
          doi = {10.3847/2041-8213/acb4ec},
       adsurl = {https://ui.adsabs.harvard.edu/abs/2023ApJ...944L...3K},
}

@ARTICLE{DerKacy_2023,
       author = {{DerKacy}, J.~M. and {Ashall}, C. and {Hoeflich}, P. and {Baron}, E. and {Shappee}, B.~J. and {Baade}, D. and {Andrews}, J. and {Bostroem}, K.~A. and {Brown}, P.~J. and {Burns}, C.~R. and {Burrow}, A. and {Cikota}, A. and {de Jaeger}, T. and {Do}, A. and {Dong}, Y. and {Dominguez}, I. and {Galbany}, L. and {Hsiao}, E.~Y. and {Karamehmetoglu}, E. and {Krisciunas}, K. and {Kumar}, S. and {Lu}, J. and {Mera Evans}, T.~B. and {Maund}, J.~R. and {Mazzali}, P. and {Medler}, K. and {Morrell}, N. and {Patat}, F. and {Phillips}, M.~M. and {Shahbandeh}, M. and {Stangl}, S. and {Stevens}, C.~P. and {Stritzinger}, M.~D. and {Suntzeff}, N.~B. and {Telesco}, C.~M. and {Tucker}, M.~A. and {Valenti}, S. and {Wang}, L. and {Yang}, Y. and {Jha}, S.~W. and {Kwok}, L.~A.},
        title = "{JWST Low-resolution MIRI Spectral Observations of SN 2021aefx: High-density Burning in a Type Ia Supernova}",
      journal = {\apjl},
         year = 2023,
        month = mar,
       volume = {945},
       number = {1},
        pages = {L2},
          doi = {10.3847/2041-8213/acb8a8},
       adsurl = {https://ui.adsabs.harvard.edu/abs/2023ApJ...945L...2D},
}

@ARTICLE{Motohara_2006,
       author = {{Motohara}, K. and {Maeda}, K. and {Gerardy}, C.~L. and {Nomoto}, K. and {Tanaka}, M. and {Tominaga}, N. and {Ohkubo}, T. and {Mazzali}, P.~A. and {Fesen}, R.~A. and {H{\"o}flich}, P. and {Wheeler}, J.~C.},
        title = "{The Asymmetric Explosion of Type Ia Supernovae as Seen from Near-Infrared Observations}",
      journal = {\apjl},
         year = 2006,
        month = dec,
       volume = {652},
       number = {2},
        pages = {L101-L104},
          doi = {10.1086/509919},
       adsurl = {https://ui.adsabs.harvard.edu/abs/2006ApJ...652L.101M},
}

@ARTICLE{Maeda_2010b,
       author = {{Maeda}, K. and {Taubenberger}, S. and {Sollerman}, J. and {Mazzali}, P.~A. and {Leloudas}, G. and {Nomoto}, K. and {Motohara}, K.},
        title = "{Nebular Spectra and Explosion Asymmetry of Type Ia Supernovae}",
      journal = {\apj},
         year = 2010,
        month = feb,
       volume = {708},
       number = {2},
        pages = {1703-1715},
          doi = {10.1088/0004-637X/708/2/1703},
       adsurl = {https://ui.adsabs.harvard.edu/abs/2010ApJ...708.1703M},
}

@ARTICLE{Silverman_2013,
       author = {{Silverman}, Jeffrey M. and {Ganeshalingam}, Mohan and {Filippenko}, Alexei V.},
        title = "{Berkeley Supernova Ia Program - V. Late-time spectra of Type Ia Supernovae}",
      journal = {\mnras},
         year = 2013,
        month = apr,
       volume = {430},
       number = {2},
        pages = {1030-1041},
          doi = {10.1093/mnras/sts674},
       adsurl = {https://ui.adsabs.harvard.edu/abs/2013MNRAS.430.1030S},
}

@ARTICLE{Black_2016,
       author = {{Black}, Christine S. and {Fesen}, Robert A. and {Parrent}, Jerod T.},
        title = "{Progressive redshifts in the late-time spectra of Type Ia supernovae}",
      journal = {\mnras},
         year = 2016,
        month = oct,
       volume = {462},
       number = {1},
        pages = {649-662},
          doi = {10.1093/mnras/stw1680},
       adsurl = {https://ui.adsabs.harvard.edu/abs/2016MNRAS.462..649B},
}

@ARTICLE{Graham_2017,
       author = {{Graham}, M.~L. and {Kumar}, S. and {Hosseinzadeh}, G. and {Hiramatsu}, D. and {Arcavi}, I. and {Howell}, D.~A. and {Valenti}, S. and {Sand}, D.~J. and {Parrent}, J.~T. and {McCully}, C. and {Filippenko}, A.~V.},
        title = "{Nebular-phase spectra of nearby Type Ia Supernovae}",
      journal = {\mnras},
         year = 2017,
        month = dec,
       volume = {472},
       number = {3},
        pages = {3437-3454},
          doi = {10.1093/mnras/stx2224},
       adsurl = {https://ui.adsabs.harvard.edu/abs/2017MNRAS.472.3437G},
}

@ARTICLE{Diamond_2018,
       author = {{Diamond}, T.~R. and {Hoeflich}, P. and {Hsiao}, E.~Y. and {Sand}, D.~J. and {Sonneborn}, G. and {Phillips}, M.~M. and {Hristov}, B. and {Collins}, D.~C. and {Ashall}, C. and {Marion}, G.~H. and {Stritzinger}, M. and {Morrell}, N. and {Gerardy}, C.~L. and {Penney}, R.~B.},
        title = "{Near-infrared Spectral Evolution of the Type Ia Supernova 2014J in the Nebular Phase: Implications for the Progenitor System}",
      journal = {\apj},
         year = 2018,
        month = jul,
       volume = {861},
       number = {2},
          eid = {119},
        pages = {119},
          doi = {10.3847/1538-4357/aac434},
       adsurl = {https://ui.adsabs.harvard.edu/abs/2018ApJ...861..119D},
}

@ARTICLE{Vallely_2020,
       author = {{Vallely}, P.~J. and {Tucker}, M.~A. and {Shappee}, B.~J. and {Brown}, J.~S. and {Stanek}, K.~Z. and {Kochanek}, C.~S.},
        title = "{Signatures of bimodality in nebular phase Type Ia supernova spectra}",
      journal = {\mnras},
         year = 2020,
        month = mar,
       volume = {492},
       number = {3},
        pages = {3553-3565},
          doi = {10.1093/mnras/staa003},
       adsurl = {https://ui.adsabs.harvard.edu/abs/2020MNRAS.492.3553V},
}

@ARTICLE{Kumar_2023,
       author = {{Kumar}, S. and {Hsiao}, E.~Y. and {Ashall}, C. and {Phillips}, M.~M. and {Morrell}, N. and {Hoeflich}, P. and {Burns}, C.~R. and others},
        title = "{Near-infrared and Optical Nebular-phase Spectra of Type Ia Supernovae SN 2013aa and SN 2017cbv in NGC 5643}",
      journal = {\apj},
         year = 2023,
        month = mar,
       volume = {945},
       number = {1},
          eid = {27},
        pages = {27},
          doi = {10.3847/1538-4357/acad73},
       adsurl = {https://ui.adsabs.harvard.edu/abs/2023ApJ...945...27K},
}

@ARTICLE{Blondin_2022,
       author = {{Blondin}, S. and {Blinnikov}, S. and {Callan}, F.~P. and {Collins}, C.~E. and {Dessart}, L. and {Even}, W. and {Fl{\"o}rs}, A. and {Fullard}, A.~G. and {Hillier}, D.~J. and {Jerkstrand}, A. and {Kasen}, D. and {Katz}, B. and {Kerzendorf}, W. and {Kozyreva}, A. and {O'Brien}, J. and {P{\'a}ssaro}, E.~A. and {Roth}, N. and {Shen}, K.~J. and {Shingles}, L. and {Sim}, S.~A. and {Singhal}, J. and {Smith}, I.~G. and {Sorokina}, E. and {Utrobin}, V.~P. and {Vogl}, C. and {Williamson}, M. and {Wollaeger}, R. and {Woosley}, S.~E. and {Wygoda}, N.},
        title = "{StaNdaRT: a repository of standardised test models and outputs for supernova radiative transfer}",
      journal = {\aap},
         year = 2022,
        month = dec,
       volume = {668},
          eid = {A163},
        pages = {A163},
          doi = {10.1051/0004-6361/202244134},
       adsurl = {https://ui.adsabs.harvard.edu/abs/2022A%26A...668A.163B},
}


\appendix

\section{Analytic Derivation of the Diffusion Solutions}
\label{app:analytical_derivation}

\subsection{General solution of the diffusion equation}
\label{app:general_diffusion}

For $\bar t > \bar t_\delta$, the scaled radiation energy density satisfies
the homogeneous diffusion equation. For the uniform velocity-space density and constant grey opacity assumed here, $D$ is constant and can be taken outside the divergence:
\begin{equation}
\frac{\partial \bar e}{\partial \bar t}
-D\,\nabla_{\mathbf v}^2\bar e
=0,
\qquad
D\equiv\frac{c}{3\kappa\bar{\rho}_v},
\label{eq:diff_eq}
\end{equation}
defined over a bounded domain $\mathcal S$ in velocity space, with boundary $\partial\mathcal S$. We impose an absorbing boundary condition,
\begin{equation}
\bar e(\mathbf v_s,\bar t)=0,
\qquad
\mathbf v_s\in\partial\mathcal S.
\label{eq:e_boundary}
\end{equation}

We consider a general initial condition at deposition time $\bar t_\delta$,
\begin{equation}
\bar e(\mathbf v,\bar t_\delta)=\bar\varepsilon_\delta(\mathbf v),
\label{eq:e_initial}
\end{equation}
where $\bar\varepsilon_\delta(\mathbf v)$ is an arbitrary integrable function on
$\mathcal S$.

For constant $D$, the solution is obtained by expanding $\bar e$ in the eigenfunctions of the Laplacian. Let $\{\psi_n(\mathbf v)\}$ satisfy
\begin{equation}
-\nabla_{\mathbf v}^2\psi_n = k_n^2\psi_n,
\label{eq:eigen_func}
\end{equation}
with boundary condition $\psi_n(\mathbf v_s)=0$ on $\partial\mathcal S$ and orthogonality
\begin{equation}
\int_{\mathcal S}\psi_n(\mathbf v)\psi_m(\mathbf v)\,dV
=
N_n\,\delta_{nm}.
\label{eq:orthonormal}
\end{equation}

The general solution for $\bar t>\bar t_\delta$ is
\begin{equation}
\bar e(\mathbf v,\bar t)
=
\sum_n A_n\,\psi_n(\mathbf v)\,
\exp\!\left[-Dk_n^2(\bar t-\bar t_\delta)\right],
\label{eq:e_full_sol}
\end{equation}
with expansion coefficients
\begin{equation}
A_n
=
\frac{1}{N_n}
\int_{\mathcal S}\bar\varepsilon_\delta(\mathbf v)\,\psi_n(\mathbf v)\,dV.
\label{eq:An_coeff}
\end{equation}

The physical radiative flux is related to the scaled solution by
\begin{equation}
\mathbf f(\mathbf v,t)
=
-\frac{c}{3\kappa\rho}\nabla e
=
-\frac{D}{t^2}\nabla_{\mathbf v}\bar e,
\label{eq:flux_vec_energy_relation}
\end{equation}
and the emergent surface flux is
\begin{equation}
F_\delta(\mathbf v_s,t)
=
-\frac{D}{t^2}\,
\hat{\mathbf n}\cdot\nabla_{\mathbf v}\bar e
\bigg|_{\mathbf v=\mathbf v_s},
\qquad
\mathbf v_s\in\partial\mathcal S,
\label{eq:flux_norm_energy_relation}
\end{equation}
which evaluates to
\begin{equation}
F_\delta(\mathbf v_s,t)
=
-\frac{D}{t^2}
\sum_n
A_n\,
\hat{\mathbf n}\cdot\nabla_{\mathbf v}\psi_n(\mathbf v_s)\,
\exp\!\left[-\frac{Dk_n^2}{2}(t^2-t_\delta^2)\right].
\label{eq:flux_full_sol}
\end{equation}

%
\subsection{Off-centre energy deposition}
\label{app:offcentre_diffusion}
\subsubsection*{Initial condition}

We specialize the general solution to the case of off-centre impulsive energy deposition. The ejecta occupy a sphere of radius $v_o$ in velocity space, so the domain boundary $\partial\mathcal S$ of Appendix~\ref{app:general_diffusion} corresponds here to $\mathbf v_s$ with $|\mathbf v_s|=v_o$. The initial condition $\bar\varepsilon_\delta(\mathbf v)$ is uniform within a displaced sphere of radius $x_s v_o$ whose centre is offset by $z_{\mathrm{off}}v_o$ along the symmetry axis:
\begin{equation}
\bar\varepsilon_\delta(\mathbf v)=
\frac{\bar E_\delta}{\tfrac{4\pi}{3}(x_s v_o)^3}
\begin{cases}
1, & \left|\mathbf v-z_{\mathrm{off}}v_o\,\hat{\mathbf z}\right|
\le x_s v_o, \\[3pt]
0, & \text{otherwise}.
\end{cases}
\label{eq:oc_initial_condition}
\end{equation}

\subsubsection*{Eigenfunctions}

Because the initial condition is axisymmetric about the $z$-axis, only $m=0$ modes contribute. The eigenfunctions reduce to
\begin{equation}
\psi_{nl}(v,\theta)
=
j_l\!\left(\frac{\alpha_{l+1/2,n}}{v_o}\,v\right)
P_l(\cos\theta),
\label{eq:oc_eigen_func}
\end{equation}
where $j_l$ are spherical Bessel functions, $P_l$ are Legendre polynomials, and $\alpha_{l+1/2,n}$ are the positive roots of $j_l(\alpha_{l+1/2,n})=0$, enforcing the boundary condition at $v=v_o$.

\subsubsection*{General solution}

Equation~\eqref{eq:e_full_sol} becomes
\begin{equation}
\begin{split}
\bar e(v,\theta,\bar t)
={}&
\sum_{n,l}
A_{nl}\,
j_l\!\left(\frac{\alpha_{l+1/2,n}}{v_o}\,v\right)
P_l(\cos\theta)\\
&\times
\exp\!\bigg[
-D\left(\frac{\alpha_{l+1/2,n}}{v_o}\right)^2
(\bar t-\bar t_\delta)
\bigg].
\end{split}
\label{eq:oc_e_full_sol}
\end{equation}

\subsubsection*{Expansion coefficients}

Projecting the initial condition onto the eigenfunctions gives
\begin{equation}
A_{nl}
=
\frac{1}{N_{nl}}
\int_{\mathcal S}\bar\varepsilon_\delta(\mathbf v)\,
j_l\!\left(\frac{\alpha_{l+1/2,n}}{v_o}\,v\right)
P_l(\cos\theta)\,dV,
\label{eq:oc_Anl_coeff}
\end{equation}
with normalization
\begin{equation}
N_{nl}
=
\frac{2\pi v_o^3}{2l+1}\,
j_{l+1}^2(\alpha_{l+1/2,n}).
\label{eq:oc_norm}
\end{equation}
Using $dV=v^2\sin\theta\,dv\,d\theta\,d\phi$ and integrating over $\phi$:
\begin{equation}
\begin{split}
A_{nl}
={}&
\frac{(2l+1)\bar E_\delta}
{v_o^3\,j_{l+1}^2(\alpha_{l+1/2,n})\,
\tfrac{4\pi}{3}(x_s v_o)^3}\\
&\times
\int_0^{\theta_{\rm max}} \!\!\sin\theta\,d\theta\,P_l(\cos\theta)
\!\!\int_{v_-^{+}(\theta)}^{v_+(\theta)}\!\!\!
j_l\!\left(\frac{\alpha_{l+1/2,n}}{v_o}\,v\right)
v^2\,dv,
\end{split}
\label{eq:oc_Anl_coeff_full}
\end{equation}
where the two radial bounds of the displaced sphere are
\begin{equation}
v_\pm(\theta)
=
v_o\left[
z_{\mathrm{off}}\cos\theta
\pm
\sqrt{x_s^2-z_{\mathrm{off}}^2\sin^2\theta}
\right],
\label{eq:oc_vpm}
\end{equation}
with $v_-^{+}(\theta)\equiv\max\left[v_-(\theta),0\right]$, and where the polar integration is restricted to the cone subtended by the source,
\begin{equation}
\theta_{\rm max}
=
\begin{cases}
\pi, & z_{\mathrm{off}}\le x_s,\\[3pt]
\arcsin\!\left(x_s/z_{\mathrm{off}}\right),
& z_{\mathrm{off}}>x_s .
\end{cases}
\label{eq:oc_thetamax}
\end{equation}
The two cases are distinguished by whether the heated region contains the origin. For $z_{\mathrm{off}}\le x_s$ it does: the square root is real for every $\theta$, $v_-\le0$, and the radial integral runs from the centre. For $z_{\mathrm{off}}>x_s$ the origin lies outside the source, the square root is real only within $\theta_{\rm max}$, and both roots are positive, so the integral runs over the chord $v_-(\theta)\le v\le v_+(\theta)$.

\subsubsection*{Surface flux}
On the boundary $v=v_o$, $\hat{\mathbf n}\cdot\nabla_{\mathbf v}\psi_{nl}
=(\alpha_{l+1/2,n}/v_o)\,j_l'(\alpha_{l+1/2,n})\,P_l(\cos\theta)$,
giving
\begin{equation}
\begin{split}
F_\delta(\theta,t)
={}&
-\frac{D}{t^2}
\sum_{n,l}
A_{nl}\,
\frac{\alpha_{l+1/2,n}}{v_o}\,
j_l'(\alpha_{l+1/2,n})\,
P_l(\cos\theta)\\
&\times
\exp\!\left[
-\frac{D}{2}
\left(\frac{\alpha_{l+1/2,n}}{v_o}\right)^2
(t^2-t_\delta^2)
\right].
\end{split}
\label{eq:oc_flux_full_sol}
\end{equation}
Here $j_l'(x)$ denotes the derivative of $j_l(x)$ with respect to its argument. This impulse response is convolved with $Q(t)$ through the convolution in Equation~\eqref{eq:convolution} to obtain the full luminosity evolution.

\subsection{Oblate spheroidal ejecta geometry}
\label{app:oblate_diffusion}

We now derive the analogous solution for oblate spheroidal ejecta containing a concentric oblate spheroidal source: we introduce oblate spheroidal coordinates, record the observer projection for this geometry, and construct the eigenfunction expansion and the resulting surface flux.
\subsubsection*{Coordinate system and scale factors}

We introduce oblate spheroidal coordinates $(\xi,\eta,\phi)$ in velocity space, related to Cartesian coordinates by
\begin{equation}
\begin{aligned}
v_x &= a\cosh\xi\sin\eta\cos\phi,\\
v_y &= a\cosh\xi\sin\eta\sin\phi,\\
v_z &= a\sinh\xi\cos\eta,
\end{aligned}
\label{eq:oblate_coords}
\end{equation}
where $\xi\ge 0$, $\eta\in[0,\pi]$, $\phi\in[0,2\pi]$, and
\begin{equation}
a=\sqrt{v_{\mathrm{eq}}^2-v_{\mathrm{pol}}^2}
\end{equation}
is the focal length, with $v_{\rm eq}$ and $v_{\rm pol}$ the equatorial and polar velocities of the outer ejecta boundary. These coordinates parametrize the same velocity-space domain as in the general solution of Appendix~\ref{app:general_diffusion}; the general boundary surface $\partial\mathcal S$ is here the single-coordinate surface $\xi=\xi_s$. The outer ejecta surface is defined by $\tanh\xi_s=v_{\mathrm{pol}}/v_{\mathrm{eq}}$, so that $v_{\mathrm{eq}}=a\cosh\xi_s$ and $v_{\mathrm{pol}}=a\sinh\xi_s$. 

The scale factors are
\begin{equation}
h_\xi=h_\eta=a\sqrt{\sinh^2\xi+\cos^2\eta},
\qquad
h_\phi=a\cosh\xi\sin\eta,
\end{equation}
and the volume element is
\begin{equation}
dV=a^3\bigl(\sinh^2\xi+\cos^2\eta\bigr)\cosh\xi\sin\eta\,d\xi\,d\eta\,d\phi.
\end{equation}

\subsubsection*{Viewing-angle projection for oblate ejecta}

For completeness, we record here the oblate-spheroidal analogue of the spherical projection (Section~\ref{sec:spherical_geom}). In the oblate spheroidal coordinate system the unit outward normal at the ejecta surface $\xi = \xi_s$ is
\begin{equation}
\hat{\mathbf{n}} = \frac{1}{\sqrt{\sinh^2 \xi_s + \cos^2 \eta}}
\begin{pmatrix}
\sinh \xi_s \sin \eta \cos \phi \\
\sinh \xi_s \sin \eta \sin \phi \\
\cosh \xi_s \cos \eta
\end{pmatrix}.
\end{equation}
The direction cosine toward an observer at
$\hat{\mathbf{k}}=(\sin\theta_k,0,\cos\theta_k)^T$ is
\begin{equation}
\mu(\eta, \phi; \theta_k) =
\frac{\sinh \xi_s \sin \eta \cos \phi \sin \theta_k
      + \cosh \xi_s \cos \eta \cos \theta_k}
     {\sqrt{\sinh^2 \xi_s + \cos^2 \eta}},
\end{equation}
and the differential surface element in velocity space (the physical surface element is $t^2dA$) is
\begin{equation}
dA = a^2 \cosh \xi_s \sin \eta
     \sqrt{\sinh^2 \xi_s + \cos^2 \eta}\, d\eta\, d\phi.
\end{equation}
The isotropic equivalent luminosity is obtained by integrating over the
visible surface $\mu>0$.

\subsubsection*{Initial condition}
We consider an oblate spheroidal source, centred on the origin and sharing the symmetry axis of the ejecta, with semi-axes $v_{\rm eq}^{\rm src} = fa\cosh\xi_1$ and $v_{\rm pol}^{\rm src} = fa\sinh\xi_1$, where $f > 0$ scales the overall source size relative to the ejecta focal length $a$ and $\xi_1$ sets its axis ratio ($\tanh\xi_1 = v_{\rm pol}^{\rm src}/v_{\rm eq}^{\rm src}$). The only constraint on $f$ is that the source lies within the ejecta. Since both spheroids are centred on the origin and share the symmetry axis, this requires $f\cosh\xi_1\le\cosh\xi_s$ and $f\sinh\xi_1\le\sinh\xi_s$. The initial condition is
\begin{equation}
\bar{\varepsilon}_\delta(\mathbf{v})
=
\frac{\bar{E}_\delta}{V_{\rm src}}
\begin{cases}
1, &
\dfrac{v_x^2+v_y^2}{\left(v_{\rm eq}^{\rm src}\right)^2}
+\dfrac{v_z^2}{\left(v_{\rm pol}^{\rm src}\right)^2}
\le 1, \\[8pt]
0, & \text{otherwise,}
\end{cases}
\label{eq:oblate_initial}
\end{equation}
where
\begin{equation}
V_{\rm src}
=
\frac{4\pi}{3}
\left(v_{\rm eq}^{\rm src}\right)^2 v_{\rm pol}^{\rm src}
=
\frac{4\pi}{3}\,f^3 a^3\cosh^2\!\xi_1\sinh\xi_1.
\end{equation}
For $f=1$ the source is confocal with the outer boundary and the membership condition simplifies to $\xi \le \xi_1$.


\subsubsection*{Eigenfunctions of the Laplacian}

For axisymmetric ($m=0$) modes, the eigenfunctions that vanish on $\xi=\xi_s$ separate as
\begin{equation}
\begin{split}
\psi_{np}(\xi,\eta)
&=
R_n^{(0)}(c_{np},\sinh\xi)\,S_n^{(0)}(c_{np},\cos\eta)\\
&\equiv R_{np}(\sinh\xi)\,S_{np}(\cos\eta),
\end{split}
\end{equation}
where $R_n^{m=0}$ and $S_n^{m=0}$ are the oblate spheroidal radial and angular functions of the first kind with spheroidal parameter $c_{np}$ \citep[following the conventions of][]{Flammer_1957}. The Dirichlet boundary condition $R_{np}(\sinh\xi_s)=0$ determines $c_{np}$ as the $p$th positive root for fixed $n$, with eigenvalue $k_{np}^2=c_{np}^2/a^2$.

\subsubsection*{General solution}

\begin{equation}
\bar e(\xi,\eta,\bar t)
=
\sum_{n,p}
A_{np}\,R_{np}(\sinh\xi)\,S_{np}(\cos\eta)\,
\exp\!\left[
-\frac{c_{np}^2}{a^2}D(\bar t-\bar t_\delta)
\right].
\label{eq:oblate_solution}
\end{equation}

\subsubsection*{Expansion coefficients}

\begin{equation}
\begin{split}
A_{np}
={}&
\frac{2\pi a^3\bar{E}_\delta}{V_{\rm src}\,N_{np}}
\int_0^\pi \int_0^{\xi_{\rm max}(\eta)}
R_{np}(\sinh\xi)\,S_{np}(\cos\eta)\\
&\times
\bigl(\sinh^2\!\xi+\cos^2\!\eta\bigr)
\cosh\xi\sin\eta\;d\xi\;d\eta,
\end{split}
\end{equation}
with norm
\begin{equation}
\begin{split}
N_{np}
={}&
2\pi a^3
\int_0^{\xi_s}\!\!\int_0^\pi
R_{np}^2(\sinh\xi)\,S_{np}^2(\cos\eta)\\
&\times
\bigl(\sinh^2\xi+\cos^2\eta\bigr)
\cosh\xi\sin\eta\,d\eta\,d\xi,
\end{split}
\end{equation}
and where $\xi_{\rm max}(\eta)$ is the $\eta$-dependent upper limit defined implicitly by the source boundary condition,
\begin{equation}
\left(\frac{\cosh\xi_{\rm max}(\eta)\,\sin\eta}{f\cosh\xi_1}\right)^2
+
\left(\frac{\sinh\xi_{\rm max}(\eta)\,\cos\eta}{f\sinh\xi_1}\right)^2
= 1,
\label{eq:xi_max}
\end{equation}
solved numerically for each $\eta$. Explicitly, with $A_s\equiv f\cosh\xi_1$ and $C_s\equiv f\sinh\xi_1$ (in units of the ejecta focal length), $\sinh^2\xi_{\rm max}(\eta)=(1-\sin^2\eta/A_s^2)/(\sin^2\eta/A_s^2+\cos^2\eta/C_s^2)$. Where the numerator is non-positive, $\sin^2\eta\ge A_s^2$ (possible only for $f\cosh\xi_1<1$), the source does not extend into that angular strip, which contributes zero to the integral; for the validation run of Section~\ref{sec:mc_validation_oblate} ($f=0.7$, $\xi_1=0.7$) this is the case for $61.5^\circ\lesssim\eta\lesssim118.5^\circ$. For $f=1$ this reduces to $\xi_{\rm max}(\eta)=\xi_1$ for all $\eta$, recovering the confocal result; the Jacobian $\sinh^2\xi+\cos^2\eta$ is then a sum of separable products, so the coefficient integral becomes a sum of products of one-dimensional integrals. Because the source is centred at the origin and symmetric about the equatorial plane, only even-$n$ modes contribute.


\subsubsection*{Surface flux}

On the surface $\xi=\xi_s$, $\hat{\mathbf n}\cdot\nabla_{\mathbf v} =(1/h_\xi)\,\partial/\partial\xi =(a\sqrt{\sinh^2\xi_s+\cos^2\eta})^{-1}\partial/\partial\xi$, so
\begin{equation}
\begin{split}
F_\delta(\eta,t)
={}&
-\frac{D\cosh\xi_s}{a\,t^2\sqrt{\sinh^2\xi_s+\cos^2\eta}}\\
&\times\sum_{n,p}
A_{np}\,R_{np}'(\sinh\xi_s)\,S_{np}(\cos\eta)\\
&\times
\exp\!\left[
-\frac{D}{2}\left(\frac{c_{np}}{a}\right)^2
(t^2-t_\delta^2)
\right],
\end{split}
\label{eq:oblate_flux}
\end{equation}
where $R_{np}'$ denotes the derivative of $R_{np}$ with respect to its argument, the factor $\cosh\xi_s$ arising from $\partial_\xi\sinh\xi$.

\section{Closed Form for the Viewing-Angle Dependence of the Peak
Luminosity}
\label{app:limb_darkening}

Section~\ref{sec:limb_darkening} states the closed forms for the viewing-angle dependence of the peak luminosity (Equations~\ref{eq:ld_sphere} and~\ref{eq:ld_oblate}); here we derive them. We first derive the observer-projection weights (Appendix~\ref{app:ld_projection}), then obtain the quasi-steady source, boundary, and curvature contributions from the time-integrated diffusion response (Appendix~\ref{app:ld_multipoles}). The final part of Appendix~\ref{app:ld_multipoles} tests their accuracy at finite time.

\subsection{The observer integral reduces to fixed weights}
\label{app:ld_projection}

For spherical ejecta of radius $R=v_ot$, the surface element is $dA=R^2d\Omega_{\hat{\mathbf n}}$, and substituting the Eddington closure into the definition of $L_{\rm iso}$ (Section~\ref{sec:viewing_angle}) gives
\begin{equation}
L_{\rm iso}(\theta_k)=3R^2\int_{\hat{\mathbf n}\cdot\hat{\mathbf k}>0}
F(\theta)\left[\mu^2+\tfrac{2}{3}\mu\right]
d\Omega_{\hat{\mathbf n}},
\label{eq:ld_Liso_raw}
\end{equation}
with $\mu=\hat{\mathbf n}\cdot\hat{\mathbf k}$. Expand the surface flux in Legendre polynomials, $F(\theta)=\sum_lF_lP_l(\cos\theta)$, and note that $\cos\theta=\hat{\mathbf n}\cdot\hat{\mathbf z}$ while the bracket depends on $\hat{\mathbf n}$ only through $\hat{\mathbf n}\cdot\hat{\mathbf k}$. The Funk--Hecke theorem then performs the angular integral exactly, for any function $g$:
\begin{equation}
\begin{split}
\int P_l(\hat{\mathbf n}\cdot\hat{\mathbf z})\,
&g(\hat{\mathbf n}\cdot\hat{\mathbf k})\,d\Omega_{\hat{\mathbf n}}\\
&=2\pi P_l(\cos\theta_k)\int_{-1}^{1}\!P_l(\mu)\,g(\mu)\,d\mu.
\end{split}
\label{eq:funk_hecke}
\end{equation}
Applying it to Equation~\eqref{eq:ld_Liso_raw} with $g(\mu)=3(\mu^2+\tfrac{2}{3}\mu)$ for $\mu>0$ and $g=0$ otherwise, and using $4\pi R^2F_0=\oint F\,dA=L$, gives Equations~\eqref{eq:ld_series} and \eqref{eq:ld_weights}, with
\begin{equation}
w_l=\left\{1,\ \frac{17}{24},\ \frac{13}{40},\ \frac{1}{16},\
-\frac{1}{48},\dots\right\}.
\label{eq:ld_weights_values}
\end{equation}
The entire $\theta_k$ dependence therefore factors into the Legendre polynomials.  The original two-dimensional integral over the visible surface is reduced to a weighted sum with fixed coefficients $w_l$.

The same reduction applies to any smooth convex boundary after one geometric substitution. The outward normal is then in one-to-one correspondence with a surface point, and the Gauss map gives $dA=d\Omega_{\hat{\mathbf n}}/K$, where $K$ is the Gaussian curvature. Convexity also ensures that the visible surface is precisely the region $\mu>0$, with no self-shadowing. Equations~\eqref{eq:ld_Liso_raw} and \eqref{eq:funk_hecke} therefore go through with $R^2F$ replaced by $\Phi\equiv F/K$, giving
\begin{equation}
L_{\rm iso}(\theta_k)=L\left[1+\sum_{l\ge1}w_l\,
\frac{\Phi_l}{\Phi_0}\,P_l(\cos\theta_k)\right],
\qquad L=4\pi\Phi_0.
\label{eq:ld_series_convex}
\end{equation}
The sphere is the special case $K=1/R^2$, for which $\Phi\propto F$ and the substitution is invisible. All the shape information is carried by $1/K(\hat{\mathbf n})$, the emitting area per unit solid angle of normal.

\subsection{Quasi-steady flux multipoles from the Poisson problem}
\label{app:ld_multipoles}
\label{app:ld_peak}
This subsection derives the angular multipoles of the surface flux without carrying the full time-dependent eigenfunction expansion. We first show that the time integral of an impulse response obeys an exact Poisson equation. We then identify the additional assumptions required to apply this zero-frequency response to continuously heated ejecta over a finite epoch. Only after making that distinction do we derive the spherical and weakly oblate multipoles. The accuracy of the quasi-steady replacement near maximum light is assessed at the end of the subsection.

The Eddington closure does not enter the derivation of the local diffusive flux; it is used only when that flux is projected toward the observer in Appendix~\ref{app:ld_projection}. Likewise, for a non-spherical boundary, we keep separate the anisotropy of the local diffusive flux and the geometrical factor $1/K$ that converts physical surface area into solid angle of the outward normal.

\subsubsection*{Time-integrated response and the quasi-steady approximation}

Consider a unit impulse deposited at transformed time $\bar t_\delta$. Its subsequent evolution obeys Equation~\eqref{eq:diff_eq},
\begin{equation}
\frac{\partial\bar e}{\partial\bar t}
-D\nabla_{\mathbf v}^2\bar e=0,
\qquad
\bar e(\mathbf v,\bar t_\delta)=\bar\varepsilon_\delta(\mathbf v),
\end{equation}
with $\bar e=0$ on $\partial\mathcal S$. Define the zeroth temporal moment of the impulse response by
\begin{equation}
\mathcal U(\mathbf v)
\equiv
D\int_{\bar t_\delta}^{\infty}
\bar e(\mathbf v,\bar t)\,d\bar t
=D\int_0^{\infty}\bar e(\mathbf v,\bar t_\delta+\bar t')\,d\bar t' .
\label{eq:ld_U_def}
\end{equation}
The second form is written in terms of the lag $\bar t'=\bar t-\bar t_\delta$. Since $D$ and $\partial\mathcal S$ do not depend on $\bar t$, $\bar e$ depends on $\bar t$ and $\bar t_\delta$ only through $\bar t-\bar t_\delta$ (each mode of Equation~\ref{eq:flux_full_sol} decays as $\exp[-Dk_n^2(\bar t-\bar t_\delta)]$), so $\mathcal U$ does not depend on $\bar t_\delta$. Integrating the diffusion equation over $\bar t$ and using $\bar e(\mathbf v,\infty)=0$ gives
\begin{equation}
-\nabla_{\mathbf v}^2\mathcal U
=\bar\varepsilon_\delta(\mathbf v),
\qquad
\mathcal U=0\quad\text{on }\partial\mathcal S .
\label{eq:ld_poisson}
\end{equation}
Equation~\eqref{eq:ld_poisson} is therefore an exact identity for the time-integrated impulse response; no quasi-steady approximation has yet been made. The integration to infinity defines the zero-frequency response of the diffusion problem.

To relate this identity to continuous radioactive heating, let $F_{\delta,l}(\bar t)$ denote the $l$th surface-flux multipole produced a transformed time $\bar t$ after a unit impulse.  At epoch $\bar t$, the corresponding continuously forced multipole follows the same convolution as Equation~\eqref{eq:convolution},
\begin{equation}
\begin{split}
F_l(\bar t)&\propto
\int_0^{\bar t}d\bar t'\,Q(\bar t')\,F_{\delta,l}(\bar t-\bar t')\\
&\simeq
Q(\bar t)\int_0^\infty F_{\delta,l}(\bar t')\,d\bar t'\\
&=-Q(\bar t)\,\big[\hat{\mathbf n}\cdot\nabla_{\mathbf v}\mathcal U\big]_l ,
\end{split}
\label{eq:ld_qs_convolution}
\end{equation}
where the last integral has been rewritten in terms of the lag $\bar t-\bar t'$. Here $F_{\delta,l}$ denotes the scaled impulse-flux multipole, $t^2$ times the physical flux, as a function of the lag; the physical multipole at the observation time follows on dividing by the observation-time $t^2$, a factor common to all $l$ that cancels in the ratios $F_l/F_0$. In the last line, $[\,\cdot\,]_l$ is the $l$th Legendre coefficient on $\partial\mathcal S$. By Equation~\eqref{eq:flux_norm_energy_relation} the scaled impulse flux is $t^2F_\delta=-D\,\hat{\mathbf n}\cdot\nabla_{\mathbf v}\bar e$ on $\partial\mathcal S$; integrating it over the lag, exchanging the time integral with the gradient and using the second form of Equation~\eqref{eq:ld_U_def} gives $\int_0^\infty F_\delta\,d\bar t'=-\hat{\mathbf n}\cdot\nabla_{\mathbf v}\mathcal U$ for a unit impulse (for general $\bar E_\delta$ both sides scale with $\bar E_\delta$, which cancels in the ratios $F_l/F_0$). The normal derivative of $\mathcal U$ is therefore the time-integrated impulse flux, not the instantaneous $F_{\delta,l}$. The second step contains the quasi-steady approximation. It requires two conditions. Let
\begin{equation}
\tau_n\equiv\frac{1}{Dk_n^2}
\end{equation}
be the relaxation time of eigenmode $n$ in $\bar t$, and define the local physical heating time by $t_Q\equiv|d\ln Q/dt|^{-1}$. Since $d\bar t=t\,dt$, the heating varies on the transformed-time scale $\bar t_Q=t\,t_Q$. Mode $n$ is quasi-steady when
\begin{equation}
\eta_{Q,n}\equiv\frac{\tau_n}{t\,t_Q}\ll1,
\qquad
\eta_{{\rm age},n}\equiv\frac{\tau_n}{\bar t}
=\frac{2\tau_n}{t^2}\ll1.
\label{eq:ld_qs_conditions}
\end{equation}
The first condition allows $Q(\bar t')$ to be treated as constant across the width of the response kernel, while the second allows the finite upper limit $\bar t$ to be extended to infinity.  When both conditions hold, the final integral in Equation~\eqref{eq:ld_qs_convolution} is given by the last line of that equation, i.e. by the surface derivative of $\mathcal U$.  The angular dependence of the quasi-steady surface flux is therefore that of $-\hat{\mathbf n}\cdot\nabla_{\mathbf v}\mathcal U$, and common time-dependent factors cancel from the multipole ratios. We denote the resulting zero-frequency ratios by $a_l^{\rm qs}\equiv F_l^{\rm qs}/F_0^{\rm qs}$; below, the superscript is occasionally suppressed when the quasi-steady meaning is unambiguous.

\subsubsection*{Equivalence to stationary diffusion in physical coordinates}

At a fixed but arbitrary epoch $t$, freeze the homologous ejecta and set $\mathbf x=t\mathbf v$. The boundary and source are then static; $t$ only fixes their physical scale. Defining
\begin{equation}
\mathcal U_x(\mathbf x;t)\equiv\frac{1}{t}\,\mathcal U\!\left(\frac{\mathbf x}{t}\right),
\qquad
\varepsilon_{\delta,x}(\mathbf x;t)\equiv\frac{1}{t^3}\,\bar\varepsilon_\delta\!\left(\frac{\mathbf x}{t}\right),
\label{eq:ld_phys_map}
\end{equation}
the factor $t^{-3}$ preserves the deposited energy, and $\nabla_{\mathbf v}=t\nabla_{\mathbf x}$ maps Equation~\eqref{eq:ld_poisson} to the Poisson form of a stationary diffusion problem,
\begin{equation}
-\nabla_{\mathbf x}^2\,\mathcal U_x=\varepsilon_{\delta,x},
\qquad
\mathcal U_x=0\quad\text{on }t\,\partial\mathcal S .
\label{eq:ld_poisson_phys}
\end{equation}
Its outward surface-flux pattern obeys
\begin{equation}
-\hat{\mathbf n}\cdot\nabla_{\mathbf x}\mathcal U_x
=-t^{-2}\hat{\mathbf n}\cdot\nabla_{\mathbf v}\mathcal U.
\end{equation}
Because the factor $t^{-2}$ is common to all angles, it cancels from the normalized multipoles. The quasi-steady counterparts of the multipole ratios $a_l\equiv F_l/F_0$ that enter the observer projection, Equation~\eqref{eq:ld_series}, are therefore
\begin{equation}
a_l^{\rm qs}
\equiv\frac{F_l^{\rm qs}}{F_0^{\rm qs}}
=\frac{\left[-\hat{\mathbf n}\cdot\nabla_{\mathbf x}
\mathcal U_x\right]_l}
{\left[-\hat{\mathbf n}\cdot\nabla_{\mathbf x}
\mathcal U_x\right]_0}
=\frac{\left[-\hat{\mathbf n}\cdot\nabla_{\mathbf v}
\mathcal U\right]_l}
{\left[-\hat{\mathbf n}\cdot\nabla_{\mathbf v}
\mathcal U\right]_0}.
\label{eq:ld_al_qs_def}
\end{equation}
Thus, $a_l^{\rm qs}$ is obtained exactly by solving this static Poisson problem for the ejecta frozen at any epoch.

\subsubsection*{Spherical outer boundary: source multipoles}

Let the outer boundary be the sphere $v=v_o$, with the deposited energy entirely contained inside it, and define
\begin{equation}
M_l\equiv\frac{1}{\bar E_\delta}
\int_{\mathcal S}
\bar\varepsilon_\delta(\mathbf v')
\left(\frac{v'}{v_o}\right)^l
P_l(\cos\theta')\,dV',
\qquad
\bar E_\delta\equiv
\int_{\mathcal S}\bar\varepsilon_\delta\,dV .
\label{eq:ld_moments}
\end{equation}
The normalization gives $M_0=1$.

It is useful to separate the free-space field generated by the source from the correction required by the absorbing outer boundary. The free-space solution of $-\nabla^2\mathcal U_{\rm free}=\bar\varepsilon_\delta$ is
\begin{equation}
\mathcal U_{\rm free}(\mathbf v)
=\frac{1}{4\pi}
\int\frac{\bar\varepsilon_\delta(\mathbf v')}
{|\mathbf v-\mathbf v'|}\,dV'.
\label{eq:ld_Ufree}
\end{equation}
For a field point outside the source, the standard multipole expansion of $|\mathbf v-\mathbf v'|^{-1}$ gives, for an axisymmetric source,
\begin{equation}
\mathcal U_{{\rm free},l}(v,\theta)
=\frac{\bar E_\delta}{4\pi v_o}M_l
\left(\frac{v_o}{v}\right)^{l+1}
P_l(\cos\theta).
\label{eq:ld_Ufree_l}
\end{equation}
This field does not in general vanish at $v=v_o$.  We therefore add a boundary correction $\mathcal U_{{\rm b},l}$.  The source is already accounted for by $\mathcal U_{\rm free}$, so the correction is harmonic, meaning that it satisfies $\nabla^2\mathcal U_{{\rm b},l}=0$.  The regular axisymmetric harmonic of angular order $l$ is proportional to $v^lP_l(\cos\theta)$; imposing $\mathcal U_{{\rm free},l}+\mathcal U_{{\rm b},l}=0$ at $v=v_o$ therefore gives
\begin{equation}
\mathcal U_{{\rm b},l}(v,\theta)
=-\frac{\bar E_\delta}{4\pi v_o}M_l
\left(\frac{v}{v_o}\right)^l
P_l(\cos\theta).
\label{eq:ld_Uboundary_l}
\end{equation}
The complete $l$th multipole in the source-free region next to the surface is thus
\begin{equation}
\mathcal U_l(v,\theta)
=\frac{\bar E_\delta}{4\pi v_o}M_l
\left[
\left(\frac{v_o}{v}\right)^{l+1}
-\left(\frac{v}{v_o}\right)^l
\right]P_l(\cos\theta).
\label{eq:ld_U_l}
\end{equation}
Differentiating at the surface gives
\begin{equation}
-\left.\frac{\partial\mathcal U_l}{\partial v}\right|_{v_o}
=\frac{\bar E_\delta}{4\pi v_o^2}
\big[(l+1)+l\big]M_lP_l(\cos\theta)
=\frac{\bar E_\delta}{4\pi v_o^2}(2l+1)M_lP_l(\cos\theta).
\label{eq:ld_flux_multipole_derivative}
\end{equation}
Since $M_0=1$, the monopole is
$\bar E_\delta/(4\pi v_o^2)$, so
\begin{equation}
a_l^{\rm qs}=(2l+1)M_l .
\label{eq:ld_al_moments}
\end{equation}
This is the electrostatic problem of a charge distribution $\bar\varepsilon_\delta$ enclosed by a grounded conducting sphere: $\mathcal U$ plays the role of the potential, $\mathcal U_{{\rm b},l}$ that of the image charges, and the surface flux $-\partial\mathcal U/\partial v$ that of the induced surface charge (with opposite sign), whose multipoles are the familiar $(2l+1)M_l$. Equation~\eqref{eq:ld_al_moments} applies specifically to a spherical outer boundary. It maps multipole moments of the radioactive source onto multipoles of the surface flux; anisotropy arising from deforming the outer boundary must be treated separately.

\subsubsection*{Off-centre source}

For the displaced uniform sphere of Equation~\eqref{eq:oc_initial_condition}, the leading asymmetry is a dipole. Explicitly,
\begin{equation}
M_1
=\frac{1}{\bar E_\delta v_o}
\int\bar\varepsilon_\delta(\mathbf v')\,v'\cos\theta'\,dV'
=\frac{\langle z'\rangle}{v_o}
=z_{\rm off},
\end{equation}
so Equation~\eqref{eq:ld_al_moments} gives
\begin{equation}
a_1^{\rm qs}=3z_{\rm off}.
\label{eq:ld_a1_offcenter}
\end{equation}
More generally, $v'^lP_l(\cos\theta')$ is a solid harmonic.  The mean-value theorem for harmonic functions states that its average over a sphere equals its value at the centre of that sphere. The displaced uniform source therefore has
\begin{equation}
M_l=z_{\rm off}^{\,l},
\qquad
a_l^{\rm qs}=(2l+1)z_{\rm off}^{\,l},
\label{eq:ld_al_sphere}
\end{equation}
provided $x_s+|z_{\rm off}|\le1$. The quasi-steady pattern is therefore independent of $x_s$: outside a uniformly charged sphere the potential is that of a point charge at its centre, so the surface flux is that induced by a point source at $z_{\rm off}v_o$. Summing Equation~\eqref{eq:ld_al_sphere} with $\sum_l(2l+1)z^lP_l(\cos\theta)=(1-z^2)/(1+z^2-2z\cos\theta)^{3/2}$ gives the quasi-steady surface flux in closed form, the charge induced on a grounded sphere by a single point charge,
\begin{equation}
\frac{F^{\rm qs}(\theta)}{F_0^{\rm qs}}
=\frac{1-z_{\rm off}^2}{\left(1+z_{\rm off}^2-2z_{\rm off}\cos\theta\right)^{3/2}},
\label{eq:ld_flux_point_image}
\end{equation}
exact for every multipole. Thus $a_1^{\rm qs}=O(z_{\rm off})$, whereas $a_2^{\rm qs}=5z_{\rm off}^2=O(z_{\rm off}^2)$. The leading off-centre anisotropy is therefore dipolar, although Equation~\eqref{eq:ld_al_sphere} is exact for every multipole of a displaced uniform sphere that remains inside the ejecta.  Inserting it into Equation~\eqref{eq:ld_series} gives Equation~\eqref{eq:ld_sphere}; with $w_1=17/24$ and $w_2=13/40$, the first two numerical coefficients are $3w_1=17/8$ and $5w_2=13/8$, as in Equation~\eqref{eq:ld_sphere_num}.

\subsubsection*{Weakly oblate source and outer boundary}

For the oblate configuration, both the source and the ejecta are centred and symmetric under $z\rightarrow-z$.  All odd multipoles therefore vanish, so the quadrupole is the leading anisotropic term. All multipoles in this and the following subsubsection are the quasi-steady ones of Equation~\eqref{eq:ld_al_qs_def}; to lighten the notation we drop the superscript qs, writing $F_l$, $\Phi_l$ and $a_2$ for $F_l^{\rm qs}$, $\Phi_l^{\rm qs}$ and $a_2^{\rm qs}$. We write the ejecta boundary $v_{\rm b}$ and the heating-source boundary $v_{\rm src}$ as weakly deformed spheres,
\begin{equation}
v_{\rm b}(\theta)=v_o\left[1+\varepsilon P_2(\cos\theta)\right],
\qquad
v_{\rm src}(\theta)=x_sv_o\left[1+\varepsilon_sP_2(\cos\theta)\right],
\label{eq:ld_weak_oblate_surfaces}
\end{equation}
with $|\varepsilon|,|\varepsilon_s|\ll1$. For an oblate ejecta, $\varepsilon<0$. If $\delta\equiv1-v_{\rm pol}/v_{\rm eq}$, then fixing $v_o$ by requiring that the deformed boundary enclose the same volume as the sphere of radius $v_o$ (so that the deformation redistributes the ejecta without changing $\bar\rho_v$) gives $\varepsilon=-2\delta/3+O(\delta^2)$. At first order, the source and boundary perturbations may be calculated independently and then added.

\medskip\noindent\emph{Source quadrupole.}
First, keep the outer boundary spherical and deform only the uniform source. Equation~\eqref{eq:ld_moments} gives
\begin{equation}
\begin{split}
M_2
&=\frac{1}{V_{\rm src}v_o^2}
\int d\Omega\,P_2(\cos\theta)
\int_0^{v_{\rm src}(\theta)}v'^4\,dv' \\
&=\frac{(x_sv_o)^5}{5V_{\rm src}v_o^2}
\int d\Omega\,P_2
\left[1+5\varepsilon_sP_2+O(\varepsilon_s^2)\right].
\end{split}
\end{equation}
Using $\int P_2\,d\Omega=0$,
$\int P_2^2\,d\Omega=4\pi/5$, and
$V_{\rm src}=4\pi(x_sv_o)^3/3+O(\varepsilon_s^2)$ yields
\begin{equation}
M_2=\frac{3}{5}\varepsilon_sx_s^2.
\label{eq:ld_M2_source}
\end{equation}
Because the outer boundary is spherical in this intermediate calculation, Equation~\eqref{eq:ld_al_moments} applies and gives
\begin{equation}
a_2^{\rm(src)}=5M_2=3\varepsilon_sx_s^2.
\label{eq:ld_a2_source}
\end{equation}
The factor $5$ is simply $2l+1$ evaluated at $l=2$.

\medskip\noindent\emph{Outer-boundary quadrupole.}
Now keep the source spherical and centred and deform only the outer boundary. In the source-free region near the surface, the unperturbed monopole solution (the $l=0$, source-free limit of Equation~\ref{eq:ld_moments}) is
\begin{equation}
\mathcal U^{(0)}(v)
=\frac{\bar E_\delta}{4\pi}
\left(\frac{1}{v}-\frac{1}{v_o}\right).
\label{eq:ld_U0_oblate}
\end{equation}
Write the perturbed solution as $\mathcal U=\mathcal U^{(0)}+\mathcal U^{(1)}$. The source has not changed, so $\mathcal U^{(1)}$ is harmonic, $\nabla^2\mathcal U^{(1)}=0$.  The Dirichlet condition must hold on the moved surface $v=v_o+v^{(1)}$, where $v^{(1)}=\varepsilon v_oP_2$. Expanding it about the spherical surface,
\begin{equation}
0=\mathcal U^{(0)}(v_o)
+v^{(1)}\,\partial_v\mathcal U^{(0)}(v_o)
+\mathcal U^{(1)}(v_o,\theta),
\end{equation}
and using $\mathcal U^{(0)}(v_o)=0$ gives
\begin{equation}
\mathcal U^{(1)}(v_o,\theta)
=-v^{(1)}\,\partial_v\mathcal U^{(0)}(v_o)
=\varepsilon\frac{\bar E_\delta}{4\pi v_o}P_2(\cos\theta).
\label{eq:ld_shape_boundary_value}
\end{equation}
The regular harmonic with this quadrupolar boundary value is unique and proportional to $v^2P_2$,
\begin{equation}
\mathcal U^{(1)}(v,\theta)
=\varepsilon\frac{\bar E_\delta}{4\pi v_o}
\left(\frac{v}{v_o}\right)^2P_2(\cos\theta).
\label{eq:ld_shape_correction}
\end{equation}
The boundary deformation modifies the surface flux through two first-order contributions. First, the unperturbed monopole gradient must be evaluated at the displaced surface,
\begin{equation}
-\left.\partial_v\mathcal U^{(0)}\right|_{v_o(1+\varepsilon P_2)}
=\frac{\bar E_\delta}{4\pi v_o^2}
\left[1-2\varepsilon P_2(\cos\theta)\right].
\label{eq:ld_shifted_monopole_flux}
\end{equation}
Second, the harmonic correction contributes
\begin{equation}
-\left.\partial_v\mathcal U^{(1)}\right|_{v_o}
=-\frac{\bar E_\delta}{4\pi v_o^2}
2\varepsilon P_2(\cos\theta).
\label{eq:ld_harmonic_flux_correction}
\end{equation}
The tilt of the surface normal relative to $\hat{\mathbf e}_v$ contributes only at second order: the zeroth-order field is radial, while the angular gradient is already first order. Adding Equations~\eqref{eq:ld_shifted_monopole_flux} and
\eqref{eq:ld_harmonic_flux_correction},
\begin{equation}
-\hat{\mathbf n}\cdot\nabla\mathcal U
=\frac{\bar E_\delta}{4\pi v_o^2}
\left[1-4\varepsilon P_2(\cos\theta)\right]
+O(\varepsilon^2),
\end{equation}
so the quadrupole generated by the outer diffusion boundary is
\begin{equation}
a_2^{\rm(bnd)}=-4\varepsilon.
\label{eq:ld_a2_boundary}
\end{equation}
The two contributions of $-2\varepsilon$ have different origins. One comes from evaluating the monopole gradient at the shifted radius; the other comes from the harmonic correction required to enforce the zero-energy-density boundary condition on the displaced surface.

\medskip\noindent\emph{Total diffusive quadrupole.}
The source and boundary corrections have the same angular dependence, $P_2(\cos\theta)$, and the Poisson equation is linear. Their first-order flux contributions therefore add:
\begin{equation}
\frac{F_2}{F_0}
\equiv a_2
=a_2^{\rm(src)}+a_2^{\rm(bnd)}
=3\varepsilon_sx_s^2-4\varepsilon .
\label{eq:ld_a2_total}
\end{equation}
The $-4\varepsilon$ term is not part of $M_2$.  The moment $M_2$ describes only the radioactive source, whereas the relation $a_l=(2l+1)M_l$ assumes a spherical outer boundary. Thus a centred spherical source has $M_2=0$, yet still produces a quadrupolar surface flux if the outer diffusion boundary is oblate.

\subsubsection*{Curvature and the quantity projected to the observer}

The previous result concerns the local diffusive flux. A second and conceptually distinct geometrical correction enters the observer integral. For a smooth convex surface, the natural variable in Equation~\eqref{eq:ld_series_convex} is $\Phi\equiv F/K$, because $dA=d\Omega_{\hat{\mathbf n}}/K$. As in the previous subsubsection, $F$ and $\Phi$ are quasi-steady; $F_\delta$ remains reserved for the impulse-response kernel in Equation~\eqref{eq:ld_qs_convolution}. For the weakly oblate spheroid with equatorial and polar semi-axes $A\equiv v_{\rm eq}$ and $C\equiv v_{\rm pol}$ (the short symbols are kept here only to keep the curvature algebra below readable),
\begin{equation}
A=v_o(1-\varepsilon/2),
\qquad
C=v_o(1+\varepsilon),
\end{equation}
the curvature measure as a function of the normal polar angle $\Theta$, defined by $\cos\Theta=\hat{\mathbf n}\cdot\hat{\mathbf z}$, is $1/K=A^4C^2/h^4$, where $h^2=A^2\sin^2\Theta+C^2\cos^2\Theta$.  Expanding to first order gives
\begin{equation}
\frac{1}{K}
=v_o^2\left[1-4\varepsilon P_2(\cos\Theta)\right]
+O(\varepsilon^2).
\label{eq:ld_curvature}
\end{equation}
This produces a second $-4\varepsilon$ quadrupole, but it is geometric rather than diffusive: it describes how much physical surface area corresponds to a fixed solid angle of outward normal.  Since the difference between $\Theta$ and the polar angle of the surface point is already $O(\varepsilon)$, it does not affect the first-order result.

To see explicitly how this curvature term enters the normalized multipole, write the diffusive surface flux from Equation~\eqref{eq:ld_a2_total} as
\begin{equation}
F(\Theta)
=F_0\left[1+a_2P_2(\cos\Theta)\right],
\qquad
a_2\equiv\frac{F_2}{F_0}
=3\varepsilon_sx_s^2-4\varepsilon .
\label{eq:ld_F_oblate_for_Phi}
\end{equation}
Using Equation~\eqref{eq:ld_curvature},
\begin{equation}
\begin{split}
\Phi(\Theta)
\equiv\frac{F(\Theta)}{K(\Theta)}
&=v_o^2F_0
\left[1+a_2P_2(\cos\Theta)\right]
\left[1-4\varepsilon P_2(\cos\Theta)\right] \\
&=v_o^2F_0
\left[1+\left(a_2-4\varepsilon\right)P_2(\cos\Theta)\right]
+O(\varepsilon^2,\varepsilon\varepsilon_s,\varepsilon_s^2).
\end{split}
\label{eq:ld_Phi_expanded}
\end{equation}
The product
$-4\varepsilon a_2P_2^2$ has been dropped because $a_2$ is already first order. Comparing Equation~\eqref{eq:ld_Phi_expanded} with the Legendre expansion
\begin{equation}
\Phi(\Theta)=\Phi_0+\Phi_2P_2(\cos\Theta)+\cdots
\end{equation}
gives, term by term,
\begin{equation}
\Phi_0=v_o^2F_0,
\qquad
\Phi_2=v_o^2\left(F_2-4\varepsilon F_0\right).
\label{eq:ld_Phi0_Phi2}
\end{equation}
The curvature term therefore survives in $\Phi_2/\Phi_0$. Its $P_2$ variation multiplies the flux monopole $F_0$ and creates an additional quadrupole in $\Phi$; only the constant factor $v_o^2$ cancels. Consequently,
\begin{equation}
\frac{\Phi_2}{\Phi_0}
=\frac{F_2}{F_0}-4\varepsilon
=3\varepsilon_sx_s^2-4\varepsilon-4\varepsilon
=3\varepsilon_sx_s^2-8\varepsilon.
\label{eq:ld_Phi2_eps}
\end{equation}
Equation~\eqref{eq:ld_Phi2_eps} contains three physically distinct terms: $3\varepsilon_sx_s^2$ from the quadrupole of the radioactive source, $-4\varepsilon$ from the deformation of the outer diffusion boundary, and a second $-4\varepsilon$ from the curvature, i.e. from the angular variation of surface area per unit solid angle of outward normal. For the source spheroid, with semi-axes $v_{\rm pol}^{\rm src}\equiv C_s=x_sv_o(1+\varepsilon_s)$ and $v_{\rm eq}^{\rm src}\equiv A_s=x_sv_o(1-\varepsilon_s/2)$, so
\begin{equation}
3\varepsilon_sx_s^2
=\frac{C_s^2-A_s^2}{v_o^2}+O(\varepsilon_s^2).
\end{equation}
Using $\varepsilon=-2\delta/3+O(\delta^2)$ and $A_s^2-C_s^2=f^2a^2$, together with $a^2/v_{\rm eq}^2=1-q^2$ and $v_o^2=v_{\rm eq}^2+O(\delta)$ in a term that is already first order, we obtain
\begin{equation}
\begin{split}
\frac{\Phi_2}{\Phi_0}
&=-8\varepsilon+\frac{C_s^2-A_s^2}{v_o^2}\\
&=\frac{16}{3}\delta-f^2(1-q^2)\equiv X .
\end{split}
\label{eq:ld_X_derivation}
\end{equation}
This is Equation~\eqref{eq:ld_Phi2_summary}.  The signs provide a useful consistency check.  For oblate ejecta, $\varepsilon<0$, so both outer-boundary terms are positive at the poles ($P_2=1$): the polar boundary lies closer to the source, and a larger physical area corresponds to each unit solid angle of polar normal.  An oblate radioactive source has $\varepsilon_s<0$ and contributes with the opposite sign because its material is preferentially distributed toward the equatorial plane.  Equation~\eqref{eq:ld_series_convex} then projects this multipole through the Eddington weight $w_2$, yielding Equation~\eqref{eq:ld_oblate}.

\subsubsection*{Finite-time accuracy and the late-time limit}

The multipoles derived above are zero-frequency, or quasi-steady, ratios. The closed forms for the peak luminosity use these ratios as approximations to the finite-time multipoles at maximum light. At maximum light the quasi-steady conditions of Equation~\eqref{eq:ld_qs_conditions} are not well satisfied (the values of $\eta_{Q}$ and $\eta_{\rm age}$ quoted below are of order $0.2$--$0.8$, not $\ll1$), so the replacement is an uncontrolled approximation at peak and must be checked against the full time-dependent diffusion solution.

For a spherical boundary, the slowest mode is the monopole, with $k_{01}=\pi/v_o$ and therefore $\tau_{01}=\tilde t^{\,2}/\pi^2$. As a representative example, model \#13 of Table~\ref{tab:example_table} has $M=1\,M_\odot$, $\tilde t=28.3\,\mathrm{d}$, and $t_0=31.1\,\mathrm{d}$. At $t=15\,\mathrm{d}$ the heating prescription of Section~\ref{sec:heating} gives $t_Q\simeq14.9\,\mathrm{d}$, so $\eta_{Q,01}\simeq0.37$ and $\eta_{{\rm age},01}\simeq0.72$ ($\bar t/\tau_{01}\simeq1.4$).  At the same epoch, the $19$ models in Table~\ref{tab:example_table} span $\eta_{Q,01}\simeq0.19$--$0.43$ and $\eta_{{\rm age},01}\simeq0.39$--$0.83$. Maximum light is therefore not an asymptotic quasi-steady regime for the monopole.

The anisotropic modes relax more rapidly.  For the dipole, the first zero of $j_1$ is $k_{11}v_o=4.493$, giving $\eta_{Q,11}\simeq0.18$ and $\eta_{{\rm age},11}\simeq0.35$ for model \#13 at $15\,\mathrm{d}$.  The multipole ratios can consequently be substantially closer to their quasi-steady values than the monopole is to its own steady response. We measure the finite-time multipoles by expanding
\begin{equation}
\frac{L_{\rm p}(\theta_k)}{L_{\rm p}^{\rm avg}}
=\sum_{l\ge0}b_lP_l(\cos\theta_k),
\qquad
b_l=\frac{2l+1}{2}\int_{-1}^{1}
\frac{L_{\rm p}(\mu)}{L_{\rm p}^{\rm avg}}
P_l(\mu)\,d\mu,
\label{eq:ld_bl_def}
\end{equation}
where $\mu=\cos\theta_k$, with peak luminosities defined in
Section~\ref{sec:mc_validation_sphere}. Because $\Lp^{\rm avg}$ is the peak of the angle-averaged light curve rather than the mean of the per-angle peaks, $b_0$ slightly exceeds unity ($1.0004$--$1.023$ for the off-centre runs below, and $1.0000$--$1.0024$ for the oblate ones), which is why the sum starts at $l=0$. For Fig.~\ref{fig:limb_darkening}, the integral is evaluated by Simpson's rule on 21 equally spaced values of $\cos\theta_k$, including both poles. The quasi-steady predictions are
\begin{equation}
b_1^{\rm closed}=\frac{17}{8}z_{\rm off},
\qquad
b_2^{\rm closed}=\frac{13}{40}X.
\label{eq:ld_bl_closed}
\end{equation}
We quote the dipole deficit as $1-b_1^{\rm closed}/b_1$, whereas the inset shows $b_1/b_1^{\rm closed}$. Evaluating all observers at the angle-averaged peak time changes $b_1$ by at most 0.4 per cent for these off-centre solutions. The accuracy at maximum light is best assessed directly from the finite-time analytic solutions, each against the quasi-steady limit of its own perturbative geometry, so that the two error sources stay separate: for a displaced spherical source the quasi-steady dipole is about $11$ per cent below the finite-time dipole (range $7$--$15$ per cent, $1-b_1^{\rm closed}/b_1$), measured over the $24$ solutions with $z_{\rm off}=0.05$--$0.30$ (six at each $z_{\rm off}$) shown in the inset of Fig.~\ref{fig:limb_darkening}, whereas for the slightly oblate calculations with $f\le1$, the quasi-steady quadrupole differs by at most $4.5$ per cent (median $1$ per cent), and stays flat in $X$ out to $X=3.3$; this holds also for the runs whose distortions lie outside the first-order regime ($|\varepsilon|$ or $|\varepsilon_s|>0.35$), which are flagged in the released table.
For strongly non-confocal sources ($f>1$) the first-order terms of Equation~\eqref{eq:ld_Phi2_summary} can nearly cancel at small $X$, in which case the measured quadrupole is dominated by higher-order terms and is not reproduced by $w_2X$ even though the normalized luminosity residual stays small (the $f=1.4$, $X=0.09$ run of Fig.~\ref{fig:limb_darkening}); $X<1.5$ alone is therefore not a sufficient accuracy criterion for arbitrary $f$.  At larger $X$, the dominant additional error is instead the first-order shape expansion, obtained directly as the ratio of the exact-oblate to the slightly-oblate quadrupole, whose median is $1.03$ at $X\le0.5$, $1.10$ for $0.5<X\le1$, $1.22$ for $1<X\le2$ and $1.39$ for $X>2$. These ratios are measured from the finite-time analytic solutions; they are not inferred from the small parameters in Equation~\eqref{eq:ld_qs_conditions}.

At asymptotically late times, the situation within the analytic diffusion model is different.  Both $\eta_{Q,n}$ and $\eta_{{\rm age},n}$ tend to zero, and the finite-time analytic multipoles converge to the zero-frequency Poisson result, by construction: the quasi-steady multipoles of Section~\ref{sec:limb_darkening} are defined as that zero-frequency limit (for the off-centre geometry, an outer-bin spread of $3w_1|z_{\rm off}|\Delta P_1=(153/40)\,|z_{\rm off}|$ between the outermost bins, i.e.\ $\myapprox38$ per cent for $z_{\rm off}=0.1$). It should not, however, be interpreted as a prediction that the RT calculation must approach the same limit.  By those epochs, the ejecta become optically thin, so the diffusion equation, the sharp diffusive boundary, and the surface-emission closure cease to provide an accurate description of the RT problem. Thus the quasi-steady approximation becomes increasingly accurate within the analytic diffusion model at late times, even while the underlying diffusion model becomes progressively less applicable to the physical radiative-transfer solution.

There is one further finite-time approximation in the peak-luminosity closed form: the angular pattern is evaluated at the angle-averaged peak time rather than at the maximum of each individual viewing-angle light curve.  This is separate from the quasi-steady replacement.  For weak anisotropy, its effect is second order; Table~\ref{tab:ld_frozen} shows how the error grows with source displacement.

\begin{table}
\centering
\caption{Error introduced by evaluating Equation~\eqref{eq:ld_sphere} at the angle-averaged peak time rather than at each observer's own maximum. Columns $2$--$3$ give the deficit relative to the true maximum of the corresponding viewing-angle light curve, for the far side ($\theta_k=\pi$) and near side ($\theta_k=0$); columns $4$--$5$ give each peak time relative to the angle-averaged peak time. The calculation uses $M=1\,M_\odot$, $v_o=10^4\,\mathrm{km\,s^{-1}}$, $\kappa=0.1\,\mathrm{cm^2\,g^{-1}}$, $M_{{}^{56}\rm Ni}=0.5\,M_\odot$, $x_s=0.3$ and $t_0=35\,\mathrm{d}$.}
\label{tab:ld_frozen}
\begin{tabular}{lcccc}
\hline
$z_{\rm off}$ & deficit ($\pi$) & deficit ($0$) &
$t_{\rm peak}(\pi)/t_{\rm peak}$ & $t_{\rm peak}(0)/t_{\rm peak}$ \\
\hline
$0.1$ & $0.7\%$ & $0.5\%$ & $1.048$ & $0.956$ \\
$0.2$ & $3.1\%$ & $1.6\%$ & $1.102$ & $0.916$ \\
$0.4$ & $18\%$  & $4.4\%$ & $1.240$ & $0.848$ \\
\hline
\end{tabular}

\end{table}

%
\section{Extended Results for the Off-Centre Ignition Models}
\label{app:extended_results}
This appendix compares the radiative-transfer light curves from \textsc{URILIGHT} against the diffusion-framework solution of the effective toy model for every combination of ejecta mass ($0.8, 0.9, 1.0, 1.1\,M_\odot$) and ignition offset $z_{\rm ig}$ ($0.00, 0.10, 0.25, 0.50, 0.75$), that is, for the $19$ models of Table~\ref{tab:example_table}. Table~\ref{tab:example_table} lists, for every sub-Chandrasekhar off-centre ignition model, the effective toy-model parameters ($M_{{}^{56}\rm Ni}$, $x_s$, $z_{\rm off}$, $\tilde{t}$, and $t_0$) used as inputs to the analytic diffusion solution. Figs~\ref{fig:zoff_Lp_Mi} and \ref{fig:zoff_L30Lp_Mi} extend the $1\,M_\odot$ comparison of Fig.~\ref{fig:zoff_L30Lp_Lp_M1sol} to all ejecta masses, showing the peak luminosity $\Lp$ and the width ratio $\Lth/\Lp$, respectively, as functions of viewing angle, one panel per model, ordered by $z_{\rm off}$. Fig.~\ref{fig:zoff_DL30Lp_DLp_Mi} shows the corresponding normalized viewing-angle variability across all masses, confirming that $z_{\rm off}$ remains the dominant driver of the scatter over the full mass range. The red line in the top panel of that figure is the dipole term of the closed form, Equation~\eqref{eq:ld_sphere_num}, evaluated at the outermost bin centres $\cos\theta_k=\pm0.9$, which gives $\Delta\Lp/\Lp^{\rm avg}=3.83\,|z_{\rm off}|$; it is a prediction of Section~\ref{sec:limb_darkening} and is not fitted to these models; it lies $\myapprox11$ per cent below the analytic points. Fig.~\ref{fig:zoff_relerr} decomposes the toy-model relative errors in $t_{\rm peak}$, $\Lp$, and $\Lth/\Lp$ across the full model set as functions of $z_{\rm off}$, ejecta mass, and viewing angle. 

The hard-sphere fit follows the high-density part of the ${}^{56}$Ni distribution, whereas the centroid weights all of its mass. For example, in models \#8, \#13 and \#18 the mean ${}^{56}$Ni density at $z>0$ is $\myapprox1.7$ times that at $z<0$, whereas the volume occupied by ${}^{56}$Ni (density above $1$ per cent of its maximum) at $z>0$ is smaller by $16$--$31$ per cent; and using only the densest $75$ per cent of the ${}^{56}$Ni mass, the centroid reproduces $z_{\rm off}$ with a median ratio of $1.0$ across the $14$ displaced models, but with a model-to-model scatter of $27$ per cent (ratios $0.50$--$1.63$).

For model \#13 (Fig.~\ref{fig:1Msun_ejecta109_sol}), the agreement at early and late times is poorer than near peak. At $t=3$--$7\,\mathrm{d}$, the angle-averaged analytic luminosity exceeds the \textsc{URILIGHT} one by $41$--$63$ per cent ($63$, $59$ and $41$ per cent over the sub-intervals $3$--$4$, $4$--$5$ and $5$--$7\,\mathrm{d}$), and at $t=7$--$10$ and $10$--$14\,\mathrm{d}$ by $19$ and $4$ per cent, respectively. The excess grows towards the ${}^{56}$Ni side, reaching a factor of $\mysim3$ at $\cos\theta_k=+0.9$ for $t<4\,\mathrm{d}$, whereas at $\cos\theta_k=-0.9$ the analytic luminosity is the lower one until $t\myapprox5.5\,\mathrm{d}$. This deviation is much larger than the early-time effects discussed in Section~\ref{sec:mc_validation_sphere}, which change the luminosity by $\mylesssim12$ per cent at these times, and therefore cannot account for it. Nor is it a truncation effect of the analytic mode sum: doubling the number of eigenmodes changes the analytic light curves of this model by less than $0.03$ per cent at every viewing angle over $t=3$--$7\,\mathrm{d}$ (Appendix~\ref{app:validation_details}). Instead, the early light curve is set by the structure near the surface, where the effective sphere most likely deviates most from the numerical ejecta. The effective sphere has a uniform density and a sharp edge at $v_o$, whereas the density of the numerical ejecta decreases smoothly outward, with $20$ per cent of the mass outside $v_o$, extending to $\myapprox1.8\,v_o$. The ${}^{56}$Ni is also distributed differently: since the numerical ejecta are centrally concentrated, half of the ${}^{56}$Ni mass lies within $0.53\,v_o$, compared with $0.64\,v_o$ in the effective sphere. The excess luminosity is not due to the constant opacity approximation: relative to \textsc{URILIGHT} run on the same ejecta with the constant opacity $\kappa=0.165\,\mathrm{cm^2\,g^{-1}}$ (test (i) of Section~\ref{sec:width_mismatch}), the analytic luminosity is still higher, by $98$, $54$ and $25$ per cent at $t=3$--$4$, $4$--$5$ and $5$--$7\,\mathrm{d}$. Moreover, \textsc{URILIGHT} run on the effective toy ejecta with the same opacity (right panel of Fig.~\ref{fig:check_geom}) is brighter than test (i) by $121$, $74$ and $36$ per cent at the same epochs. We therefore attribute the early-time deviation mainly to the reduction of the numerical ejecta to the effective sphere. At late times, $t\ge30\,\mathrm{d}$, the angle-averaged luminosity is reproduced to within $11$ per cent, but the viewing-angle contrast is not: the analytic ratio $L_{\rm iso}(\cos\theta_k=+0.9)/L_{\rm iso}(\cos\theta_k=-0.9)$ remains at $\myapprox1.4$, whereas in \textsc{URILIGHT} it decreases to $1.22$ at $t=35$--$50\,\mathrm{d}$ (and to $1.06$ at $t\myapprox60\,\mathrm{d}$ for constant opacity). This is the frozen angular pattern of the analytic solution discussed in Sections~\ref{sec:mc_validation} and \ref{sec:width_mismatch}.

Fig.~\ref{fig:Lp_L30Lp_obs_mag} shows Fig.~\ref{fig:Lp_L30Lp_obs} in magnitude space, as in fig.~C2 of \citet{sharon_2025}: the peak bolometric magnitude, $M_{\rm p}^{\rm bol}=-2.5\log_{10}(\Lp/L_0)$ with $L_0=3.0128\times10^{35}\,\mathrm{erg\,s^{-1}}$, against $M_{30}^{\rm bol}-M_{\rm p}^{\rm bol}=-2.5\log_{10}(\Lth/\Lp)$.

\begin{figure*}
    \centering
    \includegraphics[width=\linewidth]{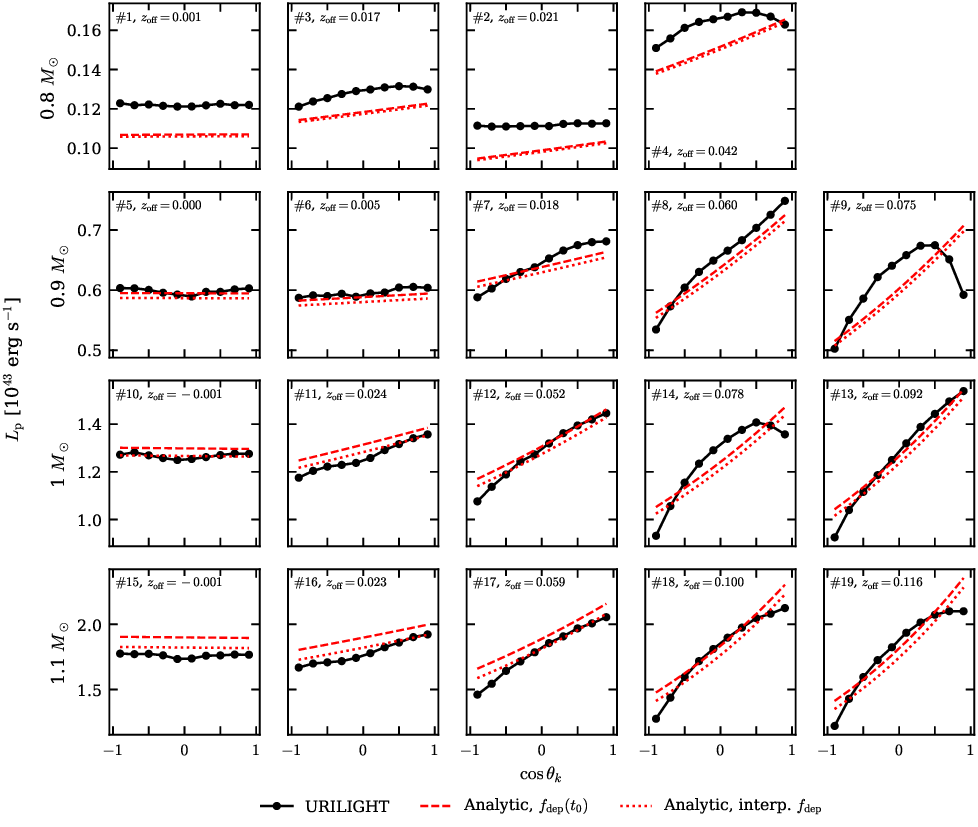}
    \caption{Peak luminosity $\Lp$ as a function of the viewing angle $\cos\theta_k$ for all $19$ off-centre ignition models, comparing the analytic toy-model predictions (red dashed and dotted lines) with the full RT calculations performed with \textsc{URILIGHT} (solid black lines with circles). As in Fig.~\ref{fig:zoff_L30Lp_Lp_M1sol}, the dashed analytic curve uses the deposition fraction of Equation~\eqref{eq:f_dep} and the dotted one the $f_{\rm dep}(t)$ tabulated by the \textsc{URILIGHT} $\gamma$-ray transport. One panel per model: each row is an ejecta mass, as labelled on the left ($0.8-1.1\,M_\odot$; the $0.8\,M_\odot$ row has four models, the others five), and the panels of a row are ordered by the nickel off-centre displacement $z_{\rm off}$, whose value is given in each panel. Rows share their vertical axis. All ten viewing-angle bins are shown, uniform in $\cos\theta_k$ over $[-1,1]$, with $\cos\theta_k=+1$ facing the ${}^{56}$Ni offset. Parameters for each model are listed in Table~\ref{tab:example_table}.}
    \label{fig:zoff_Lp_Mi}
\end{figure*}

\begin{figure*}
    \centering
    \includegraphics[width=\linewidth]{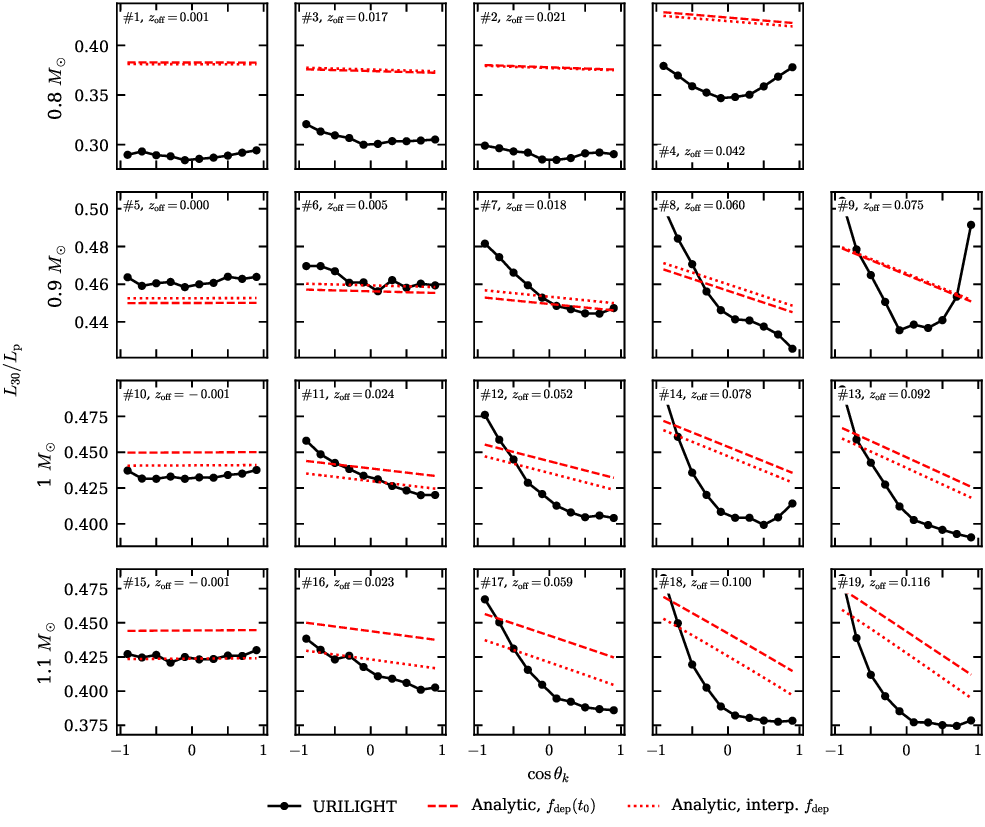}
    \caption{Same as Fig.~\ref{fig:zoff_Lp_Mi}, but showing $\Lth/\Lp$ instead of the peak luminosity $\Lp$, where $\Lth$ is the bolometric luminosity at $t=30$~days after explosion.}
    \label{fig:zoff_L30Lp_Mi}
\end{figure*}

\begin{figure*}
    \centering
    \includegraphics[width=0.9\linewidth]{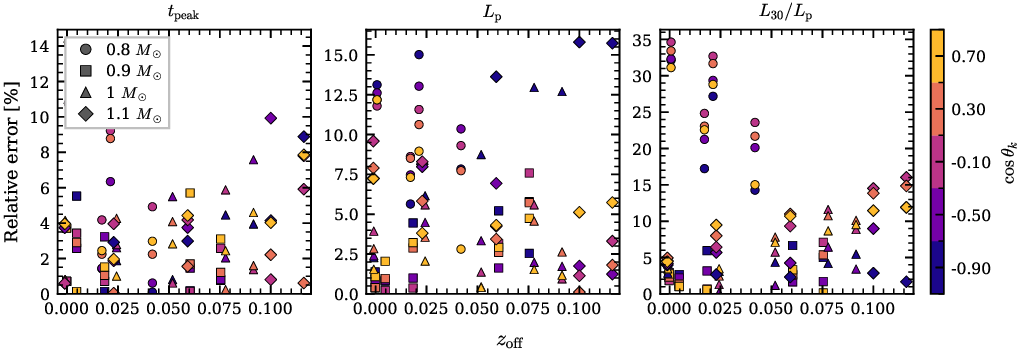}
    \caption{Relative error of the toy model against \textsc{URILIGHT} for the three peak quantities -- the peak time $t_{\rm peak}$ (left), the peak luminosity $\Lp$ (centre), and the width ratio $\Lth/\Lp$ (right) -- as a function of the nickel off-centre displacement $z_{\rm off}$. The plotted quantity is the absolute relative error referenced to \textsc{URILIGHT}, $|Y_{\rm ana}-Y_{\rm num}|/Y_{\rm num}$ for each of the three quantities $Y$, for the analytic calculation that uses the deposition fraction of Equation~\eqref{eq:f_dep}; it carries no sign, so a point does not indicate whether the toy model over- or underpredicts. Marker shape encodes the ejecta mass (circles $0.8$, squares $0.9$, triangles $1.0$ and diamonds $1.1\,M_\odot$) and marker colour the viewing angle $\cos\theta_k$ (a representative subset of five of the ten inclination bins is shown for clarity; the quoted statistics are computed over all $190$ model-angle combinations). Note the different vertical scales of the three panels.}
\label{fig:zoff_relerr}
\end{figure*}

\begin{figure*}
    \centering
    \includegraphics[width=\linewidth]{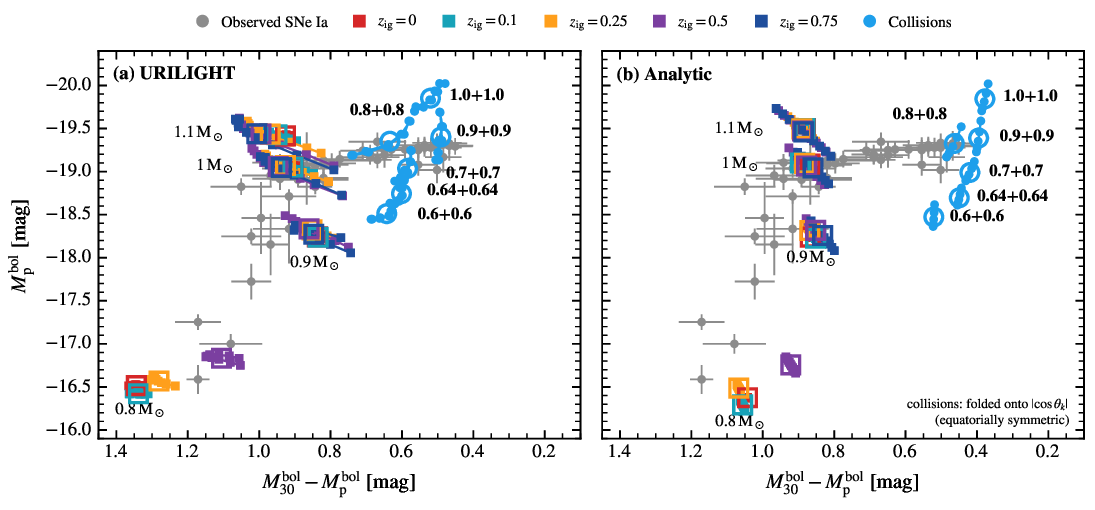}
    \caption{As Fig.~\ref{fig:Lp_L30Lp_obs}, but in magnitude space, as in fig.~C2 of \citet{sharon_2025}: the peak bolometric magnitude $M_{\rm p}^{\rm bol}$ against $M_{30}^{\rm bol}-M_{\rm p}^{\rm bol}=-2.5\log_{10}(\Lth/\Lp)$, with $M^{\rm bol}=0$ at $3.0128\times10^{35}\,\mathrm{erg\,s^{-1}}$. Both axes are reversed, so that brighter and slower events lie up and to the right. The observational errors are propagated to first order.}
    \label{fig:Lp_L30Lp_obs_mag}
\end{figure*}


\begin{table*}
    \centering
    \caption{Toy-model representation of the sub-Chandrasekhar off-centre ignition models. $M$ denotes the ejecta mass and $z_{\rm ig}$ is the ignition location (see Section~\ref{sec:models}). $M_{{}^{56}\rm Ni}$ is the total synthesized ${}^{56}$Ni mass. The parameter $x_s$ is the fractional radius of the nickel-rich region; $z_{\rm off}$ is its fractional off-centre displacement along the symmetry axis, and $z_{{\rm com},{}^{56}\rm Ni}$ is the fractional ${}^{56}$Ni-mass-weighted mean axial velocity. The outer ejecta velocity $v_o$ is computed from $M$ and the mean velocity-space density $\bar{\rho}_v$. The opacity $\kappa$ is the effective Rosseland mean opacity at peak luminosity (Equation~\ref{eq:kappa_eff}). The diffusion time-scale $\tilde{t}$, defined through $\tilde{t}^2\equiv v_o^2/D$ with $D=c/(3\kappa\bar{\rho}_v)$, is described in Section~\ref{sec:toy_construction}; $t_0$ is the characteristic $\gamma$-ray escape time-scale. Bold columns are the inputs for the analytic diffusion framework ($M_{{}^{56}\rm Ni}$, $x_s$, $z_{\rm off}$, $\tilde{t}$, $t_0$). The column $z_{\rm off}v_o$ gives the offset velocity of the ${}^{56}$Ni distribution of the toy model, and $z_{{\rm com},{}^{56}\rm Ni}v_o$ the ${}^{56}$Ni mass-centroid velocity of the numerical ejecta (Section~\ref{sec:discussion} discusses its relation to nebular-line velocity shifts).}
    \label{tab:example_table}
\setlength{\tabcolsep}{3pt}
\begin{tabular}{lllccccccccccc}
        \hline
        $\#$ & $M$ & $z_{\rm ig}$ &
        $\boldsymbol{M_{{}^{56}\rm Ni}}$ &
        $\boldsymbol{x_s}$ & $\boldsymbol{z_{\rm off}}$ &
        $z_{{\rm com},{}^{56}\rm Ni}$ &
        $v_o$ & $z_{\rm off}v_o$ & $z_{{\rm com},{}^{56}\rm Ni}v_o$ & $\bar{\rho}_v$ & $\kappa$ &
        $\boldsymbol{\tilde{t}}$ & $\boldsymbol{t_0}$ \\
         & [$M_\odot$] &  & [$M_\odot$] &  &  &  &
        [$10^4\,$km\,s$^{-1}$] & [$10^3\,$km\,s$^{-1}$] & [$10^3\,$km\,s$^{-1}$] &
        [$M_\odot\,(10^4\,$km\,s$^{-1})^{-3}$] & [cm$^2\,$g$^{-1}$] &
        [day] & [day] \\
        \hline
1 & 0.8 & 0.00 & 0.038 & 0.361 & 0.0007 & 0.0006 & 1.167 & 0.01 & 0.01 & 0.1201 & 0.106 & 21.5 & 33.3 \\
2 & 0.8 & 0.10 & 0.034 & 0.350 & 0.0213 & 0.0101 & 1.166 & 0.25 & 0.12 & 0.1206 & 0.101 & 21.0 & 33.6 \\
3 & 0.8 & 0.25 & 0.041 & 0.370 & 0.0171 & 0.0201 & 1.173 & 0.20 & 0.24 & 0.1182 & 0.099 & 20.7 & 33.8 \\
4 & 0.8 & 0.50 & 0.055 & 0.410 & 0.0419 & 0.0184 & 1.069 & 0.45 & 0.20 & 0.1563 & 0.108 & 22.6 & 36.1 \\
5 & 0.9 & 0.00 & 0.236 & 0.640 & $-$0.0002 & $-$0.0001 & 1.260 & 0.00 & 0.00 & 0.1075 & 0.148 & 26.0 & 33.0 \\
6 & 0.9 & 0.10 & 0.236 & 0.640 & 0.0047 & $-$0.0015 & 1.259 & 0.06 & $-$0.02 & 0.1076 & 0.152 & 26.3 & 33.0 \\
7 & 0.9 & 0.25 & 0.255 & 0.657 & 0.0184 & 0.0130 & 1.265 & 0.23 & 0.16 & 0.1061 & 0.152 & 26.2 & 32.7 \\
8 & 0.9 & 0.50 & 0.257 & 0.658 & 0.0604 & 0.0248 & 1.241 & 0.75 & 0.31 & 0.1124 & 0.151 & 26.4 & 33.0 \\
9 & 0.9 & 0.75 & 0.242 & 0.646 & 0.0754 & 0.0171 & 1.209 & 0.91 & 0.21 & 0.1217 & 0.146 & 26.3 & 33.9 \\
10 & 1.0 & 0.00 & 0.540 & 0.814 & $-$0.0009 & 0.0001 & 1.309 & $-$0.01 & 0.00 & 0.1063 & 0.167 & 28.5 & 31.2 \\
11 & 1.0 & 0.10 & 0.539 & 0.814 & 0.0243 & 0.0098 & 1.310 & 0.32 & 0.13 & 0.1061 & 0.160 & 27.9 & 31.2 \\
12 & 1.0 & 0.25 & 0.543 & 0.816 & 0.0518 & 0.0235 & 1.315 & 0.68 & 0.31 & 0.1050 & 0.166 & 28.3 & 31.0 \\
13 & 1.0 & 0.50 & 0.529 & 0.809 & 0.0915 & 0.0252 & 1.308 & 1.20 & 0.33 & 0.1067 & 0.165 & 28.3 & 31.1 \\
14 & 1.0 & 0.75 & 0.523 & 0.806 & 0.0778 & 0.0182 & 1.318 & 1.03 & 0.24 & 0.1044 & 0.171 & 28.8 & 31.0 \\
15 & 1.1 & 0.00 & 0.806 & 0.901 & $-$0.0011 & $-$0.0002 & 1.327 & $-$0.01 & 0.00 & 0.1125 & 0.173 & 30.2 & 30.3 \\
16 & 1.1 & 0.10 & 0.805 & 0.901 & 0.0229 & 0.0075 & 1.328 & 0.30 & 0.10 & 0.1122 & 0.173 & 30.2 & 30.3 \\
17 & 1.1 & 0.25 & 0.801 & 0.900 & 0.0592 & 0.0158 & 1.334 & 0.79 & 0.21 & 0.1107 & 0.173 & 30.1 & 30.2 \\
18 & 1.1 & 0.50 & 0.787 & 0.894 & 0.1000 & 0.0168 & 1.346 & 1.35 & 0.23 & 0.1076 & 0.176 & 30.3 & 30.0 \\
19$^{\dagger}$ & 1.1 & 0.75 & 0.783 & 0.893 & 0.1162 & 0.0142 & 1.345 & 1.56 & 0.19 & 0.1079 & 0.176 & 30.3 & 30.0 \\
        \hline
    \end{tabular}
    \begin{flushleft}
    $^{\dagger}$~For this model, $x_s+z_{\rm off}=1.0089>1$,
    so the nominal spherical source marginally extends beyond the analytic ejecta boundary. The tabulated calculation projects the nominal source onto the modes without clipping, so the protruding $0.06$ per cent of the source volume enters through the analytic continuation of the eigenfunctions. Repeating it with the source clipped at the boundary and renormalized to $M_{{}^{56}\rm Ni}$ (quadrature projection with the radial chord truncated at $v_o$; at the resolution of the tabulated calculation, $l\le100$ with $100$ radial modes per $l$) changes $\Lp$ by at most $0.2$ per cent ($0.1$ per cent for the angle average), $\Lth/\Lp$ by at most $0.04$ per cent and the peak times by at most one time-grid step ($0.03\,\mathrm{d}$), negligible compared with the several-per-cent toy-model error.
    \end{flushleft}
\end{table*}

\section{Head-on White Dwarf Collision Models}
\label{app:collision}

This appendix documents the oblate toy-model construction of Section~\ref{sec:collisions} for the head-on WD collision models \citep{Kushnir_2013}, which produce oblate ejecta rather than the approximately spherical ejecta of the off-centre ignition models. The oblate diffusion solution itself is derived in Appendix~\ref{app:oblate_diffusion} and validated against \textsc{URILIGHT} on idealized oblate ejecta in Section~\ref{sec:mc_validation_oblate}. What is tested here is the construction that reduces the collision ejecta to the six parameters of that solution representing the ellipsoid toy ejecta.

\subsection{Toy-model construction for collision ejecta}
\label{app:collision_models}
For each collision ejecta, we construct an effective toy model using a procedure similar to that used for the off-centre spherical case (Section~\ref{sec:toy_construction}), with the spherical ejecta and ${}^{56}$Ni source both replaced by oblate spheroids and all effective quantities evaluated at the epoch of bolometric peak. The model here is specified by six effective parameters: the ${}^{56}$Ni mass $M_{{}^{56}\rm Ni}$, the ejecta boundary coordinate $\xi_s$, the ${}^{56}$Ni source coordinate $\xi_1$, the confocality parameter $f$, the diffusion time-scale $\tilde{t}$, and the $\gamma$-ray escape time $t_0$. The construction of these parameters is described below.

As in the spherical case, the toy model conserves the total ejecta mass $M$ and the total ${}^{56}$Ni mass $M_{{}^{56}\rm Ni}$, both summed over the ejecta cells. The effective ejecta geometry, however, is not obtained from a mass-conserving fit, but directly from the total mass density distribution: we extract the outermost density contour at $10$ per cent of the peak density and identify its maximal radial and axial extents with the equatorial and polar velocities $v_{\rm eq}$ and $v_{\rm pol}$ of an oblate spheroid. These fix the focal length $a=(v_{\rm eq}^2-v_{\rm pol}^2)^{1/2}$ and the ejecta boundary coordinate $\xi_s=\operatorname{arctanh}(v_{\rm pol}/v_{\rm eq})$ (Appendix~\ref{app:oblate_diffusion}), while the mean velocity-space density is $\bar{\rho}_v=M/(\tfrac{4\pi}{3}v_{\rm eq}^2 v_{\rm pol})$, the total mass divided by the spheroidal volume. The choice of the $10$ per cent contour, and the reduction of the full density distribution to a single oblate boundary, are modelling choices of the effective toy-model construction; neither is uniquely determined by the underlying diffusion physics.

The ${}^{56}$Ni source region is determined in close analogy to the spherical construction, where the shape of the radioactive material is taken from a fit to the ${}^{56}$Ni distribution while its overall scale is fixed by mass conservation. Because the collision ejecta are symmetric about the equatorial plane, the ${}^{56}$Ni region is concentric with the ejecta and centred on that plane, so there is no off-centre displacement ($z_{\rm off}=0$) and the viewing-angle dependence arises entirely from the oblate geometry. We represent the source as an oblate spheroid sharing the ejecta symmetry axis. Fitting the $10$ per cent contour of the ${}^{56}$Ni density distribution, in the same way as for the ejecta boundary, would, in principle, determine both the shape and the overall scale of this spheroid, i.e.\ both its axis ratio ($\tanh\xi_1$) and its focal length $fa$. As in the spherical case, however, the fitted scale does not in general reproduce the correct ${}^{56}$Ni mass, so we retain from the fit only the shape (the axis ratio, equivalently, the source coordinate $\xi_1$) and discard the fitted scale. The overall scale, parametrized by the focal-length ratio $f$, is instead fixed by mass conservation: $f$ is chosen so that the source volume equals $V_{\rm src}=M_{{}^{56}\rm Ni}/\bar{\rho}_v$, the volume that contains the correct ${}^{56}$Ni mass at the mean ejecta density (Equation~\ref{eq:oblate_initial}). This mirrors the spherical case, in which the fitted source radius is discarded, and $x_s$ is instead fixed by $M_{{}^{56}\rm Ni}$.

The remaining two parameters are obtained exactly as in the spherical case. The diffusion time-scale is $\tilde{t}^2=a^2/D$ with $D=c/(3\kappa\bar{\rho}_v)$, where the effective opacity $\kappa$ is the mass-weighted Rosseland mean of Equation~\eqref{eq:kappa_eff}. The $\gamma$-ray escape time $t_0$ enters the radioactive heating through Equation~\eqref{eq:f_dep} and is computed from the mean column density seen by the ${}^{56}$Ni cells with an effective $\gamma$-ray opacity $\kappa_{\rm eff}=0.025\,\mathrm{cm^2\,g^{-1}}$.

As in the spherical case, none of these parameters is calibrated to reproduce the \textsc{URILIGHT} light curves, and only $\tilde{t}$ requires the completed \textsc{URILIGHT} calculation, through the peak-light opacity $\kappa$. The resulting parameters for the six equal-mass collisions are listed in Table~\ref{tab:collision_table}.

Fig.~\ref{fig:collision_effective_geom} illustrates the construction for the $0.6+0.6\,M_\odot$ collision. The total mass density (left panel) is enclosed reasonably well by a single oblate spheroid, but the ${}^{56}$Ni distribution (right panel) is not.  The mismatch is of the same character in the other equal-mass collisions. Since the closed form of Section~\ref{sec:limb_darkening} shows that the viewing-angle dependence is set by the quadrupole moment of the source and by the curvature of the boundary, an inaccurate source shape translates directly into an inaccurate angular pattern. This is a plausible reason why the oblate construction reproduces the RT calculations less well than the off-centre construction of Section~\ref{sec:toy_construction}. We emphasize that this concerns the reduction of the ejecta to a toy model, and not the oblate diffusion solution itself, which is validated on ejecta that do satisfy its assumptions in Section~\ref{sec:mc_validation_oblate}.

\subsection{Comparison with \textsc{URILIGHT} on the numerical ejecta}
\label{app:collision_equal_mass}

We compare the analytic toy model with angle-resolved \textsc{URILIGHT} light curves for six equal-mass collisions of \citet{Kushnir_2013}, $0.6+0.6$, $0.64+0.64$, $0.7+0.7$, $0.8+0.8$, $0.9+0.9$ and $1.0+1.0\,M_\odot$. Two lighter equal-mass collisions, $0.5+0.5$ and $0.55+0.55\,M_\odot$, are excluded: both their total mass density and their ${}^{56}$Ni distribution are prolate rather than oblate. Measured from the $10$ per cent contours as in Appendix~\ref{app:collision_models}, their axis ratios are $v_{\rm pol}/v_{\rm eq}=1.20$ and $1.01$ for the ejecta boundary and $1.11$ and $1.09$ for the ${}^{56}$Ni region; the diffusion solution of Appendix~\ref{app:oblate_diffusion} therefore has no prolate branch, and the construction cannot be applied to them. Table~\ref{tab:collision_table} lists the six toy-model parameters extracted from each of them by the construction of Appendix~\ref{app:collision_models}, and Fig.~\ref{fig:M06_M06_sol} compares the resulting analytic light curves with \textsc{URILIGHT} at each viewing angle for the $0.6+0.6\,M_\odot$ collision. Across the ten viewing angles, the peak luminosity $\Lp$ is reproduced with a median relative error of $3.6$ per cent, and the width ratio $\Lth/\Lp$ with a median relative error of $11.2$ per cent. Over the six collisions the per-collision medians span $0.8-5.5$ per cent in $\Lp$ and $9.6-17$ per cent in $\Lth/\Lp$, and the ordering follows neither the collision mass nor $X$: the smallest median error in $\Lp$ is that of the $1.0+1.0\,M_\odot$ collision ($0.8$ per cent) and the largest that of the $0.9+0.9\,M_\odot$ collision ($5.5$ per cent), which has at the same time the smallest median error in $\Lth/\Lp$ ($9.6$ per cent), while the largest error in $\Lth/\Lp$ ($17.0$ per cent) belongs to the $0.8+0.8\,M_\odot$ collision, whose median error in $\Lp$ is only $2.2$ per cent. As for the sub-Chandrasekhar models, the peak luminosity is captured better than the width ratio.

Two effects plausibly contribute to the poorer match of the collision toy models (Section~\ref{sec:collisions}). The first is geometric: as discussed in Appendix~\ref{app:collision_models} and shown in Fig.~\ref{fig:collision_effective_geom}, the ${}^{56}$Ni distribution in the collision ejecta is not a perfect oblate spheroid, so the reduction discards precisely the moment of the source that sets the angular pattern. A compact measure of this is the exponent $n$ of the best-fitting superellipse $|v_r/A|^n+|v_z/B|^n=1$, for which $n=2$ is an ellipse: fitted to the $10$ per cent contours of Fig.~\ref{fig:collision_effective_geom}, the total density gives $n=1.8$ for the $0.6+0.6\,M_\odot$ collision and $n\myapprox1.8$--$2.2$ over the six equal-mass collisions, close to an ellipse throughout, whereas the ${}^{56}$Ni gives $n=0.9$ and spans $n\myapprox0.9$--$3.3$, with no value near $2$: the ${}^{56}$Ni contour is alternately diamond-shaped and boxy, a deformation no spheroid can represent. The second is common to both geometries as already discussed in Section~\ref{sec:width_mismatch}. Even for idealized ejecta that do satisfy the assumptions of the oblate solution, Fig.~\ref{fig:diff_sol_oblate_fdep1} shows a residual between the analytic and RT light curves that grows after maximum, because the Dirichlet boundary condition and the grey-atmosphere closure hold the angular contrast fixed while the RT contrast decays as the ejecta become optically thin. The same mechanism, together with the constant effective opacity and the deposition prescription, was identified in Section~\ref{sec:width_mismatch} as the origin of the width-ratio mismatch of the spherical models, and $\Lth$ is measured well inside the epoch where those approximations are weakest.

The relative errors of these six equal-mass collisions are, in places, comparable to those of some off-centre sub-Chandrasekhar models, but this comparison rests on low statistics: only six equal-mass collision ejecta, which most closely resemble an oblate spheroid, versus 19 sub-Chandrasekhar models.

The construction fails more fundamentally in the case of collisions between WDs of unequal mass. About half of the unequal-mass collisions of \cite{Kushnir_2013} ($8$ of $15$) show a prolate ${}^{56}$Ni region, inside an oblate boundary in five of them and a prolate one in the other three. In the first-order expression of Equation~\eqref{eq:ld_Phi2_summary}, a prolate source ($\varepsilon_s>0$) inside an oblate boundary ($\varepsilon<0$) gives source and boundary quadrupoles of the same sign, which reinforce each other, whereas inside a prolate boundary the two are in opposition. Whether $\Lp$ rises or falls with $|\cos\theta_k|$ is then determined by the interplay between these two surfaces (although, again, a few models show no monotonicity at all). A treatment of these models is beyond the scope of this paper. For example, Fig.~\ref{fig:collision_effective_geom_M10_M05} shows the mass density of one such ejecta, the $1.0+0.5\,M_\odot$ collision, in velocity space, together with the spheroids obtained from the same $10$ per cent contour prescription used above (Appendix~\ref{app:collision_models}). Its ejecta boundary is oblate, ($v_{\rm pol}/v_{\rm eq}=0.98$), and close to an ellipse ($n\approx2.1$), whereas its ${}^{56}$Ni region is prolate, $v_{\rm pol}^{\rm src}/v_{\rm eq}^{\rm src}=1.13$, and markedly non-elliptical ($n\approx1.3$). The ${}^{56}$Ni is also not symmetric about the equatorial plane: $42$ per cent of it lies above the plane and $58$ per cent below, with a ${}^{56}$Ni-mass-weighted mean axial velocity $\langle v_z\rangle=-1.1\times10^3\,\mathrm{km\,s^{-1}}$, against $|\langle v_z\rangle|\le23\,\mathrm{km\,s^{-1}}$ ($\le0.2$ per cent of $v_{\rm pol}$) for the ${}^{56}$Ni of the six equal-mass collisions. Thus, both approximations of Appendix~\ref{app:collision_models} fail at once, and we do not attempt a quantitative comparison for these models.

\begin{table*}
    \centering
    \caption{Toy-model representation of the equal-mass head-on WD collision models of \citet{Kushnir_2013}, constructed as described in Appendix~\ref{app:collision_models}. $M$ is the total ejecta mass and $M_{{}^{56}\rm Ni}$ the synthesized ${}^{56}$Ni mass. The ejecta boundary coordinate $\xi_s$ fixes the axis ratio $\tanh\xi_s=v_{\rm pol}/v_{\rm eq}$, with $v_{\rm eq}=a\cosh\xi_s$ and $v_{\rm pol}=a\sinh\xi_s$ and $a$ the focal length; $\xi_1$ is the ${}^{56}$Ni source coordinate and $f$ the ratio of the source focal length to $a$. The mean velocity-space density is $\bar{\rho}_v=M/(\tfrac{4\pi}{3}v_{\rm eq}^2v_{\rm pol})$ and $\kappa$ is the effective Rosseland mean opacity at peak luminosity (Equation~\eqref{eq:kappa_eff}). The diffusion time-scale is $\tilde{t}^{\,2}=a^2/D$ with $D=c/(3\kappa\bar{\rho}_v)$, and $t_0$ is the $\gamma$-ray escape time. The last column is the combination $X=(16/3)\delta-f^2(1-q^2)$ of Equation~\eqref{eq:ld_Phi2_summary}, with $\delta=1-\tanh\xi_s$ and $q=\tanh\xi_s$, which is the only shape parameter entering the closed form of Section~\ref{sec:limb_darkening}; it is determined by $\xi_s$ and $f$ and is not an independent input. The running number $\#$ continues that of Table~\ref{tab:example_table}. Bold columns are the inputs to the analytic diffusion solution ($M_{{}^{56}\rm Ni}$, $\xi_s$, $\xi_1$, $f$, $\tilde{t}$, $t_0$). $M_{{}^{56}\rm Ni}$ and $t_0$ are measured directly from the two-dimensional ejecta that enter the RT calculations. Table D2 of \citet{sharon_2025} lists the same six collisions; the values here exceed those by $3.4$--$8.5$ per cent in $M_{{}^{56}\rm Ni}$ and by $0.4$--$4.5$ per cent in $t_0$, because that table is built from the spherically averaged (one-dimensional) versions of the same ejecta, which do not preserve the ${}^{56}$Ni content exactly.}
    \label{tab:collision_table}
    \small
    \setlength{\tabcolsep}{4pt}
    \begin{tabular}{llcccccccccccc}
        \hline
        $\#$ & model & $M$ & $\boldsymbol{M_{{}^{56}\rm Ni}}$ &
        $\boldsymbol{\xi_s}$ & $\tanh\xi_s$ &
        $\boldsymbol{\xi_1}$ & $\boldsymbol{f}$ & $a$ &
        $\bar{\rho}_v$ & $\kappa$ &
        $\boldsymbol{\tilde{t}}$ & $\boldsymbol{t_0}$ & $X$ \\
         &  & [$M_\odot$] & [$M_\odot$] &  &  &  &  &
        [$10^4\,$km\,s$^{-1}$] &
        [$M_\odot\,(10^4\,$km\,s$^{-1})^{-3}$] & [cm$^2\,$g$^{-1}$] &
        [day] & [day] &  \\
        \hline
        20 & $0.6+0.6$ & 1.20 & 0.332 & 1.302 & 0.862 & 1.038 & 0.838 & 0.609 & 0.191 & 0.145 & 16.5 & 40.0 & 0.56 \\
        21 & $0.64+0.64$ & 1.28 & 0.428 & 1.351 & 0.874 & 1.005 & 0.966 & 0.571 & 0.215 & 0.153 & 16.9 & 40.0 & 0.45 \\
        22$^{\dagger}$ & $0.7+0.7$ & 1.40 & 0.579 & 1.228 & 0.842 & 0.864 & 1.052 & 0.665 & 0.212 & 0.161 & 20.0 & 39.5 & 0.52 \\
        23$^{\dagger}$ & $0.8+0.8$ & 1.60 & 0.781 & 1.008 & 0.765 & 0.702 & 1.051 & 0.869 & 0.203 & 0.151 & 24.9 & 38.4 & 0.79 \\
        24 & $0.9+0.9$ & 1.80 & 0.846 & 1.109 & 0.804 & 0.983 & 0.876 & 0.863 & 0.175 & 0.152 & 23.0 & 41.0 & 0.77 \\
        25 & $1.0+1.0$ & 2.00 & 1.336 & 1.149 & 0.817 & 1.107 & 0.910 & 0.844 & 0.186 & 0.164 & 24.0 & 40.5 & 0.70 \\
        \hline
    \end{tabular}
    \begin{flushleft}
    $^{\dagger}$~For these collisions $f>1$, i.e.\ the fitted source focal length exceeds that of the ejecta; the source spheroid is nevertheless contained within the ejecta boundary. This poses no difficulty for the analytic solution, which treats a non-confocal source exactly for any $f$, requiring only that the source lie inside the boundary (Appendix~\ref{app:oblate_diffusion}).
    \end{flushleft}
\end{table*}

\begin{figure}
    \centering
    \includegraphics[width=0.9\linewidth]{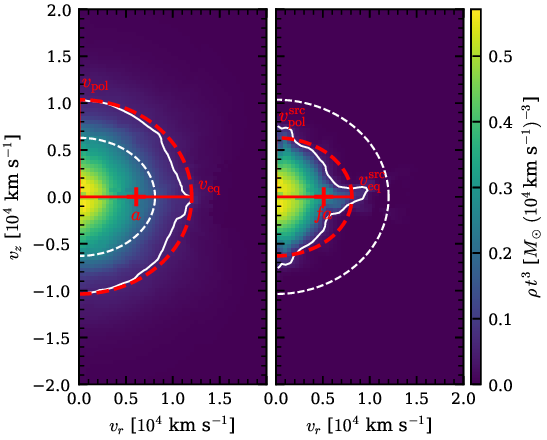}
    \caption{Effective oblate geometry of the collision toy-model construction (Appendix~\ref{app:collision_models}) for the $0.6+0.6\,M_\odot$ collision (\#20, Table~\ref{tab:collision_table}), in the $(v_r,v_z)$ velocity plane with the same units, colour scale and axes as Fig.~\ref{fig:subchandra_effective_geometry}. Left: total mass density, with the effective ejecta spheroid (semi-axes $a\cosh\xi_s$, $a\sinh\xi_s$; thick red dashed) and its focal length $a$ (red bar). Right: ${}^{56}$Ni mass density, with the effective source spheroid (semi-axes $fa\cosh\xi_1$, $fa\sinh\xi_1$). Thin white solid curves are the $10$ per cent density contours from which the shapes are fitted; thin white dashed curves show the other panel's spheroid. The ejecta spheroid circumscribes its contour, whereas the source spheroid keeps only the contour's shape, its scale being reset by ${}^{56}$Ni mass conservation; the markedly non-elliptical ${}^{56}$Ni contour is the geometric limitation of the oblate construction.}
    \label{fig:collision_effective_geom}
\end{figure}

\begin{figure}
    \centering
    \includegraphics[width=0.9\linewidth]{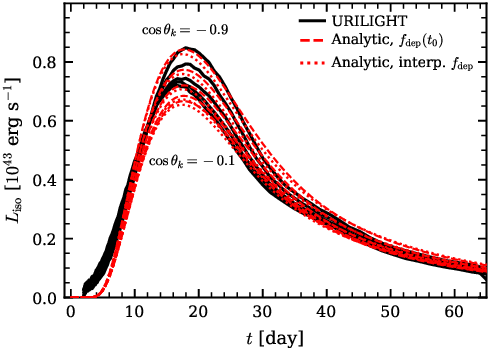}
    \caption{Bolometric light curves of the effective oblate toy model (red) compared with \textsc{URILIGHT} (solid black) for the head-on collision of two $0.6\,M_\odot$ WDs (\#20; parameters in Table~\ref{tab:collision_table}). $L_{\rm iso}$ is shown for the same five bins as in Fig.~\ref{fig:diff_sol_zoff_01_fdep1}, where $\theta_k$ is measured from the collision axis; both calculations are (approximately) symmetric about the equatorial plane, so each curve also represents its mirror bin, and the peak luminosity increases with $|\cos\theta_k|$ from edge-on ($\cos\theta_k=-0.1$, labelled below its peak) to pole-on ($-0.9$, labelled above it). As in Fig.~\ref{fig:1Msun_ejecta109_sol}, dashed uses Equation~\eqref{eq:f_dep} and dotted the tabulated $f_{\rm dep}(t)$. For this collision $\Lp$ is matched to a median of $3.6$ per cent and $\Lth/\Lp$ to $11.2$ per cent over the ten viewing angles.}
    \label{fig:M06_M06_sol}
\end{figure}

\begin{figure}
    \centering
    \includegraphics[width=0.9\linewidth]{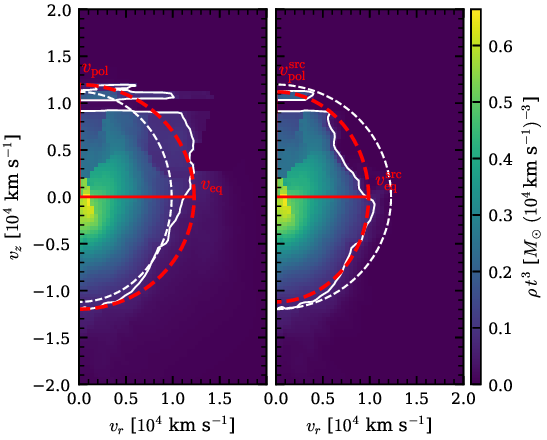}
    \caption{As Fig.~\ref{fig:collision_effective_geom}, but for the unequal-mass $1.0+0.5\,M_\odot$ head-on collision. Left: total mass density in velocity space, with the spheroid of semi-axes $v_{\rm eq}$ and $v_{\rm pol}$ (thick red dashed) read off the density contour at $10$ per cent of the peak (thin white solid). Right: ${}^{56}$Ni mass density, with the corresponding source spheroid. Unlike the six equal-mass collisions of Table~\ref{tab:collision_table}, this model admits no oblate toy model: the ${}^{56}$Ni region is prolate, with polar-to-equatorial ratio $v_{\rm pol}^{\rm src}/v_{\rm eq}^{\rm src}=1.13$. The spheroids drawn here are therefore the measured contour semi-axes, not a fit: the ${}^{56}$Ni one is the mass-conserving rescaling of its contour (similar construction of Section~\ref{app:collision_models}).}
    \label{fig:collision_effective_geom_M10_M05}
\end{figure}

\section{Early- and late-time residuals of the validation tests}
\label{app:validation_details}

The \textsc{URILIGHT} runs of Section~\ref{sec:mc_validation} use $5\times10^{7}$ UVOIR energy packets; since $f_{\rm dep}=1$ is imposed, no $\gamma$-ray packets are transported. The runs on the explosion models (Sections~\ref{sec:toy_model}--\ref{sec:collisions}) use $5\times10^7$ UVOIR energy packets for the off-centre ignition models and $4\times10^7$ for the collision models, together with $10^8$ $\gamma$-ray packets.

In the off-centre test of Fig.~\ref{fig:diff_sol_zoff_01_fdep1}, the largest discrepancies occur at early and late times. During the first few days, the discrepancy is dominated by numerical artefacts of the \textsc{URILIGHT} calculation and by light-travel-time (LTT) effects, which the analytic solution neglects. On the numerical side, the \textsc{URILIGHT} calculation starts at $t_{\rm init}=2\,\mathrm{d}$ with no escaping radiation, so the emergent luminosity is underestimated until the radiation field has built up, and packets with wavelengths above $\lambda_{\max}=3\times10^{4}$\,\AA\ are discarded; in this grey, constant-opacity calculation, packets trapped in the ejecta are redshifted beyond that limit, which biases the luminosity low. On the analytic side, neglecting LTT matters most at early times, when the light curve rises steeply. Averaged over $3$--$6\,\mathrm{d}$, the \textsc{URILIGHT} luminosity exceeds the analytic one by $\myapprox10$ per cent in the angle average and by $5$--$12$ per cent depending on the viewing angle (here and below, the average is the ratio of the time-integrated luminosities). The analytic light curves are converged in the number of eigenmodes: doubling it in each index (from $l\le100$ with $100$ radial modes per $l$ to $l\le200$ with $200$) changes them by less than $1$ per cent at every viewing angle after $t=2.2\,\mathrm{d}$, and by $0.02$ per cent averaged over $3$--$6\,\mathrm{d}$, so the truncation of the mode sum does not contribute to this difference. Repeating the \textsc{URILIGHT} calculation with $t_{\rm init}=0.5\,\mathrm{d}$ and $\lambda_{\max}=10^{6}$\,\AA, and removing LTT from the resulting light curves, reduces the difference to below $1$ per cent in the angle average and to within $4$ per cent at every viewing angle; we do not apply these changes to the other calculations. Observed light curves include LTT, and starting \textsc{URILIGHT} earlier is computationally expensive given the dense grids and large numbers of packets required to keep the MC noise low. The wavelength limit, in turn, matters only for grey, pure-scattering calculations: with realistic opacities, packets are repeatedly absorbed and re-emitted with the local thermal spectrum, so their wavelengths do not drift redwards. At most $0.4$ per cent (typically $0.1$ per cent) of the created packets are discarded in our sub-Chandrasekhar and collision models. At later times, the analytic and MC light curves diverge: the MC viewing-angle contrast decays as the ejecta become optically thin, whereas the analytic solution holds it near its quasi-steady pattern. This expected limitation of the optically thick description was discussed in Section~\ref{sec:diffusion_framework} and is quantified below.

In the oblate test of Fig.~\ref{fig:diff_sol_oblate_fdep1} the same early-time differences appear, but later, because these light curves rise more slowly (peak at $t\myapprox22\,\mathrm{d}$, compared with $\myapprox15\,\mathrm{d}$ in Fig.~\ref{fig:diff_sol_zoff_01_fdep1}). The luminosity stays below $1$ per cent of $\Lp$ until $t\myapprox6\,\mathrm{d}$, so the late start of the \textsc{URILIGHT} calculation ($t_{\rm init}=2\,\mathrm{d}$) has no noticeable effect. To compare the two models at the same stage of their evolution, we average over $8$--$12\,\mathrm{d}$, during which $L$ grows from $\myapprox0.1$ to $\myapprox0.4\,\Lp$, as it does over $3$--$6\,\mathrm{d}$ in Fig.~\ref{fig:diff_sol_zoff_01_fdep1}. Over this interval, the \textsc{URILIGHT} luminosity exceeds the analytic one by $\myapprox9$ per cent in the angle average and by $7$--$12$ per cent depending on the viewing angle, comparable to the $\myapprox10$ per cent found for Fig.~\ref{fig:diff_sol_zoff_01_fdep1}. Doubling the number of eigenmodes (from even $n\le100$ with $200$ radial modes per $n$ to $n\le200$ with $400$) changes the analytic light curves by less than $1$ per cent at every viewing angle after $t\myapprox5.3\,\mathrm{d}$, and by $0.005$ per cent averaged over $8$--$12\,\mathrm{d}$.

Taken together, Figs.~\ref{fig:diff_sol_zoff_01_fdep1} and \ref{fig:diff_sol_oblate_fdep1} establish that, for both of the asymmetries considered in this work, the analytic solution reproduces the MC light curve to better than $5$ and $12$ per cent, respectively, at every viewing angle between $t\myapprox10$ and $40\,\mathrm{d}$. In the oblate test, the largest residuals occur on the rise, at $t\myapprox10\,\mathrm{d}$, and from $t\myapprox15\,\mathrm{d}$ onward the residual is below $6$ per cent; in the off-centre test the largest residual, $4.4$ per cent, occurs at $t\myapprox35\,\mathrm{d}$ towards the ${}^{56}$Ni-facing pole. These residuals are obtained over a maximal spread of $\Lp$ with viewing angles of $56$ (off-centre) and $74$ (oblate) per cent in the two tests.

The same test for the $19$ off-centre ignition models and the six collision models gives convergence to $1$ per cent at every viewing angle after $t=1.3$--$5.2\,\mathrm{d}$ (latest for the $0.8\,M_\odot$ models) and $2.0$--$3.0\,\mathrm{d}$, respectively, well before the peak; $t_{\rm peak}$, $\Lp$ and $\Lth/\Lp$ change by less than $0.02$ per cent.

At later times, the residual is antisymmetric in viewing angle: the MC viewing-angle contrast decays as the ejecta become optically thin, whereas the analytic contrast does not. This is the expected breakdown of the optically thick description of the RT calculation discussed in Section~\ref{sec:diffusion_framework}: the analytic solution approaches a non-zero quasi-steady angular pattern, whose leading-multipole estimate is given in Appendix~\ref{app:ld_peak}. For the off-centre test, the analytic solution at $t\approx60\,\text{d}$ has an outer-bin spread of $37.8$ per cent, in close agreement with the $38.2$ per cent leading-dipole estimate. For the oblate test, however, the exact analytic solution approaches a pole-on to edge-on spread of $58.8$ per cent, rather than the $\myapprox48$ per cent predicted by the first-order quadrupole expression, $(39/100)\,|X|$ with $X=1.22$ for $\xi_s=0.9245$ and $f=0.6966$ (Section~\ref{sec:limb_darkening}). This difference is due to the first-order expansion in the flattening, rather than to the quasi-steady approximation. Those quasi-steady relations are discussed in Section~\ref{sec:limb_darkening}. This is why the comparison is most meaningful over the optically thick epochs, where the agreement remains at the few-per-cent level.

\bsp    
\label{lastpage}
\end{document}